\documentstyle[12pt]{article}
\makeatletter
\def\@maketitle{\newpage
 \null
 {\normalsize \tt \begin{flushright}
  \begin{tabular}[t]{l} \@date
  \end{tabular}
 \end{flushright}}
 \begin{center}
 \vskip 2em
 {\LARGE \@title \par} \vskip 1.5em {\large \lineskip .5em
 \begin{tabular}[t]{c}\@author
 \end{tabular}\par}
 \end{center}
 \par\vfill\leftline{{\sl \@submitted}}}

\def\abstract{\newpage\if@twocolumn
\section*{Abstract}
\else \small
\begin{center}
{\bf Abstract\vspace{-.5em}\vspace{0pt}}
\end{center}
\quotation
\fi}

\def\endabstract{\if@twocolumn\else\endquotation
\fi\newpage}

\def\@submitted{}
\def\submitted#1{\gdef\@submitted{#1}}

\def\baselinestretch{2}

\def\@setsize#1#2#3#4{\@nomath#1%
   \let\@currsize#1\baselineskip
   #2\baselineskip\baselinestretch\baselineskip
   \parskip\baselinestretch\parskip
   \setbox\strutbox\hbox{\vrule height.7\baselineskip
      depth.3\baselineskip width\z@}%
   \normalbaselineskip\baselineskip#3#4}

\skip\footins 20pt plus4pt minus4pt

\def\@xfloat#1[#2]{\ifhmode \@bsphack\@floatpenalty -\@Mii\else
   \@floatpenalty-\@Miii\fi\def\@captype{#1}\ifinner
      \@parmoderr\@floatpenalty\z@
    \else\@next\@currbox\@freelist{\@tempcnta\csname ftype@#1\endcsname
       \multiply\@tempcnta\@xxxii\advance\@tempcnta\sixt@@n
       \@tfor \@tempa :=#2\do
                        {\if\@tempa h\advance\@tempcnta \@ne\fi
                         \if\@tempa t\advance\@tempcnta \tw@\fi
                         \if\@tempa b\advance\@tempcnta 4\relax\fi
                         \if\@tempa p\advance\@tempcnta 8\relax\fi
         }\global\count\@currbox\@tempcnta}\@fltovf\fi
    \global\setbox\@currbox\vbox\bgroup
    \def\baselinestretch{1}\small\normalsize
    \boxmaxdepth\z@
    \hsize\columnwidth \@parboxrestore}
\long\def\@footnotetext#1{\insert\footins{\def\baselinestretch{1}\footnotesize
    \interlinepenalty\interfootnotelinepenalty
    \splittopskip\footnotesep
    \splitmaxdepth \dp\strutbox \floatingpenalty \@MM
    \hsize\columnwidth \@parboxrestore
   \edef\@currentlabel{\csname p@footnote\endcsname\@thefnmark}\@makefntext
    {\rule{\z@}{\footnotesep}\ignorespaces
      #1\strut}}}

\def\singlespace{%
\vskip\parskip%
\vskip\baselineskip%
\def\baselinestretch{1}%
\ifx\@currsize\normalsize\@normalsize\else\@currsize\fi%
\vskip-\parskip%
\vskip-\baselineskip%
}

\def\spacing#1{\par%
 \def\baselinestretch{#1}%
 \ifx\@currsize\normalsize\@normalsize\else\@currsize\fi}

\everydisplay{
   \abovedisplayskip \baselinestretch\abovedisplayskip%
   \belowdisplayskip \abovedisplayskip%
   \abovedisplayshortskip \baselinestretch\abovedisplayshortskip%
   \belowdisplayshortskip  \baselinestretch\belowdisplayshortskip}

\newif\if@defeqnsw \@defeqnswtrue

\def\eqnarray{\stepcounter{equation}\let\@currentlabel=\theequation
\if@defeqnsw\global\@eqnswtrue\else\global\@eqnswfalse\fi
\tabskip\@centering\let\\=\@eqncr
$$\halign to \displaywidth\bgroup\hfil\global\@eqcnt\z@
  $\displaystyle\tabskip\z@{##}$&\global\@eqcnt\@ne
  \hfil$\displaystyle{{}##{}}$\hfil
  &\global\@eqcnt\tw@ $\displaystyle{##}$\hfil
  \tabskip\@centering&\llap{##}\tabskip\z@\cr}

\@namedef{eqnarray*}{\@defeqnswfalse\global\@eqnswfalse\eqnarray}
\@namedef{endeqnarray*}{\endeqnarray}

\def\yesnumber{\global\@eqnswtrue}

\def\@@eqncr{\let\@tempa\relax\global\advance\@eqcnt by \@ne
    \ifcase\@eqcnt \def\@tempa{& & & &}\or \def\@tempa{& & &}\or
     \def\@tempa{& &}\or \def\@tempa{&}\else\fi
     \@tempa \if@eqnsw\@eqnnum\stepcounter{equation}\fi
     \if@defeqnsw\global\@eqnswtrue\else\global\@eqnswfalse\fi
     \global\@eqcnt\z@\cr}

\def\@eqnacr{{\ifnum0=`}\fi\@ifstar{\@yeqnacr}{\@yeqnacr}}

\def\@yeqnacr{\@ifnextchar [{\@xeqnacr}{\@xeqnacr[\z@]}}

\def\@xeqnacr[#1]{\ifnum0=`{\fi}\cr \noalign{\vskip\jot\vskip #1\relax}}

\def\eqalign{\null\,\vcenter\bgroup\openup1\jot \m@th \let\\=\@eqnacr
\ialign\bgroup\strut
\hfil$\displaystyle{##}$&$\displaystyle{{}##}$\hfil\crcr}
\def\endeqalign{\crcr\egroup\egroup\,}

\def\cases{\left\{\,\vcenter\bgroup\normalbaselines\m@th \let\\=\@eqnacr
    \ialign\bgroup$##\hfil$&\quad##\hfil\crcr}
\def\endcases{\crcr\egroup\egroup\right.}

\def\eqalignno{\stepcounter{equation}\let\@currentlabel=\theequation
\if@defeqnsw\global\@eqnswtrue\else\global\@eqnswfalse\fi
\let\\=\@eqncr
$$\displ@@ \tabskip\@centering \halign to \displaywidth\bgroup
  \global\@eqcnt\@ne\hfil
  $\@lign\displaystyle{##}$\tabskip\z@skip&\global\@eqcnt\tw@
  $\@lign\displaystyle{{}##}$\hfil\tabskip\@centering&
  \llap{\@lign##}\tabskip\z@skip\crcr}

\def\endeqalignno{\@@eqncr\egroup
      \global\advance\c@equation\m@ne$$\global\@ignoretrue}

\def\displ@@{\global\dt@ptrue\openup\jot\m@th % Remove \penalty from \displ@y
  \everycr{\noalign{\ifdt@p \global\dt@pfalse
      \vskip-\lineskiplimit \vskip\normallineskiplimit \fi}}}

\@namedef{eqalignno*}{\@defeqnswfalse\eqalignno}
\@namedef{endeqalignno*}{\endeqalignno}

\def\eqaligntwo{\stepcounter{equation}\let\@currentlabel=\theequation
\if@defeqnsw\global\@eqnswtrue\else\global\@eqnswfalse\fi
\let\\=\@eqncr
$$\displ@@ \tabskip\@centering \halign to \displaywidth\bgroup
  \global\@eqcnt\m@ne\hfil
  $\@lign\displaystyle{##}$\tabskip\z@skip&\global\@eqcnt\z@
  $\@lign\displaystyle{{}##}$\hfil\qquad&\global\@eqcnt\@ne
  \hfil$\@lign\displaystyle{##}$&\global\@eqcnt\tw@
  $\@lign\displaystyle{{}##}$\hfil\tabskip\@centering&
  \llap{\@lign##}\tabskip\z@skip\crcr}

\def\endeqaligntwo{\@@eqncr\egroup
      \global\advance\c@equation\m@ne$$\global\@ignoretrue}

\@namedef{eqaligntwo*}{\@defeqnswfalse\eqaligntwo}
\@namedef{endeqaligntwo*}{\endeqaligntwo}

\newtoks\@stequation

\def\subequations{\refstepcounter{equation}%
  \edef\@savedequation{\the\c@equation}%
  \@stequation=\expandafter{\theequation}%   %only want \theequation
  \edef\@savedtheequation{\the\@stequation}% %expanded once
  \edef\oldtheequation{\theequation}%
  \setcounter{equation}{0}%
  \def\theequation{\oldtheequation\alph{equation}}}

\def\endsubequations{%
  \ifnum\c@equation < 2 \@warning{Only \the\c@equation\space subequation
    used in equation \@savedequation}\fi
  \setcounter{equation}{\@savedequation}%
  \@stequation=\expandafter{\@savedtheequation}%
  \edef\theequation{\the\@stequation}%
  \global\@ignoretrue}

\newbox\strutboxa
\def\@setsize#1#2#3#4{\@nomath#1\let\@currsize#1\baselineskip
   #2\setbox\strutboxa\hbox{\vrule height.7\baselineskip
      depth.3\baselineskip width\z@}\baselineskip\baselinestretch\baselineskip
   \normalbaselineskip\baselineskip#3#4}
\def\struta{\relax\ifmmode\copy\strutboxa\else\unhcopy\strutboxa\fi}

\def\big#1{{\hbox{$\left#1\vcenter to1.428\ht\strutboxa{}\right.\n@space$}}}
\def\Big#1{{\hbox{$\left#1\vcenter to2.142\ht\strutboxa{}\right.\n@space$}}}
\def\bigg#1{{\hbox{$\left#1\vcenter to2.857\ht\strutboxa{}\right.\n@space$}}}
\def\Bigg#1{{\hbox{$\left#1\vcenter to3.571\ht\strutboxa{}\right.\n@space$}}}

\def\@eqnnum{\mbox{\rm (\theequation)}}

\makeatother
\newcommand{\unit}{1}

\newcommand{\tr}{\mbox{tr}}
\newcommand{\Tr}{\mbox{Tr}}
\newcommand{\FT}{{\cal FT}}
\newcommand{\VEV}[1]{\langle #1 \rangle}

\newcommand{\Lag}{{\cal L}}
\newcommand{\Ln}{\mbox{Ln}}

\newcommand{\dfrac}[2]{\frac{\strut \displaystyle{#1}}
                      {\strut \displaystyle{#2}}}
\newcommand{\EQ}[1]{Eq.(\ref{#1})}
\newcommand{\EQQ}[2]{Eqs.(\ref{#1}--\ref{#2})}
\newcommand{\mapright}[1]{
  \smash{\mathop{\hbox to 2em{\rightarrowfill}}\limits^{#1}}}
\newcommand{\mapleft}[1]{
  \smash{\mathop{\hbox to 2em{\leftarrowfill}}\limits_{#1}}}

\def\fsl#1{\setbox0=\hbox{$#1$}           % set a box for #1
   \dimen0=\wd0                                 % and get its size
   \setbox1=\hbox{/} \dimen1=\wd1               % get size of /
   \ifdim\dimen0>\dimen1                        % #1 is bigger
      \rlap{\hbox to \dimen0{\hfil/\hfil}}      % so center / in box
      #1                                        % and print #1
   \else                                        % / is bigger
      \rlap{\hbox to \dimen1{\hfil$#1$\hfil}}   % so center #1
      /                                         % and print /
   \fi}                                         %

\newcommand{\NJLd}{\mbox{$\mbox{NJL}_{D<4}$}}

\makeatletter
\@addtoreset{equation}{section}
\makeatother
\renewcommand{\theequation}{\thesection.\arabic{equation}}
\advance\oddsidemargin by -\textwidth
\advance\oddsidemargin by -1in
\evensidemargin=\oddsidemargin
\newcommand{\Ep}{p_{\scriptscriptstyle E}^2}
\newcommand{\Ek}{k_{\scriptscriptstyle E}^2}
\newcommand{\Eq}{q_{\scriptscriptstyle E}^2}
\newcommand{\TOP}[1]{{\rm T}\!\!\left[#1\right]}
\newcommand{\SxSB}{S$\chi$SB}
\newcommand{\tA}{{A}}
\newcommand{\tM}{{\bar M}}
\newcommand{\tR}{{\bar R}}
\newcommand{\tZ}{{\bar Z}}
\newcommand{\coeA}{\zeta_\omega}
\newcommand{\coeAd}{\zeta_D}
\newcommand{\coeC}{\xi_\omega}
\newcommand{\coeCd}{\xi_D}
\def\unit{{\mathchoice {\rm 1\mskip-4mu l} {\rm 1\mskip-4mu l} {\rm
1\mskip-4.5mu l} {\rm 1\mskip-5mu l}}}
\spacing{1.2}
\title{
  Renormalization in the Gauged Nambu-Jona-Lasinio Model
}
\author{
  {\sc Kei-ichi Kondo}\thanks{
  E-mail address: {\tt kondo@tansei.cc.u-tokyo.ac.jp}
  } \\
  {\it Department of Physics, Chiba University} \\
  {\it Chiba 263, Japan}
  \vspace{1.2em} \\
  {\sc Masaharu Tanabashi}\thanks{
    Fellow of the Japan Society for the Promotion of Science.
    }
  {}\thanks{E-mail address: {\tt tanabash@theory.kek.jp}
  } \\
  {\it National Laboratory for High Energy Physics (KEK)} \\
  {\it Tsukuba, Ibaraki 305, Japan}
  \vspace{1.2em} \\
  {\sc Koichi Yamawaki}\thanks{
    E-mail address: {\tt b42060a@nucc.cc.nagoya-u.ac.jp}
  } \\
  {\it Department of Physics, Nagoya University} \\
  {\it Nagoya 464-01, Japan}
}
\date{
  KEK-TH-344  \\
  KEK preprint 92-86 \\
  CHIBA-EP-65 \\
  DPNU-92-23 \\
  October 1992
}
\submitted{Submitted to Prog. Theor. Phys.}
\begin{document}
% title page
\begin{singlespace}
\maketitle
\newpage
\begin{abstract}
  Based on the Cornwall-Jackiw-Tomboulis effective potential, we extensively
  study nonperturbative renormalization of the gauged Nambu-Jona-Lasinio
  model in the ladder approximation with standing gauge coupling. Although
  the pure Nambu-Jona-Lasinio
  model is not renormalizable, presence of the gauge interaction makes it
  possible that the theory is renormalized as an interacting continuum
  theory at the critical line in the ladder approximation.
  Extra higher dimensional operators (``counter terms'') are not needed for
  the theory to be renormalized.
  By virtue of the effective potential approach, the renormalization
  (``symmetric renormalization'') is performed in a phase-independent
  manner both for the symmetric and the
  spontaneously broken phases of the chiral symmetry.
  We explicitly obtain $\beta$ function having a nontrivial
  ultraviolet fixed line for the renormalized coupling as well as the
  bare one. In both phases the anomalous dimension is very large ($ \ge 1$)
  without discontinuity across the fixed line. Operator product expansion
  is explicitly constructed, which is consistent with the
  large anomalous dimension owing to the appearance of the nontrivial
  extra power behavior in the Wilson coefficient for the unit operator.
The symmetric renormalization
  breaks down at the critical gauge coupling, which is cured by the
  generalized renormalization scheme (``$\tM$-dependent renormalization'').
  Also emphasized is the formal resemblance to the four-fermion theory in less
  than four dimensions which is renormalizable in $1/N$ expansion.
\end{abstract}
\end{singlespace}
\newpage

\section{Introduction}
This is an expanded version of our previous paper\cite{kn:KTY91} on the
renormalization of the gauged Nambu-Jona-Lasinio (NJL) models (gauge
theories plus NJL-type \cite{kn:NJL61} four-fermion interactions).

The gauged NJL models have recently become very popular in the context of
modern versions\cite{kn:Yama89-} of the dynamical electroweak symmetry
breaking such as the walking technicolor\cite{kn:Hold85-}, technicolor
models with strong coupling ETC\cite{kn:YBM86,kn:MY89-}, top quark condensate
model (top mode standard model)\cite{kn:MTY89-}, etc.,
and also in the context of a possible existence of nontrivial (interacting)
QED\cite{kn:Mira85,kn:Kond92-}.
Among others the most important feature of the gauged NJL model is a very
large anomalous dimension\cite{kn:MY89}
\begin{equation}
  1\leq \gamma_m< 2,
\label{eq:(1.1)}
\end{equation}
which corresponds to a very slowly damping behavior of the fermion dynamical
mass,
\begin{equation}
  \Sigma{(-p^2)} \sim (-p^2)^{-1+ \gamma_m/2}.
\label{eq:(1.2)}
\end{equation}

The simplest version of the gauged NJL model, quenched QED plus chiral
invariant four-fermion interaction
$(G/2)[(\bar \psi \psi)^2 + (\bar \psi i \gamma_5 \psi)^2]$, was first studied
by Bardeen, Leung and Love\cite{kn:BLL86} in the ladder Schwinger-Dyson (SD)
equation.
A full set of spontaneous chiral symmetry breaking (\SxSB) solutions of
the ladder SD equation and the critical line were discovered
by Kondo, Mino and Yamawaki\cite{kn:KMY89} and independently by
Appelquist, Soldate, Takeuchi and Wijewardhana\cite{kn:ASTW88}.
The critical line reads(Fig.~\ref{fig:phase});
\begin{eqnarray}
     g   &=& \frac{1}{4}\left( 1+ \sqrt{1-\frac{\alpha}{\alpha_c}} \right)^2
             \equiv g^{*} \quad  (0<\alpha<\alpha_c = \frac{\pi}{3}),\\
  \alpha &=& \alpha_c = \frac{\pi}{3} \quad (g<\frac{1}{4}),
\label{eq:(1.4)}
\end{eqnarray}
where $\alpha \equiv e^2/4\pi$ and $g \equiv G\Lambda^2/4\pi^2$, with
$\Lambda$ being the ultraviolet cutoff.
This is the line separating the \SxSB\ phase ($g > g^{*}$) and
the unbroken (symmetric) phase ($g < g^{*}$).

The critical line \EQ{eq:(1.4)} is actually the nontrivial
ultraviolet (UV) fixed line \cite{kn:KMY89,kn:NSY89,kn:KHKD90},
with the line of $\alpha = $ constant being identified
as the renormalization-group (RG) flow.
This identification is in accord with the usual expectation
that the gauge coupling $\alpha$ may not be renormalized in the absence of
the vacuum polarization in the ladder approximation.
Actually, this model, having non-running gauge coupling, may be
regarded\cite{kn:KSY91}
as the ``standing'' (non-running) limit of the ``walking'' (slowly running)
gauge theories plus four-fermion interaction in the ``improved''
ladder approximation (ladder SD equation with the gauge coupling simply
replaced by the one-loop running
one)\cite{kn:MY89,kn:MNY89,kn:Take89,kn:KSY91,kn:BKS91}.\footnote{
For the effects of the vacuum polarization in QED plus four-fermion
interaction see Ref.\cite{kn:Kond92-}}
Once the RG flow is so identified, the scaling
relation\cite{kn:KMY89,kn:ASTW88}
\begin{equation}
  \frac{M_d}{\Lambda} \sim
    \left( \frac{g - g^{*}}{g - \tilde g^{*}}
        \right)^{\frac{1}{2\sqrt{1-\alpha/\alpha_c}}}
          \qquad   (g > g^{*})
\label{eq:(1.5)}
\end{equation}
implies an explicit form of the $\beta$ function for
$g$\cite{kn:BLL89,kn:IMO89}:
\begin{equation}
  \beta_g (g,\alpha) \equiv
     \Lambda  \frac{\partial{g}}{\partial{\Lambda}}|_{\alpha,M_d}
       =-2(g - g^{*})(g - \tilde g^{*})   \qquad (g > g^{*}),
\label{eq:(1.6)}
\end{equation}
with $M_d \equiv \Sigma(0)$ and $\tilde g^{*} \equiv
\frac{1}{4} \left( 1 - \sqrt{1-\alpha/\alpha_c} \right)^2$.
\EQ{eq:(1.6)} indeed has a nontrivial UV fixed line at
$g = g^{*}$.

The very large anomalous dimension was in fact found at the UV fixed line
by Miransky and Yamawaki\cite{kn:MY89}:
\begin{equation}
  \gamma_m = 1 + \sqrt{1-\frac{\alpha}{\alpha_c}} \qquad (g=g^{*}),
\label{eq:(1.7)}
\end{equation}
which corresponds to the slowly damping \SxSB\ solutions obtained
by Refs.\cite{kn:KMY89,kn:ASTW88}:
\begin{equation}
  \Sigma{(-p^2)} \sim \left( -p^2 \right)^{-(1-
        \sqrt{1-\frac{\alpha}{\alpha_c}})/2}.
\label{eq:(1.8)}
\end{equation}
It was further suggested by Miransky and Yamawaki\cite{kn:MY89} that
such a large anomalous dimension $\gamma_m \geq 1$ would imply the
(nonperturbative) renormalizability of the four-fermion
interaction, since the four-fermion operators would then become
relevant/marginal,
$d_{(\bar \psi \psi)^2} = 2d_{\bar \psi \psi} = 2(3-\gamma_m) \leq 4$, in the
ladder approximation. Moreover, due to the {\em presence of
gauge coupling $\alpha \neq 0$ ($\gamma_m <2$)} the
theory might have a nontrivial (interacting) continuum limit
$\Lambda \rightarrow \infty$ in contrast to the pure NJL model with
$\alpha = 0$ ($\gamma_m =2$)\cite{kn:Yama90-,kn:KHKD90,kn:KSY91}.

However, the above {\em nonperturbative} renormalization procedure,
originally proposed by Miransky\cite{kn:Mira85} in the ladder QED (without
four-fermion interaction), has so far been made only for the {\em bare}
couplings and for the {\em \SxSB\ phase}.
We wish to find the RG property in terms of
the {\em renormalized} couplings and in the {\em symmetric phase} as well.
Indeed, running of the bare couplings based on such a renormalization
in the restricted coupling space may not correspond to the conventional
$\beta$ function of the renormalized
couplings in the continuum theory, unless the RG flow is
correctly identified.

Furthermore, the above renormalization procedure
was crucially based on the nontrivial (\SxSB) solution
of the SD gap equation for the fermion mass function, which is no longer
possible in the symmetric phase where
the gap equation has only a trivial solution. This would yield an
identically vanishing $\beta$ function for the bare couplings
in the symmetric phase (non-running for $g$ as well as $\alpha$).
Accordingly, the anomalous dimension in the symmetric phase of
the gauged NJL model was considered to be small
($\gamma_m = 1 - \sqrt{1-\frac{\alpha}{\alpha_c}} < 1 $)\cite{kn:BLL86}
{\em even in the vicinity of the UV fixed line}, which is contrasted with
that of the \SxSB\ phase
($\gamma_m = 1 + \sqrt{1-\frac{\alpha}{\alpha_c}} > 1$)\cite{kn:MY89},
thus implying a paradoxical discontinuity of the anomalous dimension across
the UV fixed line.

However, it was pointed out by Kikukawa and Yamawaki\cite{kn:KY90} that
such a discontinuity would be an artifact of non-running treatment
of the four-fermion coupling $g$ in the symmetric phase.\footnote{
  The nontrivial scaling behavior near the critical line in the
  symmetric phase was
  also suggested through the gap equation for the pure NJL model in
  Ref.\cite{kn:Mira90}.
}
A possible resolution was in fact demonstrated\cite{kn:KY90}
in the four-fermion theory in $D$ $(2<D<4)$ dimensions
(to be generically denoted by \NJLd\ hereafter) where nonperturbative
renormalization can be explicitly done in the $1/N$ expansion\cite{kn:RWP90}.
Through the renormalization of the fermion
four-point function (auxiliary field
propagator) as well as the two-point function (fermion propagator), one
obtains {\em running} of the coupling {\em in the symmetric phase} as well
as the \SxSB\ phase. The $\beta$ function does have a nontrivial UV fixed point
not only for the bare coupling but for the {\em renormalized} coupling.
This in fact gives rise to
{\em a large anomalous dimension ($\gamma_m=D-2$)} near the UV fixed point,
thus filling in the would-be discontinuity of the anomalous
dimension across the UV fixed point in \NJLd\cite{kn:KY90}.
The large anomalous
dimension does exist not only for the bare coupling but also
for the {\em renormalized} coupling in the continuum theory.
 Such a large anomalous dimension was in fact explicitly
shown\cite{kn:KY90} to be consistent with the operator product
expansion (OPE). Most remarkably,
the Wilson coefficient for the unit operator does have an extra nontrivial
power behavior (other than the anomalous dimension) due to nonperturbative
effects.

In the previous paper\cite{kn:KTY91} we showed that
{\em thanks to the presence of gauge interactions ($\alpha \ne 0$)},
the gauged NJL model can be {\em renormalized} in the ladder approximation
in a quite similar fashion to \NJLd.
The crucial point was that the renormalization was done through the effective
potential {\em in the symmetric as well as the \SxSB\ phase} in a
{\em phase-independent} manner, in contrast to the earlier works
based on the SD gap equation.
To demonstrate such an advantage of the effective potential approach,
we first reformulated the renormalization procedure of Ref.\cite{kn:KY90}
for \NJLd\ {\em through the effective potential}
(Similar reformulation for \NJLd\ was also made by Ref.\cite{kn:HKWY91}).
 Then, for the gauged NJL model in four dimensions we considered the
Cornwall-Jackiw-Tomboulis (CJT) effective potential\cite{kn:CJT74} and rewrote
it only in terms of the local auxiliary fields {\it \'{a} la}
Bardeen and Love\cite{kn:BL92}. This effective potential, an analogue of the
effective potential for \NJLd, was then renormalized in a very similar manner
to \NJLd\ and was explicitly written in terms of the renormalized parameters
of the continuum limit theory ($\Lambda \rightarrow \infty$).
Remarkably enough, as in \NJLd\, the auxiliary field propagator\cite{kn:ATW91}
was shown to be simultaneously renormalized through the above renormalization
of the effective potential. We explicitly computed the $\beta$ function for $g$
for the {\em renormalized} as well as the bare
coupling, which in either case has a nontrivial UV fixed point for each
$\alpha$ (fixed line in $(\alpha, g)$ plane). In either case we obtained
 {\em a large anomalous dimension both in the \SxSB\ and the symmetric phases}
  without paradoxical discontinuity across the UV fixed line, in accord with
Ref.\cite{kn:KY90}. As in \NJLd\ the OPE was explicitly given in a consistent
manner with such a large anomalous dimension {\em in both phases}.
 As in \NJLd\ the Wilson coefficient for the unit operator acquires an extra
nontrivial power
 behavior other than the anomalous dimension.

In this paper we present detailed description of the results of
Ref.\cite{kn:KTY91} on the renormalization of
the simplest gauged NJL model, gauge theories with standing gauge coupling
plus four-fermion interaction, in the ladder approximation. As a basis of
our analysis we consider the CJT effective potential written in terms of the
auxiliary fields. Here we give a more general form than the simplest
one (the BL form\cite{kn:BL92}) discussed in Ref.\cite{kn:KTY91}.
By use of this effective potential, it is shown in the \SxSB\
phase that all the amputated multi-fermion  Green functions at zero momentum
are {\em finite} at the critical line (including
the end point $\alpha = \alpha_c$) in the continuum limit $\Lambda \rightarrow
\infty$. We then give an explicit procedure to renormalize this effective
potential in that limit. This renormalization is possible owing to the
{\em presence of gauge interactions ($\alpha \ne 0$)}. The $\beta$ function
and the large anomalous dimension are obtained through this renormalization
both in the symmetric and the \SxSB\ phases. The OPE is explicitly
constructed, which is consistent with the large anomalous dimension
in both phases. Corresponding to the generalized form of the effective
potential, we consider a generalization (``$\tM$-dependent renormalization'')
of the simplest renormalization scheme (``symmetric
renormalization'')\cite{kn:KTY91} made on the symmetric vacuum. Wilson
coefficients and RG functions, etc. are calculated in the $\tM$-dependent
renormalization as well as in the symmetric renormalization.
 In particular, whereas
the symmetric renormalization breaks down at the end point
$\alpha = \alpha_c$ of the critical line,
the $\tM$-dependent renormalization still remains valid there.

The paper is organized as follows. In the next section we derive the
CJT effective potential and its variants of the gauged NJL model (in the
equivalent Yukawa form rewritten in terms of the local auxiliary fields)
in the ladder approximation. In section~3 all the amputated multi-fermion
Green functions at zero momentum are explicitly calculated from the effective
potential in the \SxSB\ phase and shown to be finite even at the end point
$\alpha = \alpha_c$ of the critical line in the continuum
limit $\Lambda \rightarrow \infty$. Section~4 critically reviews
calculation of the auxiliary field propagator made by Appelquist et
al.\cite{kn:ATW91}. Then in section~5 we present an explicit procedure
of renormalization (symmetric renormalization done on the symmetric vacuum)
which is made through the effective potential (BL effective
potential\cite{kn:BL92})
in an analogous manner to the renormalization of \NJLd\cite{kn:KY90}.
 In section~6 explicit construction of OPE is given, which is shown to be
consistent with the large anomalous dimension in both the symmetric and the
\SxSB\ phases in a quite nontrivial manner: The Wilson coefficient
for the unit operator possesses an extra power behavior other than the
anomalous dimension. Section~7 is devoted to the $\tM$-dependent
renormalization, a generalization  of the symmetric renormalization, which
remains valid at  the end point $\alpha = \alpha_c$ where the symmetric
renormalization breaks down. Section~8 is the
conclusion and discussion: We comment on the ``renormalizabilty'' of
\NJLd\ and the gauged NJL model in the language of the usual RG equation
of the equivalent Yukawa model.
In Appendix A the effective potential and auxiliary field propagators
in \NJLd\ are given.
Appendix B presents OPE in NJL$_{D<4}$.
In Appendices C and D we present RG study of the bare
parameters {\it \'{a} la} Miransky through the SD equation and through the
effective potential, respectively.

\section{Effective Potentials of Gauged NJL model}
Let us start with the lagrangian of the $SU(N)$ gauge theory plus
NJL-type four-fermion interaction:
\begin{equation}
  {\cal L} =
    \bar\psi  (i \fsl{\partial}-e \fsl{A}) \psi - m_0 \bar\psi\psi
   +\frac{G}{2N} \left[
                  (\bar\psi \psi )^2 + (\bar\psi i\gamma _5 \psi)^2
                \right]
   -\frac{1}{2} \tr(F_{\mu \nu }F^{\mu \nu }),
\label{eq:(2.1)}
\end{equation}
where $m_0$ is the bare fermion mass, $e$ the gauge coupling constant,
and $G$ the four-fermion coupling.
By using auxiliary fields $\sigma$, $\pi$,
\EQ{eq:(2.1)} is cast into an equivalent lagrangian
\begin{equation}
  {\cal L} =
    \bar\psi (i \fsl{\partial}-e \fsl{A}) \psi
    - \bar\psi (\sigma +\pi i\gamma _5) \psi - V_{\rm (cl)}(\sigma ,\pi )
    - \frac{1}{2} \tr(F_{\mu\nu}F^{ \mu \nu }),
\label{eq:(2.2)}
\end{equation}
where the classical part of the potential of auxiliary fields
 $\sigma$ and $\pi$ is given by
\begin{equation}
  V_{\rm (cl)}(\sigma ,\pi )  =
         \frac{1}{G} \left[
                     \frac{N}{2}\left(\sigma ^2+\pi ^2\right) - m_0 \sigma
                     \right].
\label{eq:(2.3)}
\end{equation}
It is more convenient to study \EQ{eq:(2.2)} than \EQ{eq:(2.1)} for the
discussions of renormalization.

The point is that as is demonstrated in \NJLd \cite{kn:KTY91},
the renormalization is studied most transparently through
the effective action (potential) written only in terms
of the auxiliary fields. Such an effective potential in the gauged NJL model
is derived from \EQ{eq:(2.2)} by integrating out the degrees of freedom of
fermion ($\psi$) and gauge boson ($A_\mu$).
An effective potential of this kind was first derived by
Bardeen and Love (BL)\cite{kn:BL92} through discussion of the vacuum
condensate of fermion composite operator.
Here we take an alternative approach\cite{kn:KTY91} using the effective
action of Cornwall, Jackiw and Tomboulis (CJT)\cite{kn:CJT74,kn:Haym91}
 and derive systematically several variants of
the CJT effective potential including the BL effective potential as a special
case.\footnote{
It was also noted in Ref.\cite{kn:Mira92} that the BL potential can be derived
from the CJT effective potential.
}
 We thus clarify the relation among various kinds of effective potentials
including the BL potential.

In the CJT formalism\cite{kn:CJT74} we introduce a bilocal external source
in a similar manner to the usual external source terms:
\begin{equation}
  \Lag_{\rm source} = - \bar\psi (x) J(x,y) \psi (y),
\end{equation}
in the generating functional:
\begin{equation}
  W[J; \sigma,\pi] =
    - i \ln \int [d\psi][d\bar \psi] [{\rm gauge}]
            \exp[i \int d^4 x ({\cal L}+{\cal L_{\rm source}})].
\label{eq:(2.5)}
\end{equation}
Corresponding to the bilocal external source term $J$,
the fermion propagator $S$ becomes a variational variable in the CJT
effective action through the Legendre transformation,
\begin{eqnarray}
 \Gamma [S; \sigma ,\pi ]
   &\equiv& W[J; \sigma,\pi] - \Tr(J \cdot S)
  \nonumber\\
   &=& -i\Tr\left(\Ln S^{-1} + S_0^{-1}  S \right) + \kappa ^{\rm 2PI}[S]
       - \int d^4 x V_{\rm (cl)}(\sigma ,\pi ),
\label{eq:(2.6)}
\end{eqnarray}
where $\kappa^{\rm 2PI}[S]$ is the sum of all the two-particle (fermion)
irreducible diagrams written in terms of $S$, and $S_0$ is
the function of $\sigma $,
\begin{equation}
  iS_0^{-1} = i \fsl{\partial} - \sigma -\pi i\gamma _5.
\end{equation}

The stationary condition for $S$ gives the SD equation of the fermion
propagator
\begin{equation}
  0 = J \equiv i{\delta \Gamma \over \delta S}
    = -S^{-1} + S_0^{-1} + i\dfrac{\delta \kappa ^{\rm 2PI}}{\delta S}.
\label{eq:(2.8)}
\end{equation}
Since the SD equation \EQ{eq:(2.8)} can be regarded as the condition for
the bilocal external source to vanish, \EQ{eq:(2.6)} reads
\begin{equation}
  \Gamma[S_{\rm sol}; \sigma, \pi] = W[J=0; \sigma, \pi],
\end{equation}
with $S_{\rm sol}$ being the solution of \EQ{eq:(2.8)}. This in fact yields
the desired effective action. Note that $S_{\rm sol}$ depends on the
auxiliary fields $\sigma$ and $\pi$.

The Yukawa-type vertex $\Gamma_S$ is given by:
\begin{equation}
  \Gamma_S(x,y; z) = -i \frac{\delta}{\delta \sigma(z)} S^{-1}_{\rm sol}(x,y),
\label{eq:(2.10)}
\end{equation}
which satisfies the following SD equation:
\begin{eqnarray}
  \Gamma_S(x,y; z) &=&
  \delta^{(4)}(x-y) \delta^{(4)}(y-z)
  \nonumber\\
  & &+\left. \dfrac{i \delta^2 \kappa^{\rm 2PI}[S]}{\delta S(y,x) \delta
S(x',y')}
     \right|_{S=S_{\rm sol}}\!\!
   S_{\rm sol} (x',x'') \Gamma_S(x'',y''; z) S_{\rm sol}(y'',y'),
\end{eqnarray}
where integration over the repeated indices $x'$,$x''$,$y'$,$y''$ is
understood.

Now, noting the translational invariance, $S(x,y)=S(x-y)$,
$\sigma(x)=\sigma$ and $\pi(x)=\pi$, we define the CJT effective potential
$V$;
\begin{equation}
  V[S; \sigma,\pi)
    = -\Gamma[S,\sigma={\rm const}, \pi={\rm const}] / \Omega,
\label{eq:(2.12)}
\end{equation}
with $\Omega$ being the space-time volume.
Then \EQ{eq:(2.12)} may be rewritten as
\begin{equation}
  V[S;\sigma,\pi)
  = V_{\rm (cl)}(\sigma,\pi) + V_{\rm (qu)}[S;\sigma,\pi),
\end{equation}
where the quantum part $V_{\rm (qu)}[S;\sigma,\pi)$ is given by
\begin{equation}
  -\Omega V_{\rm (qu)}[S;\sigma,\pi)
    = -i\Tr(\Ln S^{-1}+S_0^{-1}S)+\kappa^{\rm 2PI}[S].
\label{eq:(2.14)}
\end{equation}
\subsection{Ladder approximation}
For actual calculation we need to make an approximation
for $\kappa^{\rm 2PI}[S]$ which contains infinite number of diagrams.
Here we consider the simplest choice, namely,
the lowest diagram (two-loop diagram) depicted in Fig.~\ref{fig:CJT}
\footnote{
  Our effective potential is defined so as to keep the auxiliary fields
  not path-integrated out and hence does not include
  the ``two-loop'' graph with the $\sigma, \pi$ line sitting on the diameter
  of the fermion line circle. Or, even if we included such a diagram, it
  would vanish identically anyway, because $\sigma, \pi$ are not propagating
  at this stage.
}.
This corresponds to the {\em ladder} approximation with the
{\em fixed} gauge coupling. Although the ladder approximation is not a
systematic expansion, it picks up at least the leading log behavior of
fermion mass function which is actually important for our present purpose to
renormalize the gauged NJL model. Since the running effects of the
gauge coupling is left out of account in this approximation,
it is certainly not a good approximation for the QCD-like gauged NJL model
with a {\em normal running} gauge coupling. However, it can be
regarded\cite{kn:KSY91} as the {\em standing} limit of the {\em walking}
gauge theories (gauge theories with a slowly running coupling) plus
four-fermion interactions. For $N=1$ this approximation also corresponds to
the quenched QED plus NJL-type four-fermion interaction.

In the ladder approximation we may parameterize the fermion propagator
$S$ (in Landau gauge) by
\begin{equation}
  iS^{-1}(p) = \fsl{p} - \Sigma(-p^2) -i\gamma_5\Sigma_5(-p^2).
\end{equation}
Then we evaluate each term of the quantum part of the CJT effective
potential \EQ{eq:(2.14)}:
\begin{eqnarray}
  \dfrac{\Tr\Ln S^{-1}}{i \Omega N}
    &=& \int ^{\Lambda}\dfrac{d^4p}{(2\pi)^4 i} \tr\ln S^{-1}(p)
    \nonumber\\
    &=& \dfrac{1}{8\pi^2}\int _0^{\Lambda^2} d\Ep \Ep
        \ln(1+\dfrac{\Sigma^2(\Ep)+\Sigma_5^2(\Ep)}{\Ep}),
\label{eq:(2.16)}
    \\
  \dfrac{\Tr(S_0^{-1} S)}{i \Omega N}
    &=& -i\int^{\Lambda} \dfrac{d^4p}{(2\pi)^4 i}
       \tr\left[
          (\fsl{p}-\sigma-i\gamma_5 \pi) S(p)
       \right]
    \nonumber\\
    &=& \dfrac{1}{4\pi^2}\int  _0^{\Lambda^2} d\Ep \Ep
       \left[
          \dfrac{\sigma\Sigma(\Ep)+\pi\Sigma_5(\Ep)}
                 {\Ep+\Sigma^2(\Ep)+\Sigma_5^2(\Ep)}
         -\dfrac{\Sigma^2(\Ep)+\Sigma_5^2(\Ep)}
            {\Ep+\Sigma^2(\Ep)+\Sigma_5^2(\Ep)}
       \right],
      \hspace{1em}\phantom{a}
\label{eq:(2.17)}
\end{eqnarray}
\begin{eqnarray}
  \dfrac{\kappa^{\rm 2PI}[S]}{ \Omega N}
    &=& 2\pi C_F \alpha i \int  ^{\Lambda} \dfrac{d^4 p}{(2\pi)^4 i}
        \dfrac{d^4 k}{(2\pi)^4 i}
           \tr\left[\gamma^\mu S(p) \gamma^\nu S(k)\right]D_{\mu \nu}(p-k),
    \nonumber\\
    &=& \dfrac{1}{8\pi^2} \int_0^{\Lambda^2} d\Ep \Ep
                          \int_0^{\Lambda^2}d\Ek \Ek
        \dfrac{\Sigma(\Ep)\Sigma(\Ek)+\Sigma_5(\Ep)\Sigma_5(\Ek)}
              {(\Ep+\Sigma^2+\Sigma_5^2)(\Ek+\Sigma^2+\Sigma_5^2)} K(\Ep,\Ek),
      \hspace{1em}\phantom{a}
\label{eq:(2.18)}
\end{eqnarray}
where the Euclidean momentum integral $(p_E^2 \equiv -p^2)$ is regularized by
the ultraviolet cutoff $\Lambda$,
$C_F$ is the quadratic Casimir of the fermion representation, and
the gauge boson propagator $D_{\mu \nu}$ in Landau gauge takes the form
\(
  D_{\mu \nu}(q) = (-i/q^2)(g_{\mu \nu} - q_\mu q_\nu /q^2),
\)
and
\begin{equation}
  K(\Ep,\Ek) \equiv  \dfrac{3 C_F \alpha/ 4\pi}{\max(\Ep,\Ek)}.
\end{equation}
We have defined our effective potential by subtracting a variable-independent
divergence from \EQ{eq:(2.14)} at the origin
$\Sigma=\Sigma_5=0$, $\sigma=\pi=0$:
$
  V_{\rm (qu)} [\Sigma=0,\Sigma_5=0;\sigma=0,\pi=0) = 0.
$

Plugging \EQQ{eq:(2.16)}{eq:(2.18)} into
\EQ{eq:(2.14)}, we obtain
\begin{eqnarray}
\lefteqn{ -\dfrac{4\pi^2}{N} V [\Sigma,\Sigma_5;\sigma,\pi) }  \nonumber \\
  &=& -\dfrac{\Lambda^2}{g}
       \left[{1 \over 2}(\sigma^2+\pi^2) - m_0 \sigma \right]
  \nonumber \\
  & & + \int_0^{\Lambda^2} d\Ep \Ep \left\{
          \dfrac{1}{2} \ln\left(1+\dfrac{\Sigma^2+\Sigma_5^2}{\Ep}\right)
         -\dfrac{\Sigma^2+\Sigma_5^2}{\Ep+\Sigma^2+\Sigma_5^2}
         +\dfrac{\sigma\Sigma+\pi\Sigma_5}{\Ep+\Sigma^2+\Sigma_5^2}
        \right\}
\nonumber \\
& & + \dfrac{1}{2} \int  _0^{\Lambda^2} d\Ep \Ep \int  _0^{\Lambda^2}d\Ek \Ek
      \dfrac{\Sigma(\Ep)\Sigma(\Ek)+\Sigma_5(\Ep)\Sigma_5(\Ek)}
            {(\Ep+\Sigma^2+\Sigma_5^2)(\Ek+\Sigma^2+\Sigma_5^2)} K(\Ep,\Ek),
\label{eq:(2.20)}
\end{eqnarray}
with
\begin{equation}
 g \equiv \frac{G \Lambda^2}{4\pi^2},
\end{equation}
where we included the classical part \EQ{eq:(2.3)}.
It should be noted that our cutoff regularization does not
violate the chiral symmetry:
\begin{equation}
  \left( \begin{array}{c} \sigma \\ \pi \end{array} \right) \rightarrow
  \left(\begin{array}{cc}
    \cos\theta, & \sin\theta \\
   -\sin\theta, & \cos\theta
  \end{array}\right)
  \left(\begin{array}{c} \sigma \\ \pi \end{array} \right),
\qquad
  \left(\begin{array}{c} \Sigma \\ \Sigma_5 \end{array}\right) \rightarrow
  \left(\begin{array}{cc}
     \cos\theta, & \sin\theta \\
    -\sin\theta, & \cos\theta
  \end{array}\right)
  \left(\begin{array}{c} \Sigma \\ \Sigma_5 \end{array}\right).
\label{eq:(2.22)}
\end{equation}
\subsection{$V[\Sigma,\Sigma_5]$}
Starting with the CJT effective potential \EQ{eq:(2.20)},
we now investigate its variants in what follows.
We first derive a form of the CJT potential written only in terms of
dynamical mass $\Sigma$ and $\Sigma_5$, by eliminating $\sigma$ and $\pi$
through their stationary conditions.
The stationary condition of the effective potential for auxiliary fields gives
\begin{subequations} \label{eq:(2.23)}
\begin{eqnarray}
  0&=&   \dfrac{4\pi^2}{N}
         \dfrac{\partial}{\partial \sigma} V[\Sigma,\Sigma_5;\sigma,\pi)
  \nonumber\\
  &=& -\int_0^{\Lambda^2} d\Ep \dfrac{\Ep \Sigma}{\Ep+\Sigma^2+\Sigma_5^2}
         +\dfrac{\Lambda^2}{g}(\sigma-m_0),
\label{eq:(2.23a)}
   \\
   0&=& \dfrac{4\pi^2}{N}
   \dfrac{\partial}{\partial \pi} V[\Sigma,\Sigma_5;\sigma,\pi)
   \nonumber\\
   &=& -\int_0^{\Lambda^2} d\Ep \dfrac{\Ep \Sigma_5}{\Ep+\Sigma^2+\Sigma_5^2}
       +\dfrac{\Lambda^2}{g}\pi.
\label{eq:(2.23b)}
\end{eqnarray}
\end{subequations}
Plugging the solution of \EQ{eq:(2.23)} back into
$V[\Sigma,\Sigma_5; \sigma,\pi)$, we obtain the effective potential written
only in terms of the mass function of the fermion:
\begin{eqnarray}
\lefteqn{- \dfrac{4\pi^2}{N} V[\Sigma,\Sigma_5] } \nonumber \\
&=& \int  _0^{\Lambda^2} d\Ep \Ep \left\{
       \dfrac{1}{2} \ln(1+\dfrac{\Sigma^2+\Sigma_5^2}{\Ep})
      -\dfrac{\Sigma^2+\Sigma_5^2}{\Ep+\Sigma^2+\Sigma_5^2}
      +\dfrac{m_0 \Sigma}{\Ep+\Sigma^2+\Sigma_5^2}
    \right\}
\nonumber \\
& & + \dfrac{1}{2} \int  _0^{\Lambda^2} d\Ep \Ep \int  _0^{\Lambda^2}d\Ek \Ek
      \dfrac{\Sigma(\Ep)\Sigma(\Ek)+\Sigma_5(\Ep)\Sigma_5(\Ek)}
            {(\Ep+\Sigma^2+\Sigma_5^2)(\Ek+\Sigma^2+\Sigma_5^2)}
            \left[K(\Ep,\Ek)+\dfrac{g}{\Lambda^2}\right].
      \hspace{1em}\phantom{a}
\label{eq:(2.24)}
\end{eqnarray}

This type of CJT potential was obtained by Nonoyama, Suzuki
and Yamawaki\cite{kn:NSY89} directly from the original lagrangian
\EQ{eq:(2.1)} with the lowest $\kappa^{\rm 2PI}$ being given by
Fig.~\ref{fig:CJT2}.
Actually, the stationary condition of this effective potential \EQ{eq:(2.24)}
leads to the usual ladder SD gap equation for the fermion mass function
\EQ{eq:(C.7)} in Appendix C.

\subsection{$V[\Sigma^{\rm sol},\Sigma_5^{\rm sol};\sigma,\pi)$}
We now derive explicit form of the effective potential written solely in terms
of the auxiliary fields through several intermediate steps described in
this and the next two subsections.

As such an intermediate step we first obtain
a variant of the CJT effective potential $V[\Sigma^{\rm sol},\Sigma_5^{\rm
sol};\sigma,\pi)$ by plugging the solution
$\Sigma^{\rm sol}$ and $\Sigma_5^{\rm sol}$ of the stationary conditions,
$\delta V/\delta\Sigma = 0$, $\delta V/\delta\Sigma_5 = 0$,
back into $V[\Sigma,\Sigma_5,\sigma,\pi)$. The stationary conditions
(still not ``gap equations'') are equivalent to the ladder SD equations
\begin{subequations}
\begin{eqnarray}
 \Sigma(\Ep) &=&
   \sigma+
   \int  _0^{\Lambda^2}d\Ek\dfrac{\Ek \Sigma(\Ek)}{\Ek+\Sigma^2+\Sigma_5^2}
     K(\Ep,\Ek),
\label{eq:(2.25a)}
  \\
 \Sigma_5(\Ep) &=&
   \pi+
   \int  _0^{\Lambda^2}d\Ek \dfrac{\Ek \Sigma_5(\Ek)}{\Ek+\Sigma^2+\Sigma_5^2}
     K(\Ep,\Ek),
\label{eq:(2.25b)}
\end{eqnarray}
\end{subequations}
which are reduced to \EQ{eq:(C.5&6)},
the stationary condition of \EQ{eq:(2.24)},
when $\sigma$ and $\pi$ are eliminated by use of \EQ{eq:(2.23)}.

By using the chiral symmetry, we can always rotate
$\Sigma,\Sigma_5,\sigma$ and $\pi$
so as to set $\Sigma_5=\pi=0$, where $\Sigma$ and $\Sigma_5$ are understood
to be the solution of \EQQ{eq:(2.25a)}{eq:(2.25b)}.
Thus, it is sufficient to study the case of $\Sigma_5=0$ and
$\pi=0$. The SD equation \EQ{eq:(2.25a)} now reads
\begin{equation}
 \Sigma(\Ep) = \sigma+\int_0^{\Lambda^2}d\Ek\dfrac{\Ek \Sigma(\Ek)}
 {\Ek+\Sigma^2}K(\Ep,\Ek).
\label{eq:(2.26)}
\end{equation}

For actual evaluation of $V[\Sigma^{\rm sol}, \Sigma_5=0; \sigma, \pi=0)$
it is useful to note\cite{kn:GM87,kn:NSY89}
that $V[\Sigma, 0, \sigma,0)$ obeys a simple scaling
relation:
\begin{equation}
  V_{\rm (qu)} [\Sigma_\kappa, 0; \sigma,0; \Lambda) =
    \kappa^4 V_{\rm (qu)}[\Sigma, 0; \sigma/\kappa, 0; \Lambda/\kappa),
\label{eq:(2.27)}
\end{equation}
with
$
  \Sigma_\kappa(\Ep) \equiv  \kappa \Sigma(\Ep/\kappa^2),
$
where we made explicit the $\Lambda$-dependence of $V_{\rm (qu)}$.
Taking $\kappa$ derivative of \EQ{eq:(2.27)} at $\kappa =1$, we find
\begin{equation}
  V_{\rm (qu)} [\Sigma^{\rm sol}, 0; \sigma, 0; \Lambda)
   = \frac{1}{4} \left[2\Lambda^2 \dfrac{\partial}{\partial \Lambda^2}
                  +\sigma \dfrac{\partial}{\partial \sigma}
       \right]V_{\rm (qu)} [\Sigma^{\rm sol}, 0; \sigma, 0; \Lambda),
\label{eq:(2.28)}
\end{equation}
where we have used
$
 \delta V_{\rm (qu)}/ \delta\Sigma|_{\Sigma=\Sigma^{\rm sol}} =0.
$
By using quantum part of \EQ{eq:(2.20)} and \EQ{eq:(2.28)}, we obtain
\begin{eqnarray}
\lefteqn{- \frac{4\pi^2}{N}  V_{\rm (qu)} [\Sigma^{\rm sol},0;\sigma,0;
\Lambda)}
\nonumber \\
&=& \frac{\Lambda^4}{2} \left\{
       \dfrac{1}{2} \ln\left(1+\dfrac{\Sigma_\Lambda^2}{\Lambda^2}\right)
      -\dfrac{\Sigma_\Lambda^2}{\Lambda^2+\Sigma_\Lambda^2}
      +\dfrac{\sigma\Sigma_\Lambda}{\Lambda^2+\Sigma_\Lambda^2}
    \right\}
\nonumber \\
& & + \dfrac{3C_F }{8\pi} \alpha \dfrac{\Lambda^2
\Sigma_\Lambda}{\Lambda^2+\Sigma_\Lambda^2}
      \int  _0^{\Lambda^2}d\Ek \dfrac{\Ek \Sigma^{\rm sol}(\Ek)}
{\Ek+(\Sigma^{\rm sol})^2}
    + \dfrac{\sigma}{4} \int_0^{\Lambda^2} d\Ep
                  \dfrac{\Ep \Sigma^{\rm sol}(\Ep)}{\Ep+(\Sigma^{\rm sol})^2},
\label{eq:(2.29)}
\end{eqnarray}
where
\begin{equation}
  \Sigma_\Lambda\equiv  \Sigma^{\rm sol}(\Ep=\Lambda^2).
\end{equation}
Putting $\Ep=\Lambda^2$ in  the SD equation \EQ{eq:(2.26)},
we obtain
\begin{equation}
 \Sigma_\Lambda -\sigma
   =  \dfrac{3C_F}{4\pi} \dfrac{\alpha}{\Lambda^2}
     \int_0^{\Lambda^2}d\Ek\dfrac{\Ek \Sigma^{\rm sol}(\Ek)}
                             {\Ek+(\Sigma^{\rm sol})^2}.
\label{eq:(2.31)}
\end{equation}
Plugging \EQ{eq:(2.31)} into \EQ{eq:(2.29)},
we obtain another expression for the CJT effective potential
\begin{eqnarray}
\lefteqn{
  -\dfrac{4\pi^2}{N}  V[\Sigma^{\rm sol},0; \sigma,0; \Lambda)
} \nonumber \\
  &=&  - {\Lambda^2 \over g} \left[ {1 \over 2} \sigma^2 -m_0 \sigma \right]
       + \frac{\Lambda^4}{4}
         \ln\left(1+\dfrac{\Sigma_\Lambda^2}{\Lambda^2}\right)
       + \dfrac{4\pi}{3C_F} \dfrac{\Lambda^2}{\alpha}
         \dfrac{(\Sigma_\Lambda-\sigma)\sigma}{4},
\label{eq:(2.32)}
\end{eqnarray}
where the classical part \EQ{eq:(2.3)} was included. Note that
$\Sigma_\Lambda$ is a function of $\sigma$ as determined by the SD equation
\EQ{eq:(2.26)} and hence \EQ{eq:(2.32)} can in principle be written only
in terms of $\sigma$.

\subsection{Fermion mass function}
In order to solve $\Sigma_\Lambda$ as an explicit function of $\sigma$, we now
discuss the solution $\Sigma^{\rm sol}$ of the SD equation
\EQ{eq:(2.26)}\cite{kn:KMY89,kn:ASTW88}.
\EQ{eq:(2.26)} is equivalent to the differential equation
\begin{equation}
  \left[
     \Ep \left(\dfrac{d}{d\Ep}\right)^2
   + 2\dfrac{d}{d\Ep}
    +\frac{3C_F}{4\pi} \dfrac{\alpha}{\Ep+\Sigma^2(\Ep)}
  \right]\Sigma(\Ep) = 0,
\label{eq:(2.33)}
\end{equation}
with infrared (IR) boundary condition (BC):
\begin{equation}
  \lim_{\Ep \rightarrow 0} p_{\scriptscriptstyle E}^4
\dfrac{d}{d\Ep} \Sigma(\Ep)
= 0, \label{eq:(2.34)}
\end{equation}
and ultraviolet (UV) BC:
\begin{equation}
  \left.\left[1+\Ep \dfrac{d}{d\Ep} \right]\Sigma(\Ep)\right|_{\Ep=\Lambda^2}
= \sigma.
\label{eq:(2.35)}
\end{equation}

In high energy region $\Ep \gg \Sigma^2(\Ep)$,
the differential equation \EQ{eq:(2.33)} can be safely approximated
by the linearized equation
\begin{equation}
 0= \left[
     \Ep \left(\dfrac{d}{d\Ep}\right)^2
   + 2\dfrac{d}{d\Ep}
    +\frac{3C_F}{4\pi} \dfrac{\alpha}{\Ep}
  \right]\Sigma(\Ep)
  + {\cal O}( \dfrac{\Sigma^3(\Ep)}{(\Ep)^2}).
\label{eq:(2.36)}
\end{equation}
Then the solution of \EQ{eq:(2.36)} is written by the linear
combination of two independent solutions:
\begin{equation}
  \frac{\Sigma(\Ep)}{M} = c_1 \left(\dfrac{\Ep}{M^2}\right)^{-(1-\omega )/2}
            +d_1 \left(\dfrac{\Ep}{M^2}\right)^{-(1+\omega )/2}
            + {\cal O} (\left(\dfrac{\Ep}{M^2}\right)^{-3(1-\omega)/2-1} ),
\label{eq:(2.37)}
\end{equation}
where $M$ is an infrared scale of fermion mass function
and
\begin{equation}
  \omega \equiv \sqrt{1-\frac{\alpha}{\alpha_c}},
\end{equation}
with
\begin{equation}
  \alpha_c \equiv \dfrac{\pi}{3C_F}.
\end{equation}

The UVBC \EQ{eq:(2.35)} determines the scale $M$
{\em as a function of} $\sigma$:
\begin{equation}
  \dfrac{\sigma}{\Lambda}
  = C_1 \left( \dfrac{M}{\Lambda} \right)^{2-\omega}
  + D_1 \left( \dfrac{M}{\Lambda} \right)^{2+\omega}
  + {\cal O}(\left( {M \over \Lambda} \right)^{3(2-\omega)} ),
\label{eq:(2.40)}
\end{equation}
where $C_1$ and $D_1$ are defined by
\begin{equation}
  C_1 \equiv  \left( \dfrac{1}{2}+\dfrac{\omega }{2} \right) c_1, \qquad
  D_1 \equiv  \left( \dfrac{1}{2}-\dfrac{\omega }{2} \right) d_1.
\label{eq:(2.41)}
\end{equation}

For strong gauge coupling region $\alpha>\alpha_c$, $\omega$ becomes
pure imaginary and the fermion mass function becomes oscillating:
\begin{equation}
 \frac{\Sigma(\Ep)}{M} = \dfrac{\tA'}{\omega'}  \sqrt{\dfrac{M^2}{\Ep}}
                \sin\left[\frac{\omega'}{2}\ln\frac{\Ep}{M^2} +\omega'\delta'
                    \right]
               + {\cal O}(\left(\frac{\Ep}{M^2}\right)^{-5/2}),
\label{eq:(2.42)}
\end{equation}
where
\begin{equation}
  \omega' \equiv \sqrt{\frac{\alpha}{\alpha_c} -1}
\end{equation}
and
\begin{equation}
 \tA' = 2 \omega' \sqrt{c_1 d_1}, \qquad
 \delta' = \frac{1}{2i\omega'} \ln \left(-\frac{c_1}{d_1} \right).
\end{equation}
The coefficients $c_1$ and $d_1$ are complex conjugate to each other
so as to guarantee that the fermion mass is real.
The mass scale $M$ defined in \EQ{eq:(2.40)} becomes multivalued function of
$\sigma$ for $\alpha>\alpha_c$ due to the
oscillating behavior of the fermion mass function.
We then take $M$ with the largest absolute value (no-node solution),
since it minimizes the effective potential.
Note that the SD equation has a nontrivial solution even for $\sigma=0$ in
this region \cite{kn:MN74}.

In the weak gauge coupling region $0< \alpha <\alpha_c$, on the other hand,
the fermion mass function may be written as
\begin{equation}
  \frac{\Sigma(\Ep)}{M} =
  \frac{\tA}{\omega}  \sqrt{\dfrac{M^2}{\Ep}}
  \sinh\left[
    \frac{\omega}{2} \ln \dfrac{\Ep}{M^2} + \omega \delta
  \right]
  +{\cal O}(\left(\dfrac{\Ep}{M^2}\right)^{-3(1-\omega)/2-1}),
\label{eq:(2.45)}
\end{equation}
where
\begin{equation}
  \tA = 2\omega \sqrt{-c_1 d_1}, \qquad
  \delta = \frac{1}{2\omega} \ln \left( -\frac{c_1}{d_1} \right).
\label{eq:(2.46)}
\end{equation}

The coefficients $c_1$ and $d_1$ in \EQ{eq:(2.37)} are determined by the
IRBC \EQ{eq:(2.34)}. However, the non-linearity in the infrared region
makes it difficult to calculate them in an analytical method.
In the following, we evaluate $c_1$ and $d_1$ by using linearizing
techniques of the ladder SD equation.
Although the result varies slightly according to the choice of such a
linearizing technique, we will find in sections \ref{sec-ren}, \ref{sec-ope}
and \ref{sec-moreren} that the structure of
the renormalization  does not depend on such a detail of $c_1$ and $d_1$.

There exist two familiar linearizing techniques of the ladder
SD equation.
One is to replace the $\Sigma(\Ep)$ in the denominator of \EQ{eq:(2.26)}  by
$M$ \cite{kn:Mira80,kn:NSY89,kn:IMO89}:
\begin{equation}
 \Sigma(\Ep) = \sigma + \int_0^{\Lambda^2} d\Ek K(\Ep,\Ek)
{\Ek \Sigma(\Ek) \over \Ek+M^2}.
\label{eq:(2.47)}
\end{equation}
Such a linearization leads to the solution:
\begin{eqnarray}
 \frac{\Sigma(\Ep)}{M}
  &=& F({1 \over 2}+{\omega \over 2},{1 \over 2}-{\omega \over 2}, 2;
              {-\Ep \over M^2})
\nonumber\\
&=&  \sum_{n=1}^\infty c_n
\left({\Ep \over M^2} \right)^{-(1-\omega)/2+1-n}
+ \sum_{n=1}^\infty d_n
\left({\Ep \over M^2} \right)^{-(1+\omega)/2+1-n},
\end{eqnarray}
with
\begin{eqnarray}
  c_n &=& \dfrac{(-1)^{n-1} \Gamma(\omega )}
                {\Gamma(\frac{1}{2}+\frac{\omega
}{2})\Gamma(\frac{3}{2}+\frac{\omega }{2})}
          \dfrac{ \Gamma({1 \over 2}+{\omega \over 2}+n-1)
                  \Gamma(-{1 \over 2}+{\omega \over 2}+n-1) \Gamma(-\omega+1)}
                {\Gamma({1 \over 2}+{\omega \over 2})
                 \Gamma(-{1 \over 2}+{\omega \over 2}) \Gamma(-\omega +n)},
\nonumber\\
  d_n &=& c_n(\omega \rightarrow -\omega).
\end{eqnarray}
In particular, we find
\begin{equation}
  c_1 = \frac{1}{\omega}
        \dfrac{\Gamma(1+\omega )}
{\Gamma(\frac{1}{2}+\frac{\omega}{2})\Gamma(\frac{3}{2}+\frac{\omega}{2})},
  \qquad
  d_1 =-\frac{1}{\omega}
        \dfrac{\Gamma(1-\omega )}
{\Gamma(\frac{1}{2}-\frac{\omega}{2})\Gamma(\frac{3}{2}-\frac{\omega}{2})},
\end{equation}
which lead to
\begin{equation}
  \tA = \sqrt{8\omega \cot ({\pi \over 2}\omega) \over
  \pi (1-\omega^2)}, \qquad
  \delta = \frac{1}{2\omega} \ln\left[
             \dfrac{\Gamma(1+\omega)}{\Gamma(1-\omega)}
             \dfrac{\Gamma(\frac{1}{2}-\frac{\omega}{2})
                    \Gamma(\frac{3}{2}-\frac{\omega}{2})}
                   {\Gamma(\frac{1}{2}+\frac{\omega}{2})
                    \Gamma(\frac{3}{2}+\frac{\omega}{2})}
           \right],
\end{equation}
for $0<\alpha<\alpha_c$ and
\begin{equation}
  \tA' = \sqrt{8\omega' \coth ({\pi \over 2}\omega') \over
  \pi (1+\omega'^2)},  \qquad
  \delta' = \frac{1}{2i\omega'}\ln\left[
             \dfrac{\Gamma(1+i\omega')}{\Gamma(1-i\omega')}
             \dfrac{\Gamma(\frac{1}{2}-\frac{i\omega'}{2})
                    \Gamma(\frac{3}{2}-\frac{i\omega'}{2})}
                   {\Gamma(\frac{1}{2}+\frac{i\omega'}{2})
                    \Gamma(\frac{3}{2}+\frac{i\omega'}{2})}
           \right],
\end{equation}
for $\alpha>\alpha_c$. Here we took a normalization $\Sigma(\Ep = 0)=M$.
It should be noted that this linearizing method overestimates the order
of error in \EQ{eq:(2.37)}.

Another method is the bifurcation
technique\cite{kn:Atki87,kn:KMY89}.  As a result of the bifurcation theory of
the non-linear integral equation, the bifurcation solution from the trivial one
satisfies the integral equation which is obtained by
 ignoring the $\Sigma$ in the
denominator and placing the infrared cutoff $M$ in \EQ{eq:(2.26)}:
\begin{equation}
 \Sigma(\Ep) = \sigma + \int_{M^2}^{\Lambda^2} d\Ek K(\Ep,\Ek) \Sigma(\Ek).
\end{equation}
The normalization of the solution is given by
$
  \Sigma(\Ep=M^2) \equiv  M.
$
By using the bifurcation technique, we obtain
\begin{equation}
  c_1 = {1+\omega \over 2\omega}, \qquad
  d_1 = -{1-\omega \over 2\omega},
\end{equation}
which leads to
\begin{subequations}
\begin{equation}
  \tA = \sqrt{1-\omega^2}, \qquad
  \delta = \frac{1}{\omega} \tanh^{-1} \omega,\qquad
  (0<\alpha<\alpha_c),
\end{equation}
\begin{equation}
  \tA' = \sqrt{1+\omega'{}^2}, \qquad
  \delta' = \frac{1}{\omega'} \tan^{-1} \omega', \qquad
  (\alpha>\alpha_c).
\end{equation}
\end{subequations}

Note that $\tA$, $\delta$ evaluated by these linearizing
methods are finite in $\omega \rightarrow 0$ ($\omega' \rightarrow 0$):
\begin{subequations}
\begin{equation}
  \tA_0 \equiv \lim_{\omega\rightarrow0} \tA = \frac{4}{\pi},
  \qquad
  \delta_0 \equiv \lim_{\omega\rightarrow0} \delta = \ln4-1,
\end{equation}
for the linearization \EQ{eq:(2.47)}, and
\begin{equation}
  \tA_0 \equiv \lim_{\omega\rightarrow0} \tA = 1,
  \qquad
  \delta_0 \equiv \lim_{\omega\rightarrow0} \delta = 1,
\end{equation}
\end{subequations}
for the bifurcation method.

At $\alpha=\alpha_c$ the fermion mass function can be obtained directly from
\EQ{eq:(2.36)} plus IRBC \EQ{eq:(2.34)}. It can also be obtained by
taking $\omega \rightarrow 0$ ($\omega' \rightarrow 0$) limit of
\EQ{eq:(2.45)} (\EQ{eq:(2.42)}):
\begin{equation}
  {\Sigma(\Ep) \over M}
  = \tA_0   \sqrt{\frac{M^2}{\Ep}}
                \left[ \frac{1}{2} \ln \dfrac{\Ep}{M^2} + \delta_0 \right]
               +  {\cal O}(\left(\frac{\Ep}{M^2}\right)^{-5/2}).
\label{eq:(2.58)}
\end{equation}
In this limit \EQ{eq:(2.40)} reads
\begin{equation}
  \frac{\sigma}{\Lambda} =
    \frac{\tA_0}{2} \left(\frac{M}{\Lambda}\right)^2
    \left[ 1+\delta_0 -\ln\frac{M}{\Lambda} \right]
   +{\cal O}(\left(\frac{M}{\Lambda}\right)^6),
\label{eq:(2.59)}
\end{equation}
where we have used the following relation derived from \EQ{eq:(2.46)}
and \EQ{eq:(2.41)}:
\begin{equation}
  \lim_{\omega\rightarrow 0} \omega C_1
  = -\lim_{\omega\rightarrow 0} \omega D_1 = \frac{\tA_0}{4},
  \qquad
  \lim_{\omega\rightarrow 0} (C_1+D_1) = \frac{\tA_0}{2}(\delta_0+1).
\label{eq:(2.60)}
\end{equation}

\subsection{$V(M)$}
Let us now return to the CJT effective potential \EQ{eq:(2.32)}.
By using the explicit solution $\Sigma^{\rm sol}(\Ep)$ of the SD equation
\EQ{eq:(2.37)}, we find:
\begin{equation}
  \dfrac{\Sigma_\Lambda}{\Lambda}
  = c_1 \left( \dfrac{M}{\Lambda} \right)^{2-\omega }
  + d_1 \left( \dfrac{M}{\Lambda} \right)^{2+\omega }
  + {\cal O}(\left({M \over \Lambda}\right)^{3(2-\omega)}).
\label{eq:(2.61)}
\end{equation}
Plugging \EQ{eq:(2.61)} and \EQ{eq:(2.40)} into \EQ{eq:(2.32)},
we obtain an effective potential solely expressed in terms of $M$:
\begin{eqnarray*}
  - 8\pi^2 \dfrac{V(M)}{N \Lambda^4}
  &=& \left( \dfrac{1}{g^*}-\dfrac{1}{g} \right)
      C_1\left[
        C_1 \left(\dfrac{M}{\Lambda}\right)^{4-2\omega }
       +\dfrac{2+\omega }{2} D_1 \left(\dfrac{M}{\Lambda}\right)^4
      \right]
  \\
  & &+\left( \dfrac{1}{\tilde g^*}-\dfrac{1}{g} \right)
      D_1\left[
        D_1 \left(\dfrac{M}{\Lambda}\right)^{4+2\omega }
       +\dfrac{2-\omega }{2} C_1 \left( \dfrac{M}{\Lambda} \right)^4
      \right] +{\cal O}( \left(\dfrac{M}{\Lambda} \right)^{4(2-\omega)})
  \\
  & & + \dfrac{2m_0}{\Lambda g}
      \left[
        C_1 \left(\dfrac{M}{\Lambda}\right)^{2-\omega }
       +D_1 \left(\dfrac{M}{\Lambda}\right)^{2+\omega }
       +{\cal O}(\left(\dfrac{M}{\Lambda}\right)^{3(2-\omega)}) \right],
  \yesnumber
  \label{eq:(2.62)}
\end{eqnarray*}
where $C_1$ and $D_1$ are defined in \EQ{eq:(2.41)} and
\begin{eqnarray}
         g^* \equiv \dfrac{1}{4} (1+\omega)^2, \qquad
  \tilde g^* \equiv \dfrac{1}{4} (1-\omega)^2.
\end{eqnarray}

It should be stressed again that $M$ is {\em a function}
of $\sigma$ determined by \EQ{eq:(2.40)}.

\subsection{$V(\sigma,\pi)$}
In this subsection we make more explicit the $\sigma$-dependence of
the effective potential \EQ{eq:(2.62)}.
To this end it is convenient to rewrite \EQ{eq:(2.62)} so as to
leave $M$ partly unsolved:
\begin{eqnarray}
  -\dfrac{4\pi^2}{N} \dfrac{V(\sigma,\pi=0)}{\Lambda^4}
  &=& \frac{1}{g} \dfrac{m_0 \sigma}{\Lambda^2}
     + \left( \frac{1}{g^*}-\frac{1}{g} \right) \dfrac{\sigma^2}{2\Lambda^2}
  \nonumber\\
  & &+ \left( \frac{1}{\tilde g^*}-\frac{1}{g^*} \right)
       \dfrac{\sigma^2}{2\Lambda^2}
          \dfrac{D_1 \left(\frac{M}{\Lambda}\right)^{\omega} \left[
                    D_1 \left(\frac{M}{\Lambda}\right)^{\omega}
                  +\frac{2-\omega}{2} C_1
                    \left(\frac{M}{\Lambda}\right)^{-\omega}
                  \right]}
                 {\left[ C_1 \left(\frac{M}{\Lambda}\right)^{-\omega}
                 +D_1 \left(\frac{M}{\Lambda}\right)^\omega \right]^2}
  \nonumber\\
  & & + {\cal O}(\left(\frac{\sigma}{\Lambda}\right)^4),
\label{eq:(2.64)}
\end{eqnarray}
where we have used \EQ{eq:(2.40)}. This effective potential is actually the
basis for studying the renormalization in this paper, particularly in
section~7.

As a special case of our effective potential \EQ{eq:(2.64)}, we obtain the BL
potential\cite{kn:BL92} derived through a different method, which is valid
only for $1> \omega >0$ ($0< \alpha <\alpha_c$). In this region
\EQ{eq:(2.40)} can be solved in a recursive way,
\begin{equation}
  \left( \dfrac{M}{\Lambda} \right)^{2-\omega}
 = \dfrac{1}{C_1} \dfrac{\sigma}{\Lambda}
  -\dfrac{D_1}{C_1}
   \left( \dfrac{1}{C_1}\dfrac{\sigma}{\Lambda} \right)^{(2+\omega)/(2-\omega)}
  + {\cal O} ( \left( \dfrac{\sigma}{\Lambda} \right)^{\eta-1} ),
\label{eq:(2.65)}
\end{equation}
with $\eta$ being
\begin{equation}
  \eta \equiv \min( 4, \dfrac{2(2+\omega)}{2-\omega} ).
\end{equation}
Plugging \EQ{eq:(2.65)} into \EQ{eq:(2.64)} (or \EQ{eq:(2.62)}), we obtain
\begin{equation}
  -\dfrac{4\pi^2}{N} \dfrac{V(\sigma,\pi=0)}{\Lambda^4} =
     \frac{1}{g} \dfrac{m_0 \sigma}{\Lambda^2}
    +\left( \frac{1}{g^*} - \frac{1}{g} \right) \dfrac{\sigma^2}{2\Lambda^2}
    -\frac{4\coeA}{\alpha/\alpha_c} \dfrac{2-\omega}{4}
     \left(\dfrac{\sigma}{\Lambda}\right)^{4/(2-\omega)}
    + {\cal O}( \left(\frac{\sigma}{\Lambda}\right)^\eta ),
\label{eq:(2.67)}
\end{equation}
where $\coeA$ is defined by:
\begin{equation}
  \coeA \equiv -\omega \left( \frac{2}{1+\omega}\right)^{4/(2-\omega)}
           c_1^{-(2+\omega)/(2-\omega)} d_1 >0.
\end{equation}
It is easy to recover the pseudoscalar auxiliary field $\pi$ in
the effective potential \EQ{eq:(2.67)}:
\begin{eqnarray}
  -\dfrac{4\pi^2}{N} \dfrac{V(\sigma,\pi)}{\Lambda^4}
  &=&
     \frac{1}{g} \dfrac{m_0 \sigma}{\Lambda^2}
    +\left( \frac{1}{g^*} - \frac{1}{g} \right)
     \dfrac{\sigma^2+\pi^2}{2\Lambda^2}
    -\frac{4\coeA}{\alpha/\alpha_c} \dfrac{2-\omega}{4}
     \left(\dfrac{\sigma^2+\pi^2}{\Lambda^2}\right)^{2/(2-\omega)}
  \nonumber\\
  & &\qquad
   + {\cal O}( \left(\frac{\sigma^2+\pi^2}{\Lambda^2}\right)^{\eta/2} ).
\label{eq:(2.69)}
\end{eqnarray}

The expansion \EQ{eq:(2.65)} obviously breaks down at $\alpha=\alpha_c$ where
${\cal O}((\sigma/\Lambda)^{\eta-1})$ term
in \EQ{eq:(2.65)} gives the same order contribution as others.
A remarkable feature of our expression \EQ{eq:(2.64)} for the effective
potential is that it has a definite value in the limit
$\omega \downarrow 0$ $(\alpha \uparrow \alpha_c)$:
\begin{equation}
  - \dfrac{4\pi^2}{N} \dfrac{V(\sigma,\pi=0)}{\Lambda^4}
  = \frac{1}{g}\dfrac{m_0 \sigma}{\Lambda^2}
   +\left( \dfrac{1}{g^{*}} - \frac{1}{g} \right) \dfrac{\sigma^2}{2\Lambda^2}
   -8 \dfrac{\sigma^2}{2\Lambda^2}
    \dfrac{\left[ \frac{3}{4}+\delta_0-\ln\frac{M}{\Lambda} \right]}
          {\left[1+\delta_0 -\ln\frac{M}{\Lambda} \right]^2}
   +{\cal O}(\left(\frac{\sigma}{\Lambda}\right)^4),
\label{eq:(2.70)}
\end{equation}
where we have used \EQ{eq:(2.60)}. Since
$M$ is written in terms of $\sigma$ through
the UVBC at $\alpha=\alpha_c$ \EQ{eq:(2.59)},
the above effective potential can be further expressed in terms of
the auxiliary field:\footnote{
This expression actually coincides with the one obtained directly
from the solution at $\alpha = \alpha_c$.}
\begin{equation}
  - \dfrac{4\pi^2}{N} \dfrac{V(\sigma,\pi=0)}{\Lambda^4}
  = \frac{1}{g}\dfrac{m_0 \sigma}{\Lambda^2}
   +\left( \dfrac{1}{g^{*}} - \frac{1}{g} \right) \dfrac{\sigma^2}{2\Lambda^2}
   -16 \dfrac{\sigma^2}{\Lambda^2}
       \dfrac{1}{\ln\left(\frac{\Lambda^2}{\sigma^2}\right)}
   +{\cal O}\left(
       \dfrac{\ln\ln \left(\frac{\Lambda^2}{\sigma^2}\right)}
             {\left(\ln\left(\frac{\Lambda^2}{\sigma^2}\right)\right)^2}
    \right),
\end{equation}
where we have used the relation
\begin{equation}
  -\ln\frac{M}{\Lambda}
  = \frac{1}{4} \ln\left(\frac{\Lambda^2}{\sigma^2}\right)
   + {\cal O}(\ln\ln \frac{\Lambda^2}{\sigma^2}),
\end{equation}
derived from \EQ{eq:(2.59)}.
Actually, in the limit $\Lambda \gg \sigma$ this potential agrees with
the effective potential derived by Bardeen and Love\cite{kn:BL92}
for $\alpha=\alpha_c$.

\EQ{eq:(2.64)} is also applicable to the strong gauge coupling region
$\alpha>\alpha_c$ by performing the analytic continuation
$\omega=i\omega'$, $\omega' =\sqrt{\alpha/\alpha_c-1}$.

Finally, we comment on the relation of our derivation of the effective
potential \EQ{eq:(2.67)} to that by Bardeen and Love\cite{kn:BL92}
which was derived  based on the observation:
\begin{equation}
  \VEV{\bar\psi\psi} =
  \frac{d}{d\sigma} V_{\rm (qu)}(\sigma,\pi=0).
\label{eq:(2.73)}
\end{equation}
Actually, \EQ{eq:(2.73)} is manifest in the CJT formalism, since
$V_{\rm (qu)}(\sigma,\pi)$ is identified as
$V_{\rm (qu)}[\Sigma^{\rm sol},\Sigma_5^{\rm sol},\sigma,\pi)$ which satisfies
\begin{eqnarray}
\lefteqn{
  \dfrac{d}{d\sigma} V_{\rm (qu)} [\Sigma^{\rm sol}, 0, \sigma, 0)
} \nonumber \\
  &=& \dfrac{\partial}{\partial\sigma}
      V_{\rm (qu)} [\Sigma^{\rm sol}, 0, \sigma, 0)
   +\int_0^{\Lambda^2} d\Ep
    \dfrac{\partial \Sigma^{\rm sol}(\Ep)}{\partial \sigma}
    \left. \dfrac{\delta}{\delta\Sigma(\Ep)} V_{\rm (qu)}[\Sigma, 0, \sigma, 0)
    \right|_{\Sigma=\Sigma^{\rm sol}}
  \nonumber \\
  &\equiv& \VEV{ \bar \psi\psi}
   = -\dfrac{N}{4\pi^2} \int_0^{\Lambda^2} d\Ep \Ep
    \dfrac{\Sigma^{\rm sol}(\Ep)}{\Ep+(\Sigma^{\rm sol})^2},
\label{eq:(2.74)}
\end{eqnarray}
where in the last line we have used the stationary condition
\[
 \left. \dfrac{\delta}{\delta\Sigma(\Ep)} V_{\rm (qu)}[\Sigma, 0, \sigma, 0)
 \right|_{\Sigma=\Sigma^{\rm sol}} =
 \left. \dfrac{\delta}{\delta\Sigma(\Ep)} V[\Sigma, 0, \sigma, 0)
 \right|_{\Sigma=\Sigma^{\rm sol}} = 0.
\]

\subsection{Gap equation}
Now, we are interested in the solution $\sigma=\sigma_{\rm sol}$
of the stationary condition
$0 = \partial V(\sigma,\pi=0)/\partial \sigma$
of the effective potential \EQ{eq:(2.69)}:
\begin{equation}
  0 = \frac{1}{g}\frac{m_0}{\Lambda}
     +\left(\frac{1}{g^*}-\frac{1}{g}\right)\frac{\sigma}{\Lambda}
     -\frac{4\coeA}{\alpha/\alpha_c}
      \left(\frac{\sigma}{\Lambda}\right)^{(2+\omega)/(2-\omega)}
     +\cdots.
\label{eq:(2.75a)}
\end{equation}

The instability of the symmetric vacuum $\VEV{\sigma}=\VEV{\pi}=0$
for $g>g^*$ in the chiral symmetric limit $m_0=0$ is manifest
in the effective potential.
Thus it is readily seen that
\begin{equation}
g =g^{*} \equiv \frac{1}{4}(1+\omega)^2  \qquad (0<\alpha<\alpha_c)
\end{equation}
is the critical line\cite{kn:KMY89,kn:ASTW88}.
Actually, we can read off the \SxSB\ solution
$\sigma=\sigma_{\rm spont}$ from \EQ{eq:(2.75a)} at $m_0=0$.
The scaling at $g \simeq g^*$ of the \SxSB\ solution to \EQ{eq:(2.75a)}
is given by
\begin{equation}
  \frac{\sigma_{\rm spont}}{\Lambda}
  = \left[\frac{\alpha/\alpha_c}{4\coeA}
          \left( \frac{1}{g^*} - \frac{1}{g} \right)
    \right]^{(2-\omega)/2\omega}
   + \cdots,
\label{eq:(2.77)}
\end{equation}
where $\cdots$ stands for terms with higher power of $(1/g^* - 1/g)$.
\EQ{eq:(2.77)} combined with \EQ{eq:(2.65)}
leads to the scaling of dynamical mass of fermion
$M_d = \VEV{M}$:\cite{kn:KMY89,kn:ASTW88}
\begin{equation}
  \frac{M_d}{\Lambda}
  \sim \left( \frac{1}{g^*} - \frac{1}{g} \right)^{1/2\omega}.
\label{eq:(2.78)}
\end{equation}

However, \EQ{eq:(2.69)} is not valid
in the $\omega\rightarrow0$ ($\alpha\rightarrow\alpha_c$) limit,
in which case we need to return to \EQ{eq:(2.64)} or \EQ{eq:(2.62)},
the form before the expansion \EQ{eq:(2.65)} is applied.
The stationary condition of \EQ{eq:(2.62)} with $m_0=0$ leads to
\begin{equation}
  \dfrac{M_d}{\Lambda}
  = \left( -\frac{C_1}{D_1}
            \dfrac{\frac{1}{g^*}-\frac{1}{g}}
                  {\frac{1}{\tilde g^*}-\frac{1}{g}}
    \right)^{1/2\omega}
   + {\cal O} (
    \left( \frac{1}{g^*}-\frac{1}{g}
    \right)^{(5-4\omega)/2\omega}).
\label{eq:(2.79)}
\end{equation}
In the $\omega\rightarrow0$ ($g^{*} \rightarrow 1/4$) limit we find the
 essential
singularity-type scaling:\cite{kn:KMY89,kn:ASTW88}
\begin{equation}
  \frac{M_d}{\Lambda}
  = \exp\left[ 1+\delta_0 - \dfrac{8}{4 - 1/g} \right].
\label{eq:(2.80)}
\end{equation}

\subsection{Yukawa-type vertex}
As shown in \EQ{eq:(2.10)}, the Yukawa-type vertex is calculated by the
$\sigma$ derivative of $S^{-1}$.
Since we have already evaluated the fermion mass function $\Sigma$ under
the ansatz of constant $\sigma$, it is easy to determine the Yukawa-type
vertex at $q=0$:
\begin{equation}
  \Gamma_S(-p^2) \equiv \Gamma_S(p,q=0)
    = \frac{\partial}{\partial \sigma} \Sigma(-p^2),
\end{equation}
where the momentum assignment of Yukawa-type vertex $\Gamma_S(p,q)$
is depicted in Fig~\ref{fig:yukawa}.
Plugging \EQ{eq:(2.65)} into \EQ{eq:(2.37)}, we find
\begin{equation}
  \dfrac{\Sigma(-p^2)}{\Lambda}
  = \frac{2}{1+\omega} \left(\dfrac{-p^2}{\Lambda^2}\right)^{-(1-\omega)/2}
    \frac{\sigma}{\Lambda}
   + {\cal O}(\left(\frac{\sigma}{\Lambda}\right)^{(2+\omega)/(2-\omega)}),
\label{eq:(2.82)}
\end{equation}
which leads to the Yukawa-type vertex on the chiral symmetric vacuum
$\VEV{\sigma}=0$:
\begin{equation}
  \Gamma_S(-p^2) = \left. \frac{\partial}{\partial\sigma} \Sigma(-p^2)
                   \right|_{\sigma=0}
                 = \frac{2}{1+\omega}
                   \left(\dfrac{-p^2}{\Lambda^2}\right)^{-(1-\omega)/2}.
\label{eq:(2.83)}
\end{equation}

In the $\omega\rightarrow0$ ($\alpha\rightarrow \alpha_c$), however,
the expansion of the fermion mass function around the symmetric vacuum
\EQ{eq:(2.82)} loses its validity.
Thus, we need to evaluate the Yukawa-type vertex without employing
such an expansion around the symmetric vacuum at $\omega\rightarrow0$.
Here we calculate the Yukawa-type vertex corresponding to the expansion
around non-zero $\sigma$:
\begin{equation}
  \Gamma_S(-p^2; M)
  \equiv \frac{\partial}{\partial \sigma} \Sigma(-p^2)
  = \dfrac{  \frac{\partial}{\partial M} \Sigma(-p^2) }
          {  \frac{\partial \sigma}{\partial M} }.
\end{equation}

Combining the $M$ derivative of $\sigma$ from \EQ{eq:(2.40)},
\begin{equation}
  \frac{\partial \sigma}{\partial M}=
  (2-\omega) C_1 \left(\frac{M}{\Lambda}\right)^{1-\omega}
           +(2+\omega) D_1 \left(\frac{M}{\Lambda}\right)^{1+\omega}
  +{\cal O}(\left(\frac{M}{\Lambda}\right)^{5-3\omega}),
\label{eq:(2.85)}
\end{equation}
and the $M$ derivative of $\Sigma(-p^2)$ from \EQ{eq:(2.37)},
\begin{equation}
  \frac{\partial \Sigma(-p^2)}{\partial M} =
  \frac{2(2-\omega)}{1+\omega} C_1
    \left(\dfrac{-p^2}{M^2}\right)^{-(1-\omega)/2}
    \hspace{-1.0em}
   +\frac{2(2+\omega)}{1-\omega} D_1
    \left(\dfrac{-p^2}{M^2}\right)^{-(1+\omega)/2}
    \hspace{-1.0em}
   +{\cal O}(\left(\dfrac{-p^2}{M^2}\right)^{-(5-3\omega)/2}
    \hspace{-1em}),
\end{equation}
we obtain
\begin{equation}
 \Gamma_S(-p^2; M) =
      \frac{2}{1+\omega} \left(\dfrac{-p^2}{\Lambda^2}\right)^{-(1-\omega)/2}
      \dfrac{ \left[
             1 +\frac{1+\omega}{1-\omega} K
                 \left(\dfrac{-p^2}{M^2}\right)^{-\omega}
             +{\cal
O}(\left(\dfrac{-p^2}{M^2}\right)^{-2+\omega}\hspace{-0.5em})
              \right]
            }{1 +K \left(\frac{M}{\Lambda}\right)^{2\omega}
          +{\cal O}(\left(\frac{M}{\Lambda}\right)^{2(2-\omega)})},
\label{eq:(2.87)}
\end{equation}
where $K$ is defined by
\begin{equation}
  K\equiv \frac{(2+\omega)D_1}{(2-\omega)C_1}.
\label{eq:(2.88)}
\end{equation}
As expected, \EQ{eq:(2.87)} is indeed applicable to $\omega\rightarrow0$ limit:
\begin{equation}
  \Gamma_S(-p^2; M)
  = 2 \left(\dfrac{-p^2}{\Lambda^2}\right)^{-1/2}
      \dfrac{ -1+2\delta_0 + \ln\left(\frac{-p^2}{M^2}\right)
              + {\cal O}(\left(\frac{-p^2}{M^2}\right)^{-2})
      }{       1+2\delta_0 - \ln\left(\frac{M^2}{\Lambda^2}\right)
      }.
\label{eq:(2.89)}
\end{equation}

\section{Finiteness of Amputated Green Functions}
In this section we calculate from the effective potential all the (amputated)
multi-fermion Green functions at zero momentum of $\sigma$
in the \SxSB\ phase and
show that they are {\em finite} in the continuum limit
$\Lambda\rightarrow \infty$,
once the four-fermion coupling $g$ is fine tuned so as to fix the fermion
dynamical mass  $M_d$.
This strongly suggests the existence of an
explicit renormalization scheme to make the effective potential finite at
the full critical line including the end point $\alpha=\alpha_c$.

\subsection{Green functions for $0<\alpha<\alpha_c$}
Let us first calculate the second derivative of the effective potential
\EQ{eq:(2.64)},
\begin{equation}
  V^{(2)}(\sigma) \equiv
  \left(\dfrac{\partial}{\partial \sigma}\right)^2 V(\sigma,\pi=0),
\label{eq:(3.1)}
\end{equation}
so as to evaluate the auxiliary field propagator at zero-momentum.
By using \EQ{eq:(2.85)}, we obtain
\begin{eqnarray}
 - {4\pi^2 \over N} {V^{(2)}(\sigma) \over \Lambda^2}
 =  {1 \over g^*}-{1 \over g}
  + \left({1 \over \tilde g^*}-{1 \over g^*} \right)
    \dfrac{K \left({M \over \Lambda}\right)^{2\omega}}
          {1 + K \left({M \over \Lambda}\right)^{2\omega}},
\label{eq:(3.2)}
\end{eqnarray}
where $K$ is defined in \EQ{eq:(2.88)} and we have neglected
higher order in $M/\Lambda$.
The solution of the gap equation in chiral limit, \EQ{eq:(2.79)}, reads
\begin{equation}
  \frac{1}{g^*}-\frac{1}{g} =
  -\left(\frac{1}{\tilde g^*}-\frac{1}{g^*}\right)
   \dfrac{(2-\omega)K\left(\frac{M_d}{\Lambda}\right)^{2\omega}}
         {(2+\omega)+(2-\omega)K\left(\frac{M_d}{\Lambda}\right)^{2\omega}}.
\label{eq:(3.3)}
\end{equation}
Thus, \EQ{eq:(3.2)} reads
\begin{equation}
  -\frac{4\pi^2}{N} \dfrac{V^{(2)}(\sigma_{\rm spont})}{\Lambda^2}
  =
   \dfrac{2\omega K\left(\frac{M_d}{\Lambda}\right)^{2\omega}
          \left(\frac{1}{\tilde g^*}-\frac{1}{g^*}\right)}
   {\left[1+K\left(\frac{M_d}{\Lambda}\right)^{2\omega}\right]
    \left[2+\omega+(2-\omega)
K\left(\frac{M_d}{\Lambda}\right)^{2\omega}\right]
   }.
\label{eq:(3.4)}
\end{equation}
The cutoff dependence of $V^{(2)}\sim\Lambda^{2(1-\omega)}$
cancels exactly that of the Yukawa-type vertex \EQ{eq:(2.87)}
and we obtain {\em finite} amputated four-point Green function
in $\Lambda/M_d \rightarrow \infty$ limit:
\begin{eqnarray}
\lefteqn{
  \Gamma_4(p_{(1)},-p_{(1)},p_{(2)},-p_{(2)})
} \nonumber\\
 &=& V^{(2)}(\sigma_{\rm spont})
     \prod_{j=1}^{2}
     \left[
       \dfrac{1}{V^{(2)}(\sigma_{\rm spont})} \Gamma_S(-p_{(j)}^2, M_d)
     \right]
 \nonumber\\
 &=& -\frac{4\pi^2}{N}
      \dfrac{
      \prod_{j=1}^2 \left[
\frac{2}{1+\omega}\left(\dfrac{-p_{(j)}^2}{M_d^2}\right)^{-(1-\omega)/2}
       \hspace{-1.5em}
+\frac{2K}{1-\omega}\left(\dfrac{-p_{(j)}^2}{M_d^2}\right)^{-(1+\omega)/2}
       \hspace{-1.5em}
       +{\cal O}(\left(\frac{-p_{(j)}^2}{M_d^2}\right)^{-(5-3\omega)/2})
       \hspace{-0.2em}
      \right]
      }{\frac{2\omega M_d^2 K}{2+\omega}
        \left( \frac{1}{\tilde g^*}-\frac{1}{g^*}\right)}.
      \hspace{2em}
\label{eq:(3.5)}
\end{eqnarray}

To discuss higher-point Green functions (see Fig.~\ref{fig:amputated}),
we need
to evaluate multi-$\sigma$ vertices.
At zero-momentum of $\sigma$, they are calculated from
$n=m+2$ ($m\ge1$)-th derivative of effective potential:
\begin{equation}
  V^{(n)}(\sigma) \equiv
  \left(\frac{\partial}{\partial\sigma}\right)^n V(\sigma,\pi=0).
\label{eq:(3.6)}
\end{equation}
After a straightforward calculation using \EQ{eq:(2.85)}, we find
\begin{equation}
  -\frac{4\pi^2}{N} V^{(m+2)}(\sigma) =
  \dfrac{ 2\omega K M^{2-m} \left(\frac{M}{\Lambda}\right)^{-(2+m)(1-\omega)}
          \left(\frac{1}{\tilde g^*}-\frac{1}{g^*}\right)
          P_m(\omega,K(M/\Lambda)^{2\omega})
       }{ [(2-\omega) C_1]^m
         \left[1+K\left(\frac{M}{\Lambda}\right)^{2\omega}\right]^{2m+1}},
\label{eq:(3.7)}
\end{equation}
where $P_m(\omega,z)$ is given by
\begin{equation}
  P_m(\omega,z)
  = \sum_{\ell=0}^{m-1}
    Q_{m,\ell}(\omega) (2\omega z)^{\ell} (1+z)^{m-\ell-1},
\label{eq:(3.8)}
\end{equation}
with $Q_{m,\ell}(\omega)$ being certain polynomial in $\omega$ with degree
$m-\ell-1$, e.g.,
\begin{equation}
  Q_{m,0}(\omega) = \begin{cases}
             1                                                     & ($m=1$) \\
            \displaystyle
            \prod_{\ell=2}^m \left((\ell+1)\omega-2(\ell-1)\right) & ($m\ge 2$)
           \end{cases}.
\label{eq:(3.9)}
\end{equation}

Now, we are ready to evaluate the amputated $2n$-point Green function
$(n \ge 3)$ at zero momentum.
It is remarkable to see the cancellation of $\Lambda$ dependence of
\EQ{eq:(2.87)}, \EQ{eq:(3.4)} and \EQ{eq:(3.7)} in the
fermion amputated Green functions.
Actually,
we find the Green functions remain {\em finite} in continuum limit
($\Lambda/M_d\rightarrow\infty$):
\begin{eqnarray}
\lefteqn{
  \Gamma_{2n}(p_{(1)},-p_{(1)},p_{(2)},-p_{(2)}, \cdots, p_{(n)},-p_{(n)})
} \nonumber\\
 &=& V^{(n)}(\sigma_{\rm spont})
     \prod_{j=1}^{n}
     \left[
       \dfrac{1}{V^{(2)}(\sigma_{\rm spont})} \Gamma_S(-p_{(j)}^2, M_d)
     \right]
 \nonumber\\
 &=& \left(-\frac{4\pi^2}{N} \right)^{n-1} Q_{n-2,0}(\omega)
 \nonumber\\
  & & \quad\times
      \dfrac{
      \prod_{j=1}^n \left[
\frac{2}{1+\omega}\left(\dfrac{-p_{(j)}^2}{M_d^2}\right)^{-(1-\omega)/2}
       \hspace{-1.5em}
+\frac{2K}{1-\omega}\left(\dfrac{-p_{(j)}^2}{M_d^2}\right)^{-(1+\omega)/2}
       \hspace{-1.5em}
       +{\cal O}(\left(\frac{-p_{(j)}^2}{M_d^2}\right)^{-(5-3\omega)/2})
       \hspace{-0.2em}
      \right]
      }{
        \left(2\omega C_1\right)^{n-2}
        \frac{2\omega(2-\omega)^{n-2} }{(2+\omega)^{n}}
         M_d^{3n-4} K^{n-1}
        \left(\frac{1}{\tilde g^*}-\frac{1}{g^*}\right)^{n-1}
      }.
      \hspace{2em}
\label{eq:(3.10)}
\end{eqnarray}

\subsection{Green functions at $\alpha=\alpha_c$}
We next consider Green functions at $\alpha=\alpha_c$, i.e., $\omega=0$.
This limit should be taken carefully, since
the Yukawa-type vertex \EQ{eq:(2.89)} and the effective potential
depend on $\Lambda$ not only in power of the cutoff but also in logarithm
of the cutoff.
As shown in the following, however, finiteness of the amputated Green functions
persists also in this limit, thanks to the cancellation
of logarithmic dependence on $\Lambda$.

Taking $\omega\rightarrow0$ limit of \EQ{eq:(3.4)} and \EQ{eq:(3.7)},
we find
\begin{equation}
  \frac{4\pi^2}{N} V^{(2)}(\sigma_{\rm spont})
   =\dfrac{16\Lambda^2}{
      \left[1+2\delta_0-\ln \left(\frac{M_d^2}{\Lambda^2}\right)\right]
      \left[2+2\delta_0-\ln \left(\frac{M_d^2}{\Lambda^2}\right)\right]
    }
\label{eq:(3.11)}
\end{equation}
and
\begin{equation}
  \frac{4\pi^2}{N} V^{(m+2)}(\sigma)
  = \dfrac{32 M^{2-m} \left(\frac{M}{\Lambda}\right)^{-(m+2)}}
          {\left(\frac{A_0}{2}\right)^m
\left[1+2\delta_0-\ln\left(\frac{M^2}{\Lambda^2}\right)\right]^{m+2}}
    \left[ Q_{m,0}(0)
        + \frac{2Q_{m,1}(0)}{1+2\delta_0-\ln\left(\frac{M^2}{\Lambda^2}\right)}
        + \cdots
    \right],
\label{eq:(3.12)}
\end{equation}
respectively.
Thus, it is clear that the logarithmic divergence of
\EQ{eq:(2.89)} cancels out that of \EQ{eq:(3.11)} and \EQ{eq:(3.12)} in the
amputated Green functions. Actually, by noting
$Q_{m,0}(0)=(-2)^{m-1} (m-1)!$, we obtain {\em finite} Green functions in the
continuum limit  $\Lambda/M_d\rightarrow\infty$:
\begin{equation}
  \Gamma_4(p_{(1)},-p_{(1)},p_{(2)},-p_{(2)})
  = \frac{\pi^2}{N} \frac{1}{M_d^2}
    \prod_{j=1}^2 \left[
      \dfrac{
         -1+2\delta_0+\ln \left({-p_{(j)}^2 \over M_d^2}\right)
           + {\cal O}(\left({-p_{(j)}^2 \over M_d^2}\right)^{-2})
      }{\sqrt{-p_{(j)}^2/M_d^2}}
    \right]
\label{eq:(3.13)}
\end{equation}
for the four-point Green function and
\begin{eqnarray}
\lefteqn{
 \Gamma_{2n}(p_{(1)},-p_{(1)},p_{(2)},-p_{(2)},\cdots,p_{(n)},-p_{(n)}) =
} \nonumber\\
& & \left(-\frac{4\pi^2}{N}\right)^{n-1}
    \dfrac{(n-3)! \tA_0^2}{2^n M_d^{3n-4}}
    \prod_{j=1}^n \left[
      \dfrac{
         -1+2\delta_0+\ln \left({-p_{(j)}^2 \over M_d^2}\right)
           + {\cal O}(\left({-p_{(j)}^2 \over M_d^2}\right)^{-2})
      }{\sqrt{-p_{(j)}^2/M_d^2}}
    \right]
\label{eq:(3.14)}
\end{eqnarray}
for the $2n$ ($n\ge3$)-point Green functions.

\section{Auxiliary Field Propagators}
We have shown in the previous section that all the
multi-fermion Green functions are finite at zero momentum of $\sigma$
in the \SxSB\ phase. This strongly suggests that the effective potential can
be renormalized. We further wish to
show finiteness of the Green functions also at {\em non-zero momentum}.
This requires knowledge of the {\em effective action} not restricted to
the {\em effective potential}, which is, however,
a far-reaching problem even in the ladder
approximation.
Directly relevant to such a problem are the propagators of
the auxiliary fields. If the renormalization of the effective potential
can simultaneously renormalize these propagators, it would be very
promising for the validity of such a renormalization. This is actually the
case in the \NJLd\ model\cite{kn:KY90,kn:KTY91,kn:HKWY91} which is
renormalizable in $1/N$ expansion.

Thus we are interested in the calculation of the propagators of the
auxiliary fields. Difficulty in such a calculation resides in the
lack of our knowledge on the Yukawa-type vertex function
$\Gamma _S(p,q)$ at non-vanishing momentum of the auxiliary fields
$q\not=0$. Recently, Appelquist, Terning and Wijewardhana\cite{kn:ATW91}
made an interesting resummation technique based on a further approximation
(besides the ladder approximation) to
evaluate the auxiliary field propagator without knowing $\Gamma _S(p,q\not=0)$.
Most amazingly, as we will show in the next section, their $\sigma$
propagator is in fact renormalized through our renormalization conditions
of the {\em effective potential}.\cite{kn:KTY91}

Here we briefly review the calculation of Ref.\cite{kn:ATW91}
and discuss the validity and its
possible modifications.
Let us consider the propagator of auxiliary field $\sigma$
in the symmetric vacuum $\sigma_{\rm sol}=0$ (see Fig.\ref{fig:auxprop}):
\begin{equation}
 i D^{-1}_{\sigma \sigma}(-q^2) = - \int {d^4 k \over (2\pi)^4 i}
  \tr \left[\Gamma_S(k,q) {1 \over \fsl{k}}
   {1 \over \fsl{k}-\fsl{q}} \right]
  -V''_{\rm (cl)}(\sigma=0).
\label{eq:(4.1)}
\end{equation}
For $q^2=0$, it is given by the second derivative of
the effective potential:
\begin{equation}
 i D^{-1}_{\sigma \sigma}(-q^2=0) = - V''(\sigma_{\rm sol}=0)
= {N \Lambda^2 \over 4\pi^2} \left( {1 \over g^*}-{1 \over g} \right).
\end{equation}
Now we consider the second derivative of \EQ{eq:(4.1)} on $q_\mu$
(see Fig.\ref{fig:auxprop2})\cite{kn:GM91,kn:Mira92,kn:ATW91}.
By using the self-consistent equation for $\Gamma_S$ (see
Fig.\ref{fig:yukawaSD}),  we can easily show that Fig.\ref{fig:auxprop2} can be
expressed as \begin{eqnarray}
\lefteqn{
 \frac{1}{N} \dfrac{\partial^2}{\partial q_\mu \partial q_\nu}
   i D_{\sigma \sigma}^{-1}(-q^2)
} \nonumber\\
  &=&  -\int \dfrac{d^4 k}{(2\pi)^4 i}
         \tr \left[\Gamma_S(k,q) {1 \over \fsl{k}} \Gamma_S(k-q,-q)
         \left(\dfrac{\partial^2}{\partial q_\mu \partial q_\nu}
           \frac{1}{\fsl{k}-\fsl{q}}\right) \right]
  \nonumber\\
  & & -\int {d^4 k \over (2\pi)^4 i}
         \tr\left[\left(\frac{\partial}{\partial q_\nu} \Gamma_S(k,q)\right)
            \frac{1}{\fsl{k}} \Gamma_S(k-q,-q)
            \left(\frac{\partial}{\partial q_\mu}
\frac{1}{\fsl{k}-\fsl{q}}\right)
      + (\mu \leftrightarrow \nu) \right],
\label{eq:(4.3)}
\end{eqnarray}
which is diagramatically depicted in Fig.\ref{fig:auxprop3}.
In the same fashion, we find a symbolic expression of $N$-th
derivative:
\begin{eqnarray}
\lefteqn{
 \frac{1}{N} \left({\partial \over \partial q} \right)^{\ell}
   i D_{\sigma \sigma}^{-1}(-q^2)
} \nonumber\\
&=& -\int {d^4 k \over (2\pi)^4 i}
     \sum_{n=0}^{\ell-1} {}_\ell C_n
     \tr \left[
       \left({\partial^n \over \partial q^n} \Gamma_S(k,q)\right)
       {1 \over \fsl{k}} \Gamma_S(k-q,-q)
       \dfrac{\partial^{\ell-n}}{\partial q^{\ell-n}}
         \frac{1}{\fsl{k}-\fsl{q}}
     \right].
\label{eq:(4.4)}
\end{eqnarray}

It was then observed\cite{kn:ATW91} that the contribution from
$\partial^n \Gamma_S(k,q)/\partial q^n|_{q=0}$ is negligible for
$n\not=0$, which implies
\begin{equation}
  \frac{1}{N}\left({\partial \over \partial q}\right)^{\ell}
    i D_{\sigma \sigma}^{-1}(-q^2) \Big|_{q=0}
   = -\int {d^4 k \over (2\pi)^4 i}
      \tr \left[
        \Gamma_S(-k^2) \frac{1}{\fsl{k}}
        \Gamma_S(-k^2) (-1)^{\ell}
        \dfrac{\partial^{\ell}}{\partial k^{\ell}}
          \frac{1}{\fsl{k}}
      \right].
\label{eq:(4.5)}
\end{equation}
Thus, the resummation of the Taylor expansion around $q=0$ leads to a
compact formula:
\begin{equation}
 \frac{1}{N} i D_{\sigma \sigma}^{-1}(-q^2)
= - \int {d^4 k \over (2\pi)^4 i}
\tr \left[ \Gamma_S(-k^2)
{1 \over \fsl{k}} \Gamma_S(-k^2) {1 \over \fsl{k}-\fsl{q}}  \right]
+ {\rm constant}.
\label{eq:(4.6)}
\end{equation}
After subtraction at $q^2=0$, the integral of \EQ{eq:(4.6)} yields
finally:\cite{kn:ATW91}
\begin{eqnarray}
 & & i D_{\sigma \sigma}^{-1}(\Eq) - i D_{\sigma \sigma}^{-1}(\Eq=0)
\nonumber\\
& &= {N \over 8\pi^2} \int_0^{\Lambda^2} d\Ek \Ek \Gamma_S^2(\Ek)
\left[ \left({\Ek \over \Eq}-2 \right) \theta(\Eq-\Ek) - {\Eq \over \Ek}
\theta(\Ek-\Eq) \right]
\label{eq:(4.7)}\\
& &= - {N \over 8\pi^2} \Eq \left[  {2\alpha_c \over \omega \alpha}
\Gamma_S^2(\Eq)
- {1 \over 1-\omega} \Gamma_S^2(\Lambda^2)  \right]
\nonumber\\
& &= - {N \over 8\pi^2} \Eq {4 \over (1+\omega)^2}
\left[  {2\alpha_c \over \omega \alpha} ({\Eq \over \Lambda^2})^{-1+\omega}
- {1 \over 1-\omega}  \right] .
\label{eq:(4.8)}
\end{eqnarray}

Now, let us discuss the validity of \EQ{eq:(4.8)}.
It was argued\cite{kn:ATW91} that the contributions coming from
\begin{equation}
 {\partial^n \Gamma_S(k,q) \over \partial q^n} \Big|_{q=0}
\end{equation}
are to be in higher order of $\alpha$:
\begin{equation}
 {\partial^n \Gamma_S(k,q) \over \partial q^n} \Big|_{q=0}
\sim {1 \over k^n} {\alpha \over 4\alpha_c} \Gamma_S(-k^2).
\end{equation}
It was also shown numerically\cite{kn:ATW91}
that these terms give smaller contributions
in $D_{\sigma \sigma}^{-1}$ than that from the zero-derivative of $\Gamma_S$,
{\em when the integral is regularized by an IR cutoff}.

However, we note that the dominant integral in \EQ{eq:(4.6)} comes from
IR region where $\Gamma_S(k,q=0)$ does have a singularity which is
actually absent in the original $\Gamma_S(k,q\ne0)$. Thus we must be
careful
about the effect of higher derivatives which compensates such a singularity
around $k=0$. Here we evaluate the order of magnitude of
such an effect, by considering
possible modifications of $\Gamma_S$ in \EQ{eq:(4.6)} which could change the
IR behavior of the integral of \EQ{eq:(4.6)} at $q\not=0$.

The simplest such modification of $\Gamma_S$ would be the replacement
in \EQ{eq:(4.7)}:
\begin{equation}
 \Gamma_S^2(\Ek) \rightarrow \Gamma_S^2(\Ek) \theta(\Ek-\Eq)
+ \Gamma_S^2(\Eq) \theta(\Eq-\Ek) .
\end{equation}
Then we obtain
\begin{equation}
 i D_{\sigma \sigma}^{-1}(\Eq) -  i D_{\sigma \sigma}^{-1}(0)
= -\dfrac{N \Eq}{8\pi^2}
   \left( {3 \over 2} + {1 \over 1-\omega} \right) \Gamma_S^2(\Eq)
   + {\cal O}(\Eq).
\label{eq:(4.12)}
\end{equation}
Another possible modification is the symmetric
calculation under the exchange of $k$ and $k-q$ in \EQ{eq:(4.6)}:
\begin{equation}
 i D_{\sigma \sigma}^{-1}(\Eq)
= -N \int {d^4k \over (2\pi)^4 i} \tr\left[ \Gamma_S(-k^2)
{1 \over \fsl{k}} \Gamma_S(-(k-q)^2) {1 \over \fsl{k}-\fsl{q}}
\right] + {\rm constant}.
\label{eq:(4.13)}
\end{equation}
Such a modification leads to
\begin{eqnarray}
 i D_{\sigma \sigma}^{-1}(\Eq) - i D_{\sigma \sigma}^{-1}(0)
= {N \over 4\pi^2} {\Gamma(-\omega) \over \Gamma(2+\omega)}
\left[ {\Gamma({3 \over 2}+{\omega \over 2}) \over \Gamma({3 \over 2}
-{\omega \over 2})}
\right]^2  \Eq \Gamma_S^2(\Eq) + {\cal O}(\Eq).
\label{eq:(4.14)}
\end{eqnarray}

Although these modifications are somewhat arbitrary, the results seem to
suggest validity of the {\em functional form} of the $\sigma$ propagator:
\begin{equation}
  i D_{\sigma \sigma}^{-1}(-q^2)
   = \frac{N \coeC}{\alpha/\alpha_c} q^2 \Gamma_S^2(-q^2) - V''(\sigma=0),
\label{eq:(4.15)}
\end{equation}
with the precise form of $\coeC$ being subject to the details of
the IR treatment.
Whereas various modifications agree with each other near the pure
NJL model $\alpha=0$, the deviation becomes significant at $\alpha=\alpha_c$.
Actually, $\coeC$ diverges at $\alpha=\alpha_c$ for \EQ{eq:(4.8)},
while \EQ{eq:(4.12)} does not.
In the next section we shall renormalize the generic form of
\EQ{eq:(4.15)} not restricted to \EQ{eq:(4.8)}, the original form
of Ref.\cite{kn:ATW91}.

\section{Symmetric Renormalization}\label{sec-ren}
In this section we formulate the renormalization of the gauged NJL model,
based on the effective potential \EQ{eq:(2.64)} expanded around the symmetric
vacuum (``symmetric renormalization'').  This is done
in a similar manner to the pure NJL model in $D(2<D<4)$ dimensions
(NJL$_{D<4}$) \cite{kn:KTY91}.
We first reformulate \cite{kn:KTY91,kn:HKWY91} the renormalization of
NJL$_{D<4}$ \cite{kn:KY90} at the leading  order of $1/N$ through the effective
potential approach.
Then in the gauged NJL model we apply the same method through the effective
potential written in terms of the auxiliary fields.

\subsection{\NJLd\ model}
The lagrangian of \NJLd\ model
is given by: \begin{equation}
  {\cal L} = \bar\psi i \fsl{\partial} \psi -m_0 \bar\psi \psi
           + \frac{G}{2N}
             \left[(\bar\psi\psi)^2 +(\bar\psi i\gamma_5 \psi)^2 \right],
\label{eq:(5.1)}
\end{equation}
where the fermion field $\psi $ belongs to the fundamental
representation of  $SU(N)$, with the summation of $SU(N)$ indices being
understood.  In \EQ{eq:(5.1)}, the fermion spinor is given by that in
four dimensional space-time, so that the model possesses
$U(1)_L \times U(1)_R$ symmetry for $m_0=0$ besides the $SU(N)$ symmetry.

We first demonstrate the symmetric renormalization through effective
potential.
Introducing the auxiliary fields $\sigma$, $\pi$ in the same way as
the gauged NJL model, we obtain the effective potential $V(\sigma,\pi)$
\cite{kn:KTY91,kn:HKWY91}:
\begin{eqnarray}
\lefteqn{
 - \dfrac{(4\pi)^{D/2} \Gamma (D/2)}{4N \Lambda^D} V(\sigma, \pi)
} \nonumber\\
  &=& \dfrac{1}{g} \dfrac{m_0 \sigma}{\Lambda^2}
     +\left(\dfrac{1}{g^*}-\dfrac{1}{g}\right)
      \dfrac{\sigma^2+\pi^2}{2\Lambda^2}
     -\dfrac{1}{2-\frac{D}{2}} \dfrac{\coeAd}{D}
      \left(\dfrac{\sigma ^2+\pi^2}{\Lambda ^2}\right)^{D/2}
     +{\cal O}(\left(\frac{\sigma^2+\pi^2}{\Lambda^2}\right)^{2}),
  \hspace{1em}
\label{eq:(5.2)}
\end{eqnarray}
with $\Lambda$ being the ultraviolet cutoff in the loop integral,
and $g^*$ and $\coeAd$ are defined by $g^{*}\equiv D/2-1$
and $\coeAd \equiv B(D/2-1,3-D/2)$, respectively.
Note that $g^*\rightarrow 0$ and $\coeAd^{-1}\rightarrow 0$
as $D\rightarrow 2$, so
that the divergence in the second and the third terms cancel each other to
give a well-known logarithmic factor in the Gross-Neveu model
\cite{kn:Coleman}.
For a detailed derivation of
\EQ{eq:(5.2)}, see Appendix~\ref{sec-appA}.

Propagators of the auxiliary fields are calculated as \EQ{eq:(A.24)}.
In the symmetric vacuum $\VEV{\sigma }=0$  they read
\begin{eqnarray}
\lefteqn{
  -iD_{\sigma\sigma}^{-1}(-q^2) = -iD_{\pi \pi }^{-1}(-q^2)
} \nonumber\\
 &=& V''(\sigma =0, \pi=0)
     +  \dfrac{4N}{(4\pi )^{D/2}\Gamma
(D/2)}\dfrac{\coeCd}{2-D/2}(-q^2)^{D/2-1},
\label{eq:(5.3)}
\end{eqnarray}
where $\coeCd\equiv B(3-D/2,D/2-1)/\Gamma (D-1)$ and
\begin{equation}
  V''(\sigma=0,\pi=0)
  \equiv \left. \dfrac{\partial^2 V(\sigma ,\pi=0)}{\partial \sigma ^2}
         \right|_{\sigma=0}
   =  \dfrac{4N \Lambda ^{D-2}}{(4\pi )^{D/2}\Gamma (D/2)}
      \left(\frac{1}{g}-\frac{1}{g^{*}}\right).
\label{eq:(5.4)}
\end{equation}
\EQ{eq:(5.4)} is negative for $g>g^{*}$, implying
  appearance of a tachyon pole in the auxiliary field
propagators,  another signal of  instability of the symmetric
vacuum.

Let us next consider the renormalization of \NJLd\ model
at the $1/N$ leading order.
Due to   absence of the divergence in the fermion propagator and
the vertex function $\Gamma _S$,
wave function renormalizations of the fermion and the auxiliary fields
are not required at this stage:
\begin{equation}
 \psi=\psi_R, \ \sigma_R = \sigma.
\end{equation}
Thus we   concentrate our attention to the renormalization of
the effective potential \EQ{eq:(5.2)}.
The divergence $\Lambda ^{D-2}$ in \EQ{eq:(5.2)} can be absorbed into
the redefinition of renormalized parameters $g_R, m_R$:
\begin{equation}
  \Lambda ^{D-2} \left(\frac{1}{g}-\frac{1}{g^{*}}\right)=
  \mu ^{D-2} \left(\frac{1}{g_R}-\frac{1}{g_R^{*}}\right)
\label{eq:(5.6)}
\end{equation}
and
\begin{equation}
  \dfrac{\Lambda ^{D-2}}{g} m_0 = \dfrac{\mu ^{D-2}}{g_R} m_R,
\label{eq:(5.7)}
\end{equation}
where $\mu $ is the renormalization scale.
These are precisely the same renormalization conditions as those
imposed by Kikukawa and Yamawaki \cite{kn:KY90} through the renormalization of
the propagator of $\sigma$.
\par
With the above definition of renormalized parameters
\EQQ{eq:(5.6)}{eq:(5.7)} we obtain a renormalized effective
potential:
\begin{eqnarray}
\lefteqn{
 - \dfrac{(4\pi )^{D/2} \Gamma (D/2)}{4N \mu ^D} V_R(\sigma _R,\pi _R)
} \nonumber\\
  &=& \dfrac{1}{g_R} \dfrac{m_R \sigma _R}{\mu ^2}
   +\left(\dfrac{1}{g_R^*}-\dfrac{1}{g_R}\right)
\dfrac{\sigma _R^2+\pi _R^2}{2\mu ^2}
   - \dfrac{1}{2-\frac{D}{2}}\dfrac{\coeAd}{D}\left(\dfrac{\sigma _R^2+\pi
_R^2}{\mu ^2}\right)^{D/2},
   \hspace{2em}
\label{eq:(5.8)}
\end{eqnarray}
where higher power terms ${\cal O}(\sigma ^4)$ disappear  in $\Lambda
\rightarrow \infty $ (even at $D=2$).
This renormalization breaks down at $D=4$, which is signaled by the
singularity $(2-D/2)^{-1}$ in the last term of \EQ{eq:(5.8)},
corresponding to the logarithmic divergence of $(\sigma^2+\pi^2)^2$
coupling in $D=4$ NJL model.
(We would need extra ``counter term'' such as the
eight-fermion operators $(\bar \psi \psi)^4$ to ``renormalize'' the model in
$D=4$.)

In our renormalization \EQQ{eq:(5.6)}{eq:(5.7)},
the critical value of {\em renormalized}
 four-fermion coupling is left undetermined.
Actually, it is a free parameter which corresponds to   varieties
of renormalization schemes.
In the $D\rightarrow 2$ limit, however, we need to define the singular
part of  $g_R^*$ as a function of $D$:
\begin{equation}
  \dfrac{1}{g_R^*} = \dfrac{1}{D/2-1} + \mbox{regular function of $D$},
\end{equation}
so that we can reproduce the usual renormalization of the Gross-Neveu
model \cite{kn:Coleman}.
Again, the choice of the regular part corresponds to the choice of
renormalization scheme. In the following, we take the simplest choice
$g_R^*=D/2-1=g^*$  (different from
that in Ref.\cite{kn:KY90}).

It is straightforward to get RG functions
$\beta _g$ and $\gamma _m$ from \EQQ{eq:(5.6)}{eq:(5.7)}:
\footnote{The next-to-leading order corrections to this result have been
calculated \cite{kn:HKWY91,kn:HKK92,kn:IT92}.}
\begin{eqnarray}
  \beta _g(g_R)   &=& (D-2) g_R
\left(1 - \dfrac{g_R}{g_R^{*}} \right),
\label{eq:(5.10)}  \\
  \gamma _m(g_R) &=& (D-2) \dfrac{g_R}{g_R^{*}},
\label{eq:(5.11)}
\end{eqnarray}
where $\beta_g \equiv\mu \partial g_R/\partial\mu$ and
$\gamma _m m_R \equiv -\mu \partial m_R/\partial \mu $.
This agrees with Ref.\cite{kn:KY90}.
It is evident that we also obtain the same form as
\EQQ{eq:(5.10)}{eq:(5.11)} for
$\beta _g(g) \equiv \Lambda \partial g/\partial \Lambda$ and
$\gamma_m(g) \equiv -(\Lambda/m_0) \partial m_0/\partial \Lambda$
in the bare coupling $g$, with the UV fixed point $g_R^*$ simply
replaced by $g^*$.
As was pointed out by Kikukawa and Yamawaki \cite{kn:KY90},
it is clear from \EQQ{eq:(5.10)}{eq:(5.11)} that \NJLd\ model has
a nontrivial ultraviolet fixed point on which the RG  functions are
continuous when approached from both phases.

It is also obvious that the propagators of $\sigma$ and $\pi$ \EQ{eq:(5.3)}
 can also be renormalized by the same condition as \EQ{eq:(5.6)},
as was originally done in Ref.\cite{kn:KY90}:
\begin{eqnarray}
\lefteqn{
  -i D_{\sigma \sigma}^{(R)}{}^{-1}(-q^2)
= -i D_{\pi \pi}^{(R)}{}^{-1}(-q^2)
} \nonumber \\
& &= V''_R(\sigma_R=0,\pi_R=0)
   +\dfrac{4N}{(4\pi)^{D/2} \Gamma(D/2)}\frac{\coeCd}{2-D/2} (-q^2)^{D/2-1}.
\label{eq:(5.12)}
\end{eqnarray}

It should be stressed again that this renormalization breaks down at
$D=4$, signaled by the appearance of $(2-{D \over 2})^{-1}$
singularity in the term of $(\Eq)^{{D \over 2}-1}$ ( $\rightarrow
(\Eq)$ as $D \rightarrow 4$), which corresponds to the logarithmic
divergence of the kinetic term of $\sigma$ and $\pi$ in $D=4$ NJL model
(We would need higher dimensional operator
$\partial_\mu (\bar \psi \psi) \partial^\mu (\bar \psi \psi)$ as a
``counter term'' to ``renormalize'' the model in $D=4$).

The fact that we can renormalize the theory without higher dimensional
operators $(\bar\psi\psi)^4$ and
\(
  \partial_\mu (\bar\psi\psi) \partial^\mu (\bar\psi\psi)
\) at $1/N$ leading order simply reflects   the following fact:
$(\bar \psi \psi)^2$ is a relevant operator due to a large anomalous
dimension $\gamma_m=D-2$ at $g_R=g_R^{*}$, i.e.,
$\dim(\bar \psi \psi)^2=2(D-1-\gamma_m)=2<D$,
while the would-be ``counter terms'' $(\bar \psi \psi)^4$ and
\(
  \partial_\mu (\bar \psi \psi) \partial^\mu (\bar \psi \psi)
\) are irrelevant operators,
\(
  \dim(\bar \psi \psi)^4=4(D-1-\gamma_m)=4>D
\),
\(
 \dim[\partial_\mu (\bar \psi \psi) \partial^\mu (\bar \psi \psi)]
  =2(D-\gamma_m)=4>D
\).
At $D=4$, however, all these operators equally have dimension $4(=D)$
and become marginal operators.  Hence they should be included in order
to make the theory renormalizable, in which case the NJL model in its
renormalizable version becomes identical to the Higgs-Yukawa system
("standard model") \cite{kn:Hasenfratz,kn:Zinn-Justin}.
%????
%

\subsection{Symmetric renormalization of the gauged NJL model}
Now, we are ready to study the renormalization properties of
the gauged NJL model,
 based on the simplest effective potential \EQ{eq:(2.69)} expanded
around the symmetric vacuum (symmetric renormalization)\cite{kn:KTY91}.
The gauge coupling
$\alpha $ does not get renormalized, in accord with the absence of vacuum
polarization
 in the gauge boson  propagator in this approximation.
Such an approximation becomes realistic in the {\em standing} gauge
theory as a limit of {\em walking} gauge theory \cite{kn:Hold85-}.
Thus the renormalization is operative only on the four-fermion coupling
$g$.  Actually, this renormalization $(0<\omega<1)$ is done in a very similar
manner to NJL$_{D<4}$$(2<D<4)$.
Both cases break down in the pure NJL limit $(D=4, \omega=1)$.
However it should be noted that while
the renormalization of NJL$_{D<4}$   is valid
even at $D=2$ where  \EQ{eq:(5.2)} still remains valid, the renormalization
of the gauged NJL model
in this scheme breaks down at $\omega=0$ where \EQ{eq:(2.69)} becomes no longer
valid.
Renormalization scheme valid also at $\omega=0$ should be based on the general
form of the effective potential \EQ{eq:(2.64)},
which will be discussed in section~7.

In our definition of (bare) auxiliary
field, the (bare)  Yukawa-type vertex $\Gamma_S$ does not depend on the
four-fermion coupling and vanishes in the $\Lambda \rightarrow \infty$ limit.
Thus, it should be renormalized via redefinition of auxiliary fields to
obtain a finite interacting theory in the  $\Lambda \rightarrow \infty $
limit: \begin{equation}
  \sigma_R \equiv \left(\dfrac{\Lambda}{\mu}\right)^{1-\omega} \sigma,
  \qquad
  \pi_R \equiv \left(\dfrac{\Lambda}{\mu}\right)^{1-\omega} \pi ,
\label{eq:(5.13)}
\end{equation}
with $\mu $ being ``renormalization point''.\footnote{
  Similar redefinition of the auxiliary fields was also made \cite{kn:BL92},
  with $\mu$ being taken as the dynamical fermion mass $\mu=M_d$.
}
According to this definition of renormalized fields,
the Yukawa-type vertex \EQ{eq:(2.83)} is renormalized:
\begin{equation}
  \Gamma_S^{(R)}(-p^2)
  = \left( \frac{\mu}{\Lambda} \right)^{1-\omega} \Gamma_S(-p^2)
  = \dfrac{2}{1+\omega} \left(\dfrac{-p^2}{\mu^2}\right)^{-(1-\omega)/2}
\label{eq:(5.14)}
\end{equation}
for the symmetric vacuum.
It should be noted that this definition of the renormalized fields
$\sigma_R$, $\pi_R$ simultaneously renormalizes the Yukawa-type vertex in
\SxSB\ vacuum \EQ{eq:(2.87)}:
\begin{eqnarray}
\lefteqn{
 \Gamma_S^{(R)}(-p^2; M_d)
} \nonumber\\
  &=& \left( \frac{\mu}{\Lambda} \right)^{1-\omega} \Gamma_S(-p^2; M_d)
  \nonumber\\
  &=& \dfrac{2}{1+\omega} \left(\dfrac{-p^2}{\mu^2}\right)^{-(1-\omega)/2}
    \left[ 1 + \frac{(1+\omega)(2+\omega)}
                    {(1-\omega)(2-\omega)}\frac{D_1}{C_1}
               \left(\dfrac{-p^2}{M_d^2}\right)^{-\omega}
               \hspace{-0.8em}
             +{\cal O}(\left(\frac{-p^2}{M_d^2}\right)^{-2+\omega}
              \hspace{-0.3em})
    \right].
    \hspace{2em}
\label{eq:(5.15)}
\end{eqnarray}

Then the effective potential \EQ{eq:(2.69)}  is
expressed in terms of the renormalized auxiliary fields:
\begin{eqnarray}
\lefteqn{
  - \dfrac{4\pi^2}{N} \dfrac{V_R(\sigma_R, \pi_R)}{\mu^4}
} \nonumber\\
 &=& \left(\dfrac{\Lambda}{\mu}\right)^{1+\omega}\frac{1}{g}
     \dfrac{m_0 \sigma_R}{\mu^2}
   + \left(\dfrac{\Lambda}{\mu}\right)^{2\omega}
     \left(\dfrac{1}{g^{*}}-\frac{1}{g}\right)
     \dfrac{\sigma_R^2+\pi_R^2}{2\mu^2}
   - \dfrac{4\coeA}{\alpha/\alpha_c}\dfrac{2-\omega}{4}
     \left(\dfrac{\sigma_R^2+\pi_R^2}{\mu^2}\right)^{2/(2-\omega)}
     \hspace{-2em},
 \hspace{2em}
\end{eqnarray}
where we have dropped out the contributions which vanish in the
 $\Lambda \rightarrow \infty$ limit.

Now, the parallelism between \NJLd\ and gauged NJL model is
manifest. The effective potential \EQ{eq:(2.69)} can be renormalized
by the definition of renormalized four-fermion coupling
\begin{equation}
  \Lambda ^{2\omega } \left(\dfrac{1}{g^*} - \dfrac{1}{g}\right)
  = \mu ^{2\omega }\left(\dfrac{1}{g_R^*} - \dfrac{1}{g_R}\right),
\label{eq:(5.17)}
\end{equation}
and renormalized current mass of the fermion:
\begin{equation}
  \Lambda ^{1+\omega } \dfrac{m_0}{g} = \mu ^{1+\omega }
\dfrac{m_R}{g_R}.
\label{eq:(5.18)}
\end{equation}
Again, the ambiguity of $g_R^*$ corresponds to the choice of
renormalization scheme.
According to this renormalization,
we find the renormalized effective potential
\begin{eqnarray}
\lefteqn{
  - \dfrac{4\pi^2}{N} \dfrac{V_R(\sigma_R, \pi_R)}{\mu^4}
} \nonumber\\
 &=& \frac{1}{g_R}\dfrac{m_R \sigma_R}{\mu^2}
   + \left(\dfrac{1}{g_R^{*}} - \frac{1}{g_R} \right)
     \dfrac{\sigma_R^2+\pi_R^2}{2\mu^2}
   - \dfrac{4\coeA}{\alpha/\alpha_c}\dfrac{2-\omega}{4}
     \left(\dfrac{\sigma_R^2+\pi_R^2}{\mu^2}\right)^{\frac{2}{2-\omega }}.
\label{eq:(5.19)}
\end{eqnarray}
Thus, {\em all} the multi-fermion Green functions
have been renormalized {\em at zero momentum} of the auxiliary field.

How about the renormalization at non-zero momentum, then?
Remarkably enough, the auxiliary field propagators \EQ{eq:(4.15)} are
also renormalized via the above definition of the renormalized parameters:
\begin{eqnarray}
\lefteqn{
  -iD_{\sigma\sigma}^{(R)-1}(-q^2) = -iD_{\pi\pi}^{(R)-1}(-q^2)
} \nonumber\\
& &=  V_R''(\sigma_R=0,\pi_R=0)
     + \dfrac{N \coeC}{\alpha/\alpha_c } (-q^2) \Gamma^{(R)2}_S(-q^2).
\label{eq:(5.20)}
\end{eqnarray}
Thus the four-fermion Green function can also be  renormalized
at {\em non-zero momentum} in a similar manner to \NJLd.
In spite of the formal resemblance, however, one should note that the gauged
NJL model has only been shown to be renormalized at the level of the ladder
approximation, in sharp contrast to \NJLd\ which is shown to be
{\em renormalizable} in the systematic $1/N$ expansion\cite{kn:RWP90}.

We should also emphasize that the renormalization
\EQQ{eq:(5.17)}{eq:(5.18)} is based on the effective potential
and thus it holds both in the chiral symmetric ($g_R<g_R^*$)
and the \SxSB\ ($g_R>g_R^*$) phases.

The definition of the renormalized parameters $g_R$ and $m_R$
\EQQ{eq:(5.17)}{eq:(5.18)} leads to the RG
functions  $\beta _g$ and $\gamma _m$ (Fig.~\ref{fig:rgefunc}):
\begin{eqnarray}
  \beta _g(g_R,\alpha )   &=& 2\omega g_R
\left(1-\dfrac{g_R}{g_R^*}\right),
\label{eq:(5.21)} \\
  \gamma _m(g_R,\alpha ) &=& 1-\omega +2\omega \dfrac{g_R}{g_R^*}.
\label{eq:(5.22)}
\end{eqnarray}
Thus the theory does have a nontrivial ultraviolet fixed line
$g_R=g_R^*$, on which the mass operator of fermion acquires a large
anomalous dimension $\gamma _m=1+\omega $.
It should also be noted that
the anomalous dimension $\gamma _m$ is
{\em continuous across the critical line} $g_R=g_R^*$, in contrast to the
earlier phase-dependent calculation through the fermion propagator
based on the solution of the SD gap equation.
This is actually in accord with the suggestion\cite{kn:KY90}
that the renormalization through the four-fermion Green function
(auxiliary field propagator) in the gauged NJL model may lead to the
large anomalous dimension  in the symmetric phase as well as in the \SxSB\
phase.
\par
At $\alpha=\alpha_c$ ($\omega=0$) \EQQ{eq:(5.21)}{eq:(5.22)} would yield
$\beta_g \equiv 0$ and $\gamma_m \equiv 1$.  This may be an artifact of the
symmetric renormalization based on the effective potential \EQ{eq:(2.69)}
which is no longer valid at $\alpha=\alpha_c$.  A possible modification at
$\alpha=\alpha_c$ will be given in Section 7.

We thus have found that the gauged NJL model is renormalized
within the ladder approximation for non-vanishing gauge coupling
$\alpha>0$. It is well known, however,
that the pure (non-gauged) NJL model cannot be renormalized in
$D=4$ due to uncontrollable logarithmic
divergence  even in the leading approximation of $1/N$ expansion (or
chain approximation). We here discuss  how the existence of
gauge interaction improves the structure of divergence.
The anomalous dimension of composite operator $\bar\psi \psi $
at critical point of the pure NJL model is given by $\gamma _m = 2$.
Such a large anomalous dimension makes the scaling dimension
of $\bar\psi \psi $ very small: $\dim \bar\psi \psi =3-\gamma _m=1$,
and hence higher dimensional operators such as $(\bar\psi \psi )^4$
and $\partial_\mu (\bar\psi \psi ) \partial^\mu (\bar\psi \psi )$
become marginal operators.
As a result,
logarithmic divergence associated with these operators appear.
On the other hand,
the gauge interaction makes the scaling behavior of  $\bar\psi \psi $
softer: $\dim \bar\psi \psi =3-\gamma _m=2-\omega >1$.
Thus, the above higher dimensional operators  becomes
{\em irrelevant} operators and there are no uncontrollable divergence.

In the diagrammatic picture,
the softness of the scaling dimension of $\bar\psi \psi $ corresponds
to  the softness of high energy behavior of Yukawa-type vertex
function $\Gamma ^{(R)}_S$ \EQ{eq:(5.14)}.
Thanks to such a soft behavior of $\Gamma ^{(R)}_S$,
the structure of divergence is improved and the logarithmic
divergence disappears in the presence of gauge interaction.
%????
%******
\section{Operator Product Expansion}\label{sec-ope}
Now that we have obtained a renormalized theory having a nontrivial UV fixed
line with a very large anomalous dimension, $\gamma_m=1+\omega \ (g_R=g_R^*)$,
we can explicitly construct OPE both in the symmetric and the \SxSB\ phases
and  see how such a large anomalous dimension fits in the general framework
of OPE \cite{kn:KTY91}.
\par
The OPE relevant to the fermion mass function takes the form
\begin{eqnarray}
  -i \FT \TOP{\psi(x)\bar\psi(0)}
      &=&  c_{\unit}(p;g_R,\alpha, m_R, \mu ) \unit
      \nonumber \\
      & &+ c_{\bar\psi\psi}(p;g_R,\alpha, m_R, \mu)
         \left[ (\bar\psi\psi)_R +\gamma_5(\bar\psi\gamma_5\psi)_R \right]
      + \cdots .
\label{eq:(6.1)}
\end{eqnarray}
We explicitly calculate the Wilson coefficients $c_{\unit}$ and
$c_{\bar\psi\psi}$ on the nonperturbative solution
which we know already.
Taking the vacuum expectation value of \EQ{eq:(6.1)}, we write
\begin{equation}
 -iS(p) = c_{\unit} + c_{\bar\psi\psi}  \VEV{(\bar\psi\psi)_R} + \cdots.
\label{eq:(6.2)}
\end{equation}
The fermion propagator in LHS takes the form
\begin{equation}
  -i S(p) = \frac{\fsl{p}}{p^2}+{\Sigma(-p^2) \over p^2} + \cdots.
\label{eq:(6.3)}
\end{equation}
Comparing \EQ{eq:(6.3)} with \EQ{eq:(6.2)}, we write
\begin{equation}
  \Sigma(-p^2) = p^2 m_R(\mu) c'_{\unit}(p; g_R,\alpha,\mu)
               + p^2 \VEV{(\bar\psi\psi)_R}
                   c_{\bar\psi\psi}(p; g_R,\alpha, 0,\mu)
               + \cdots,
\label{eq:(6.4)}
\end{equation}
where we have expanded the
Wilson coefficients around the chiral symmetric limit $m_R=0$:
\begin{subequations}
\begin{eqnarray}
  c_\unit(p; g_R, \alpha, m_R, \mu)
    &=& \frac{\fsl{p}}{p^2}+m_R (\mu) c'_{\unit}(p; g_R,\alpha, \mu) + \cdots,
  \\
  c_{\bar\psi\psi}(p; g_R, \alpha, m_R, \mu)
    &=& c_{\bar\psi\psi}(p; g_R, \alpha, 0, \mu) + \cdots.
\end{eqnarray}
\end{subequations}

We denote by $\Sigma_{\rm explicit}(-p^2)$ the part of the
fermion mass function owing to the explicit chiral symmetry breaking
$m_R\not=0$:
\begin{equation}
  \Sigma_{\rm explicit}(-p^2)
   = p^2 m_R(\mu) c'_\unit(p; g_R, \alpha, \mu) + \cdots ,
\label{eq:(6.6)}
\end{equation}
which actually takes the form
\begin{equation}
\Sigma_{\rm explicit}(-p^2) =
  \begin{cases}
     \Gamma_S(-p^2; M_d) \sigma_{\rm explicit},
      & (\SxSB\ vacuum) \cr\cr
     \Gamma_S(-p^2) \sigma_{\rm explicit},
      & (symmetric~vacuum)
  \end{cases}
\label{eq:(6.7)}
\end{equation}
where $\sigma_{\rm explicit}$ is defined by
\begin{equation}
  \sigma_{\rm explicit} \equiv \sigma_{\rm sol} - \sigma_{\rm spont}
     = \left. \dfrac{\partial \sigma _{\rm sol}}{\partial m_0}
    \right|_{m_0=0} m_0 + \cdots,
\label{eq:(6.8)}
\end{equation}
with $\sigma_{\rm sol}$ being the solution of the stationary condition
of the effective potential \EQ{eq:(2.64)},
\(
  \partial V(\sigma,\pi=0) /\partial \sigma= 0.
\)
For $0<\alpha<\alpha_c$
it is straightforward to calculate $\sigma_{\rm explicit}$  from
\EQ{eq:(2.75a)}
as a Taylor series in the fermion bare mass $m_0$:
\begin{equation}
  \sigma_{\rm explicit} =
  \begin{cases}
     \dfrac{2-\omega}{2\omega} \dfrac{m_0}{\frac{g}{g^*}-1}
      & (\SxSB\ vacuum)
     \\
     \dfrac{m_0}{1-\frac{g}{g^*}}
      & (symmetric~vacuum)
\end{cases}.
\label{eq:(6.9)}
\end{equation}

By comparing \EQ{eq:(6.6)} with \EQ{eq:(6.7)} and
\EQ{eq:(6.9)}, we find
\begin{equation}
  c_{\unit}'(p;g_R,\mu ) =
  \begin{cases}
    \dfrac{1}{p^2}\dfrac{2-\omega}{2\omega }
    \dfrac{1}{\dfrac{g_R}{g_R^*}-1} \Gamma^{(R)}_S(-p^2; M_d)
    & (\SxSB\ vacuum)
    \\
    \dfrac{1}{p^2}\dfrac{1}{1-\dfrac{g_R}{g_R^*}} \Gamma_S^{(R)}(-p^2)
    & (symmetric~vacuum)
  \end{cases},
\label{eq:(6.10)}
\end{equation}
with the renormalized vertex $\Gamma^{(R)}_S$ being given by \EQ{eq:(5.15)},
where we have used \EQ{eq:(5.18)};
\begin{equation}
   \dfrac{m_0}{m_R}
      = \left( \dfrac{\mu}{\Lambda} \right)^{1+\omega} \dfrac{g}{g_R} \
      (\equiv Z_m).
\end{equation}

Now, it is easy to show that the Wilson coefficients satisfy the RG equation:
\begin{subequations}
\begin{eqnarray}
  0&=&\left[ {\cal D} + 2 +\gamma_m(g_R) \right]
      c_{\unit}'(\kappa p; g_R,\alpha,\mu ),
   \label{eq:(6.12a)}
   \\
  0&=&\left[ {\cal D} + 4 -\gamma _m(g_R) \right]
      c_{\bar\psi \psi }(\kappa p; g_R,\alpha, 0,\mu ),
\label{eq:(6.12b)}
\end{eqnarray}
\end{subequations}
where
\begin{eqnarray}
  {\cal D} \equiv
      \kappa \dfrac{\partial}{\partial \kappa}
        - \beta _g(g_R) \dfrac{\partial}{\partial g_R}.
\end{eqnarray}
\EQ{eq:(6.12a)} is readily solved to yield
\begin{equation}
  c'_\unit(\kappa p; g_R, \alpha; \mu)
  \simeq \kappa^{-(2+\gamma_m^*)} c'_\unit(p; \bar g(\kappa), \alpha; \mu)
\end{equation}
where $\gamma_m^* \equiv \gamma_m(g_R^*)=1+\omega$
and $\bar g(\kappa)$ is the solution of
\begin{equation}
  \kappa \frac{d}{d\kappa}  \bar g(\kappa) = \beta_g(\bar g), \qquad
  \mbox{with} \quad \bar g(\kappa=1) = g_R(\mu).
\label{eq:(6.15)}
\end{equation}
\EQ{eq:(6.15)} can be solved
using the  $\beta$ function \EQ{eq:(5.21)}:
\begin{equation}
  \dfrac{1}{1-\frac{\bar g(\kappa)}{g_R^*}}
    = 1 + \kappa^{2\omega} \dfrac{g_R/g_R^*}{1-\frac{g_R}{g_R^*}}.
\end{equation}
Then \EQ{eq:(6.10)} implies a quite unusual situation, i.e.,
the Wilson coefficient
$c'_{\unit}$ does have a strong momentum dependence
\begin{equation}
  c'_{\unit}(p; \bar g(\kappa), \mu ) \sim \kappa^{2\omega}.
\end{equation}
This is combined with \EQ{eq:(6.9)}, yielding finally the high energy
behavior of $\Sigma_{\rm explicit}$:
\begin{equation}
  \Sigma_{\rm explicit}(-\kappa^2 p^2)
  \simeq m_R(\mu) p^2 \kappa^{-(1+\omega)}
         c'_\unit( p; \bar g(\kappa), \alpha; \mu)
  \sim \kappa^{-(1-\omega)}.
\label{eq:(6.18)}
\end{equation}
\par
Thus the Wilson coefficient for the explicit chiral symmetry breaking term
has a nontrivial factor \EQ{eq:(6.10)}, yielding additional momentum dependence
$\kappa^{2\omega}$, which actually compensates the momentum dependence
$\kappa^{-(1+\omega)}$ arising from the anomalous dimension.
As a result we obtain $\kappa^{-(1-\omega)}$ behavior of the fermion
mass function
$\Sigma_{\rm explicit}(-\kappa^{2}p^2)$, in accord with the solutions of
the SD equation (see Appendix C):
\begin{equation}
  \Sigma_{\rm explicit}(-\kappa^{2} p^2)\simeq
    \kappa^{-(1-\omega)} \Sigma_{\rm explicit}(-p^2).
\label{eq:(6.19)}
\end{equation}

This peculiar phenomenon with the Wilson coefficient is precisely the same as
that
in NJL$_{D<4}$ which was discovered by Kikukawa and Yamawaki\cite{kn:KY90}.
Explicit calculation in \NJLd\ is given in Appendix~\ref{sec-appO}
where the Wilson coefficient $ c'_{\unit}$ takes the form
\begin{eqnarray}
  c'_\unit(p; g_R,\mu)
    &=& \begin{cases}
          \dfrac{1}{D-2} \dfrac{1}{\frac{g_R}{g_R^*}-1} \dfrac{1}{p^2} +\cdots
                 & (\SxSB\ vacuum)   \cr\cr
          \dfrac{1}{1-\frac{g_R}{g_R^*}} \dfrac{1}{p^2} + \cdots
                 & (symmetric vacuum)
        \end{cases},
  \label{eq:(6.20)}
\end{eqnarray}
which yields
$c'_\unit(p; \bar g(\kappa), \mu) \simeq \kappa^{D-2}=\kappa^{\gamma_m^*}$.
Solving RG equation for $c'_\unit$, we obtain
\begin{eqnarray}
  c'_\unit(\kappa p; g_R, \mu) \simeq
  c'_\unit(p; \bar g(\kappa), \mu) \kappa^{-(2+\gamma_m^*)}.
\label{eq:(6.21)}
\end{eqnarray}
Then in \NJLd\ we obtain
\begin{eqnarray}
\Sigma_{\rm explicit}(-\kappa^{2}p^2)
  &\simeq& \kappa^{2} p^2 m_R(\mu) c'_\unit(\kappa p; g_R, \mu)
  \nonumber\\
  &\simeq& p^2 m_R(\mu) c'_\unit(p; \bar g(\kappa), \mu) \kappa^{-\gamma_m^*}
  \nonumber\\
  &=& \mbox{constant},
\label{eq:(6.22)}
\end{eqnarray}
which is actually in accord with the explicit solution of
the SD gap equation given in Appendix~A.

Let us next turn to the Wilson coefficient of $(\bar\psi\psi)_R$,
$c_{\bar\psi\psi}$.
Such a coefficient function can be determined from
the fermion four-point function by taking $x\rightarrow 0$ limit:
\begin{equation}
  -i \VEV{\TOP{\psi (x)\bar\psi (0)\psi (y)\bar\psi (z)}}_{\rm connected}
   = c_{\bar\psi \psi }(x)
     \VEV{\TOP{(\bar\psi \psi )_R(0)\psi (y)\bar\psi (z)}}_{\rm connected}
   +\cdots,
\label{eq:(6.23)}
\end{equation}
where $c_\unit$ does not appear in the RHS of \EQ{eq:(6.23)},
since it contributes
only to the disconnected diagrams.

In the following calculation $x$,$y$ and $z$ are Fourier transformed
to  $p$,$q$ and $k$, respectively.
It is sufficient to evaluate the case of $q=k$ to obtain
 $c_{\bar\psi \psi }$.
Hereafter,
the external legs for $q$ and $k$ are understood to be amputated.

We first calculate the OPE coefficient function on the symmetric
vacuum.
The $\sigma$-exchange diagram (Fig.~\ref{fig:fourfunc}a) is given
by: \begin{equation}
  S(p) i\Gamma_S(-p^2) S(p) D_{\sigma\sigma}(0) \Gamma_S(-q^2)
  = \dfrac{1}{p^2} \Gamma_S(-p^2) \dfrac{1}{-V''(\sigma_{\rm sol})}
    \Gamma_S(-q^2) + \cdots,
\label{eq:(6.24)}
\end{equation}
where we have used a relation
$D_{\sigma\sigma}(0)=-i/V''(\sigma_{\rm sol})$.
In addition to the above
there  exist ``pure ladder'' diagrams (Fig.~\ref{fig:fourfunc}b)
contributing to the LHS of \EQ{eq:(6.23)}, which are hard to be calculated.
Here we {\em assume} that such diagrams have softer high energy behavior
than  that of \EQ{eq:(6.24)} and ignore them in the following calculations.

The RHS of \EQ{eq:(6.23)} can be calculated by (Fig.~\ref{fig:OPE}).
Since a bubble diagram is given by the second derivative of
$V_{\rm (qu)} \equiv V - V_{\rm (cl)}$, we obtain (Fig.~\ref{fig:OPE})
\begin{eqnarray}
\lefteqn{
  Z_m c_{\bar\psi\psi}(p; g_R, m_R=0, \mu)
  \left( 1 - i V_{\rm (qu)}''(\sigma_{\rm sol}) D_{\sigma\sigma}(0) \right)
  \Gamma_S(-q^2) }
\nonumber \\
& & =-c_{\bar\psi\psi}(p; g_R, m_R=0, \mu) V_{\rm (cl)}''(\sigma_{\rm sol})
      \dfrac{Z_m}{-V''(\sigma_{\rm sol})} \Gamma_S(-q^2),
\label{eq:(6.25)}
\end{eqnarray}
where we have used the renormalization of the composite operator
\begin{equation}
  (\bar\psi \psi )_R = Z_m (\bar\psi \psi ).
\end{equation}

Equating \EQ{eq:(6.24)} and \EQ{eq:(6.25)}, we finally obtain
the OPE coefficient function $c_{\bar\psi\psi}$:
\begin{eqnarray}
  c_{\bar\psi \psi }(p;g_R,\mu )
  &=&  -\dfrac{\Gamma_S(-p^2)}{p^2}
        \dfrac{Z_m^{-1}}{V''_{\rm (cl)}(\sigma_{\rm sol})}
  \nonumber\\
  &=& -\dfrac{\Gamma_S^{(R)}(-p^2)}{p^2} \dfrac{G_R}{N},
\label{eq:(6.27)}
\end{eqnarray}
where $G_R$ is defined as
\begin{equation}
  G_R\equiv 4\pi^2 \dfrac{g_R}{\mu ^2}.
\end{equation}
We can easily see that \EQ{eq:(6.27)} does satisfy the RG
 equation \EQ{eq:(6.12b)}.
Thus
\begin{eqnarray}
  c_{\bar\psi \psi }(\kappa p;g_R,\mu )
  \simeq c_{\bar\psi \psi }(p;\bar g(t),\mu )  \kappa^{-(4-\gamma_m^*)} .
\label{eq:(6.29)}
\end{eqnarray}

As for the \SxSB\ vacuum, the OPE coefficient
$c_{\bar\psi\psi}$ can be evaluated by
replacing the Yukawa-type vertex in the symmetric vacuum in \EQ{eq:(6.27)} with
that in the broken vacuum: \begin{equation}
  c_{\bar\psi \psi }(p;g_R,\mu )
  = -\dfrac{\Gamma_S^{(R)}(-p^2; M_d)}{p^2} \dfrac{G_R}{N}.
\label{eq:(6.30)}
\end{equation}
It is easy to see from \EQ{eq:(6.29)} that the coefficient
$ c_{\bar\psi \psi }$, having no extra momentum dependence, yields a correct
high
energy behavior of
$\Sigma_{\rm dyn}(-p^2)
=p^2 \VEV{(\bar\psi \psi )_R} c_{\bar\psi \psi }(p;g_R,\alpha,\mu)$:
\begin{equation}
  \Sigma_{\rm dyn}(-\kappa^2 p^2)
  = \kappa^2 p^2 \VEV{(\bar\psi \psi )_R}
  c_{\bar\psi \psi }(p;\bar g(\kappa),\mu )
  \sim \kappa^{-(2-\gamma_m^*)} \sim \kappa^{-(1-\omega)},
 \label{eq:(6.31)}
\end{equation}
which indeed agrees with the solution of the SD equation  given
in Appendix C.

This result is also similar to that of NJL$_{D<4}$ \cite{kn:KY90}.
See Appendix B for details.

\section{$\tM$-Dependent Renormalization}\label{sec-moreren}
In the previous section, we have carried out the symmetric renormalization of
the gauged NJL model with $\alpha<\alpha_c$ ($\omega\ne0$),
based on the effective potential around the symmetric vacuum \EQ{eq:(2.69)}.
Such a renormalization cannot directly apply to
$\alpha=\alpha_c$ ($\omega=0$),
since the expansion \EQ{eq:(2.69)} loses its validity at this point.
However, breakdown of the expansion \EQ{eq:(2.69)} does not necessarily
imply the non-renormalizability of the gauged NJL model at $\alpha=\alpha_c$.
Actually, as we have studied in section~{3},
the fermion scattering amplitudes remain finite even at $\alpha=\alpha_c$
as well as at $\alpha<\alpha_c$, once the bare four-fermion coupling is
fine-tuned  so as to make the fermion mass finite.
This fact suggests that the renormalization may be possible by a suitable
definition of renormalized parameters even at $\alpha=\alpha_c$.
Actually, the general form of the effective potential \EQ{eq:(2.64)}
remains
valid even at $\alpha=\alpha_c$.

In this section we present yet another
renormalization scheme which  renormalizes the effective potential
\EQ{eq:(2.64)} instead of \EQ{eq:(2.69)}.
The renormalization described in this
section introduces a redundant mass parameter $\tM$
($\tM$-dependent renormalization) and $\tM=0$ corresponds to the symmetric
renormalization.
The $\tM$-dependent renormalization is
related to the symmetric renormalization via a finite renormalization
at $0<\alpha<\alpha_c$ and remains valid at $\alpha=\alpha_c$.

The Yukawa-type vertex  \EQ{eq:(2.87)} at the non-trivial $M\ne0$ is
renormalized by the wave function renormalization: \begin{equation}
  \dfrac{\sigma}{ C_1 \left(\frac{\tM}{\Lambda}\right)^{1-\omega}
                 +D_1 \left(\frac{\tM}{\Lambda}\right)^{1+\omega} }
  = \dfrac{\sigma_\tR}{ C_1 \left(\frac{\tM}{\mu}\right)^{1-\omega}
                       +D_1 \left(\frac{\tM}{\mu}\right)^{1+\omega} },
\label{eq:(7.1)}
\end{equation}
which leads to the renormalized Yukawa-type vertex in the \SxSB\ vacuum
$M=M_d$:
\begin{equation}
  \Gamma_S^{(\tR)}(-p^2; M_d) =
   \dfrac{ \partial \Sigma(-p^2)/ \partial M \bigg|_{M=M_d}
         }{(2-\omega)\left[
             C_1\left(\frac{\tM}{\mu}\right)^{1-\omega}
           + D_1\left(\frac{\tM}{\mu}\right)^{1+\omega}
           \right] \left(\frac{M_d}{\tM}\right)^{1-\omega}
         }.
\end{equation}
Note that the same renormalization condition simultaneously makes finite
the Yukawa-type vertex in the symmetric vacuum:
\begin{equation}
  \Gamma_S^{(\tR)}(-p^2)
  = \Gamma_S^{(\tR)}(-p^2; M=0)
  =\frac{2}{1+\omega}
    \left(\frac{-p^2}{\mu^2}\right)^{-(1-\omega)/2}
    \left[ 1 + \frac{D_1}{C_1} \left(\frac{\tM}{\mu}\right)^{2\omega}
    \right]^{-1}.
\end{equation}

We next define the renormalized four-fermion coupling $g_{\tR}$ and
renormalized fermion mass $m_{\tR}$ by:
\begin{eqnarray*}
\lefteqn{
  \left[C_1\left(\frac{\tM}{\Lambda }\right)^{-\omega }
      + D_1 \left(\frac{\tM}{\Lambda }\right)^{\omega }\right]^2
  \left(\frac{1}{g^*}-\frac{1}{g}\right)
} \\
\lefteqn{
 +D_1 \left(\frac{\tM}{\Lambda }\right)^{\omega }
  \left[D_1\left(\frac{\tM}{\Lambda }\right)^{\omega } + \frac{2-\omega }{2}
  C_1 \left(\frac{\tM}{\Lambda }\right)^{-\omega }\right]
  \left(\frac{1}{\tilde g^*}-\frac{1}{g^*}\right)
} \\
& & =
  \left[C_1\left(\frac{\tM}{\mu }\right)^{-\omega}
      + D_1\left(\frac{\tM}{\mu} \right)^{\omega }
  \right]^2
  \left(\frac{1}{g_{\tR}^*}-\frac{1}{g_{\tR}}\right)
  \\
& & \quad + D_1\left(\frac{\tM}{\mu }\right)^{\omega }
  \left[D_1\left(\frac{\tM}{\mu}\right)^{\omega}
    + \frac{2-\omega}{2} C_1 \left(\frac{\tM}{\mu }\right)^{-\omega }
  \right]
  \left(\frac{1}{\tilde g_{\tR}^*}-\frac{1}{g_{\tR}^*}\right),
\yesnumber
\label{eq:(7.4)}
\end{eqnarray*}
and
\begin{equation}
  \frac{m_0}{g}
  \left[ C_1\left(\frac{\tM}{\Lambda }\right)^{-(1+\omega)}
       + D_1\left(\frac{\tM}{\Lambda }\right)^{-(1-\omega)}
  \right]
  = \frac{m_{\tR}}{g_{\tR}}
    \left[ C_1 \left(\frac{\tM}{\mu }\right)^{-(1+\omega)}
         + D_1\left(\frac{\tM}{\mu }\right)^{-(1-\omega)}
    \right] .
\label{eq:(7.5)}
\end{equation}
Here $g_{\tR}^*$ and $\tilde g_{\tR}^*$ are arbitrary parameters
corresponding to the choice of $\tM$ in this renormalization scheme.
It is easy to see that the \EQ{eq:(7.4)} reduces to the symmetric
renormalization \EQ{eq:(5.17)} in the $\tM \rightarrow 0$ limit. Also
note that unlike the symmetric
renormalization, $g_{\tR}^*$ in \EQ{eq:(7.4)}  does not corresponds to a
``critical'' coupling, unless $\tM/\mu=0$.

Starting from the bare effective potential \EQ{eq:(2.64)},
 we obtain a renormalized
effective potential in the $\Lambda\rightarrow\infty$ limit:
\begin{eqnarray*}
  -\dfrac{4\pi^2}{N} \dfrac{V_{\tR}(\sigma_{\tR},\pi_{\tR}=0)}{\mu^4}
  &=& \dfrac{1}{g_{\tR}} \dfrac{m_{\tR} \sigma_{\tR}}{\mu^2}
     +\frac{1}{2}\left(\frac{1}{g_{\tR}^*}-\frac{1}{g_{\tR}}\right)
                 \left(\frac{\sigma_{\tR}}{\mu}\right)^2
  \\
  & & +\frac{1}{2}\left(\frac{1}{\tilde g_{\tR}^*}-\frac{1}{g_{\tR}^*}\right)
                  \left(\frac{\sigma_{\tR}}{\mu}\right)^2
       \dfrac{D_1\left[D_1\left(\frac{\tM}{\mu}\right)^{2\omega}
                     +\frac{2-\omega }{2}C_1\right]}
             {\left[C_1\left(\frac{\tM}{\mu }\right)^{-\omega }
               +D_1\left(\frac{\tM}{\mu }\right)^{\omega }
              \right]^2}
  \\
  & & +\frac{1}{2}\left(\frac{1}{\tilde g^*}-\frac{1}{g^*}\right)
                  \left(\frac{\sigma _{\tR}}{\mu }\right)^2
       \dfrac{ \frac{2-\omega }{2} C_1 D_1
               \left[\left(\frac{M}{\tM}\right)^{2\omega }-1\right]}
             {\left[C_1 \left(\frac{\tM}{\mu }\right)^{-\omega }
              + D_1 \left(\frac{\tM}{\mu }\right)^{\omega }
              \right]^2}.
  \label{eq:(7.6)} \yesnumber
\end{eqnarray*}
\EQ{eq:(2.40)} and \EQ{eq:(7.1)} lead to the relation between $M$ and the field
$\sigma_{\tR}$:
\begin{equation}
  \dfrac{\sigma _{\tR}}{\mu }
  =\left[C_1 + D_1\left(\frac{\tM}{\mu }\right)^{2\omega }\right]
   \left(\frac{M}{\mu }\right)^{2-\omega}.
\end{equation}
Note that $\tM$ is a certain constant but not a field like $M$.

We  choose the parameters $g_{\tR}^*$ and $\tilde g_{\tR}^*$;
\begin{equation}
  \dfrac{1}{g_{\tR}^* - \tilde g_{\tR}^*} =
  \frac{1}{\omega } + \mbox{regular function of $\omega $},
\end{equation}
so as to take a sensible
$\omega \rightarrow 0$ ($\alpha \rightarrow \alpha _c$) limit.
Hereafter we take the simplest choice:
\begin{equation}
  g_{\tR}^* = g^* = \frac{(1+\omega )^2}{4}, \qquad
  \tilde g_{\tR}^* = \tilde g^* = \frac{(1-\omega )^2}{4}.
\end{equation}
Such a choice of $g_{\tR}^*$ and $\tilde g_{\tR}^*$ leads to the
following effective potential at $\omega=0$ ($\alpha=\alpha_c$):
\begin{eqnarray}
\lefteqn{
  -\dfrac{4\pi^2}{N} \dfrac{V_{\tR}(\sigma_{\tR},\pi_{\tR}=0)}{\mu^4}
} \nonumber\\
  & &
  = \dfrac{1}{g_{\tR}} \dfrac{m_{\tR} \sigma_{\tR}}{\mu^2}
   +\frac{1}{2} \left(\frac{1}{g_{\tR}^*} - \frac{1}{g_{\tR}}\right)
    \left(\dfrac{\sigma _{\tR}}{\mu }\right)^2
   -4 \left(\dfrac{\sigma _{\tR}}{\mu }\right)^2
     \dfrac{\frac{3}{4}+\delta _0+\ln\frac{\mu M}{\tM^2}}
           {\left[1+\delta _0-\ln\frac{\tM}{\mu }\right]^2}.
\end{eqnarray}

The stationary condition of the effective potential \EQ{eq:(7.6)}
yields the (renormalized) gap equation.
Solving the gap equation in the chiral symmetric limit $m_{\tR}=0$,
we obtain the scaling relation of the dynamical mass of the fermion
$M_d\equiv \VEV{M}$:
\begin{equation}
  \left(\frac{M_d}{\tM}\right)^{2\omega }
  = -\frac{D_1}{C_1} \left(\frac{\tM}{\mu }\right)^{2\omega }
    -\dfrac{\frac{1}{g_{\tR}^*}-\frac{1}{g_{\tR}}}
           {\frac{1}{\tilde g_{\tR}^*}-\frac{1}{g_{\tR}^*}}
     \dfrac{\left[C_1\left(\frac{\tM}{\mu }\right)^{-\omega }
              +D_1\left(\frac{\tM}{\mu }\right)^{\omega } \right]^2
          }{C_1 D_1} .
\label{eq:(7.10)}
\end{equation}
The critical NJL coupling $g_{\tR}^{\rm crit}$ is determined from
\EQ{eq:(7.10)} at $M_d=0$:
\begin{equation}
  \frac{1}{g_{\tR}^{\rm crit}}
  = \frac{1}{g_{\tR}^*}
  + \dfrac{\left(\frac{1}{\tilde g_{\tR}^*} - \frac{1}{g_{\tR}^*}\right)
            D_1^2 \left(\frac{\tM}{\mu }\right)^{2\omega }
         }{\left[C_1\left(\frac{\tM}{\mu }\right)^{-\omega }
             +D_1\left(\frac{\tM}{\mu }\right)^{\omega } \right]^2} .
\end{equation}
Note here that the critical coupling can be calculated as
\begin{equation}
  g_{\tR}^{\rm crit} =
  \begin{cases}
    g_{\tR}^*,        & for $\tM/\mu = 0$ \\
    \tilde g_{\tR}^*, & for $\tM/\mu \rightarrow \infty$
  \end{cases} .
\end{equation}
If we take $\tM=M_d$ (a natural choice in the \SxSB\ phase)
in \EQ{eq:(7.10)}, we find:
\begin{equation}
  \left(\frac{\tM}{\mu}\right)^{2\omega}
  = \left(\frac{M_d}{\mu}\right)^{2\omega}
  = -\frac{C_1}{D_1} \dfrac{\frac{1}{g_{\tR}^*} -\frac{1}{g_{\tR}}}
                           {\frac{1}{\tilde g_{\tR}^*} -\frac{1}{g_{\tR}}}.
\label{eq:(7.13)}
\end{equation}

Now, we are ready to discuss the RG properties.
It is straightforward to derive the $\beta$ function from the definition of
renormalized four-fermion coupling \EQ{eq:(7.4)}:
\begin{equation}
  \beta _g(g_{\tR},\alpha ; \tM/\mu )
  = -2\omega g_{\tR}^2
    \dfrac{
      \left(\frac{1}{g_{\tR}^*}-\frac{1}{g_{\tR}}\right)C_1^2
      \left(\frac{\tM}{\mu}\right)^{-2\omega }
     -\left(\frac{1}{\tilde g_{\tR}^*}-\frac{1}{g_{\tR}}\right)D_1^2
      \left(\frac{\tM}{\mu}\right)^{2\omega }
    }{
      \left[C_1\left(\frac{\tM}{\mu }\right)^{-\omega }
       +D_1\left(\frac{\tM}{\mu }\right)^{\omega }\right]^2
    } ,
\label{eq:(7.14)}
\end{equation}
and the anomalous dimension $\gamma _m$ from \EQ{eq:(7.5)}:
\begin{equation}
  \gamma_m(g_{\tR}, \alpha ; \tM/\mu )
  = 1+\omega
    \dfrac{
      \left(2\frac{g_{\tR}}{g_{\tR}^*}-1\right)C_1^2
      \left(\frac{\tM}{\mu }\right)^{-2\omega }
     -\left(2\frac{g_{\tR}}{\tilde g_{\tR}^*}-1\right)D_1^2
      \left(\frac{\tM}{\mu }\right)^{2\omega }
     }{
      \left[C_1\left(\frac{\tM}{\mu }\right)^{-\omega }
       +D_1\left(\frac{\tM}{\mu }\right)^{\omega } \right]^2
     } .
\label{eq:(7.15)}
\end{equation}

Putting $\tM=0$ in \EQ{eq:(7.14)} and \EQ{eq:(7.15)}, we find
(Fig.~\ref{fig:rgefunc})
\begin{eqnarray}
   \beta _g(g_{\tR}, \alpha ; \tM=0)
  &=& 2\omega g_{\tR}\left(1-\frac{g_{\tR}}{g_{\tR}^*}\right), \\
   \gamma _m(g_{\tR}, \alpha ; \tM=0)
  &=& 1-\omega +2\omega \frac{g_{\tR}}{g_{\tR}^*},
\end{eqnarray}
which coincide with  \EQQ{eq:(5.21)}{eq:(5.22)} as they should.
\par
Plugging \EQ{eq:(7.13)} into \EQ{eq:(7.14)} and \EQ{eq:(7.15)}, on the
other hand, we obtain (Fig.~\ref{fig:rgefunc})
\begin{eqnarray}
   \beta _g(g_{\tR}, \alpha ; \tM=M_d)
  &=& -2(g_{\tR}-\tilde g_{\tR}^*)(g_{\tR}-g_{\tR}^*), \\
   \gamma _m(g_{\tR}, \alpha ; \tM=M_d)
  &=& 2g_{\tR}+\frac{\alpha }{2\alpha _c},
\end{eqnarray}
which remain valid even at $\alpha=\alpha_c$
and coincide with the form of the RG functions of
bare parameters,  \EQ{eq:(C.19)} and \EQ{eq:(C.24)},
given in Appendix C.

Finally, we make a brief comment on the relation of the symmetric
renormalization and the $\tM$-dependent renormalization.
 From \EQ{eq:(5.13)} and \EQ{eq:(7.1)}, we obtain
\begin{equation}
  \sigma_R = \tZ_\sigma \sigma_\tR,
  \qquad
  \tZ_\sigma^{-1}
  \equiv 1+\frac{D_1}{C_1} \left(\frac{\tM}{\mu}\right)^{2\omega}
\end{equation}
in the $\Lambda\rightarrow\infty$ limit.
Using \EQ{eq:(5.17)} and \EQ{eq:(7.4)}, we obtain a relation between
the renormalized four-fermion coupling of the symmetric renormalization and
that of the $\tM$-dependent one:
\begin{equation}
  \frac{1}{g^*}-\frac{1}{g_R} =
  \tZ_\sigma^{-2} \left(\frac{1}{g^*}-\frac{1}{g_\tR}\right)
 +\frac{D_1^2}{C_1^2} \left(\frac{\tM}{\mu}\right)^{4\omega}
  \left(\frac{1}{\tilde g^*}-\frac{1}{g^*}\right),
\end{equation}
where we have used $g^*=g_R^*=g_{\tR}^*$ and $\tilde g^* = \tilde g_{\tR}^*$.
It is easy to show that the current mass in the $\tM$-dependent renormalization
scheme is expressed by
\begin{equation}
  m_R = \tZ_m m_\tR, \qquad
  \tZ_m = \tZ_\sigma^{-1} \dfrac{g_R}{g_\tR}
\end{equation}
in the $\Lambda\rightarrow\infty$ limit.
Thus, the symmetric renormalization and
the $\tM$-dependent renormalization are
{\em connected by a finite renormalization unless $\omega=0$}.

The OPE coefficient functions in $\tM$-dependent renormalization
$\tilde c_\unit'$ and $\tilde c_{\bar\psi\psi}$ are calculated from
\begin{subequations} \label{eq:(7.23)}
\begin{eqnarray}
  \tilde c_\unit'(p; g_\tR, \alpha, \tM/\mu; \mu)
    &=&  \tZ_m c_\unit'(p; g_\tR, \alpha; \mu),
    \\
  \tilde c_{\bar\psi\psi}(p; g_\tR, \alpha, \tM/\mu; \mu)
    &=&  \tZ_m^{-1} c_{\bar\psi\psi}(p; g_\tR, \alpha; \mu).
\end{eqnarray}
\end{subequations}
By using \EQ{eq:(6.10)}, \EQ{eq:(6.27)} and \EQ{eq:(6.30)}, we find
\begin{subequations}
\begin{eqnarray}
\tilde c_\unit'(g_\tR, \alpha, \tM/\mu; \mu)
  &=& \dfrac{\frac{1}{p^2}\frac{2-\omega}{2\omega} \Gamma_S^{(\tR)}(-p^2; M_d)}
            {\frac{g_\tR}{g^*}-1
              + \tZ_\sigma^2 \frac{D_1^2}{C_1^2}
                \left(\frac{\tM}{\mu}\right)^{4\omega}
                \left(\frac{1}{\tilde g^*}-\frac{1}{g^*}\right)
            },
  \\
\tilde c_{\bar\psi\psi}(g_\tR, \alpha, \tM/\mu; \mu)
  &=& -\frac{1}{p^2} \frac{4\pi^2 g_\tR}{N\mu^2} \Gamma_S^{(\tR)}(-p^2; M_d)
\end{eqnarray}
\end{subequations}
for the \SxSB\ vacuum, and
\begin{subequations}
\begin{eqnarray}
\tilde c_\unit'(g_\tR, \alpha, \tM/\mu; \mu)
  &=& \dfrac{\frac{1}{p^2} \Gamma_S^{(\tR)}(-p^2; M=0)}
            {1-\frac{g_\tR}{g^*}
              - \tZ_\sigma^2 \frac{D_1^2}{C_1^2}
                \left(\frac{\tM}{\mu}\right)^{4\omega}
                \left(\frac{1}{\tilde g^*}-\frac{1}{g^*}\right)
            },
  \\
\tilde c_{\bar\psi\psi}(g_\tR, \alpha, \tM/\mu; \mu)
  &=& -\frac{1}{p^2} \frac{4\pi^2 g_\tR}{N\mu^2} \Gamma_S^{(\tR)}(-p^2; M=0)
\end{eqnarray}
\end{subequations}
for the symmetric vacuum, where we have used
\begin{equation}
  \Gamma_S^{(R)}(-p^2; M) = \tZ_\sigma^{-1} \Gamma_S^{(\tR)}(-p^2; M) .
\end{equation}
We notice that
\begin{equation}
 \tZ_\sigma \rightarrow
 {1 \over -2\omega \ln {\tM \over \mu} + {\cal O}(\omega^2)}
\end{equation}
as $\omega \rightarrow 0$.   Therefore $\tilde c_{\bar\psi\psi}$
on \SxSB\ vacuum and
$\tilde c'_\unit$  remain finite in $\omega\rightarrow0$ limit, while
$\tilde c_{\bar\psi\psi}$ in the symmetric vacuum diverges.

Special attention should be paid to the $\omega\rightarrow0$ limit
where two renormalization schemes
are not connected by a finite renormalization.
The above calculation implies  that the OPE coefficient function
$\tilde c_{\bar\psi\psi}$ cannot be made finite in a vacuum-independent manner.
Such a peculiar phenomenon may be an artifact coming from the
infrared singularity in the symmetric vacuum at $\alpha=\alpha_c$.

\section{Conclusion and Discussion}
We have presented renormalization of
the simplest gauged NJL model, gauge theories with standing gauge coupling
plus four-fermion interaction, in the ladder approximation. Through the CJT
effective potential written in terms of the auxiliary fields, we have
established, in an analogous manner to \NJLd, a phase-independent
renormalization (symmetric renormalization) valid both for the symmetric and
the S$\chi$SB phases as far as $\alpha\ne\alpha_c$.
Accordingly, the $\beta$ function and the anomalous dimension were
obtained phase-independently in both phases: The theory has a nontrivial
UV fixed line and a large anomalous dimension there. The OPE was explicitly
constructed, which is consistent with the large
anomalous dimension in both phases.
The Wilson coefficient for the unit operator has an extra power behavior
other than the anomalous dimension.
The symmetric renormalization done on the symmetric
vacuum breaks down at the end point $\alpha = \alpha_c$ of the critical line,
while the $\tM$-dependent renormalization  still remains valid.

The reason why the renormalization is possible in \NJLd\ and gauged NJL
model is very simple. In \NJLd\ the four-fermion operators
become relevant/marginal:
\(
  \dim \left(( \bar \psi \psi)^2 \right)= 2 \dim (\bar \psi \psi)
  = 2(D-1 - \gamma_m) = 2 \le D
\), while other fermion operators become irrelevant:
\(
  \dim \left( (\bar \psi \psi)^4 \right)=
  \dim \left( \partial_\mu (\bar \psi \psi)
    \partial^\mu (\bar \psi \psi) \right)
  = 4 >D
\), etc. for $2 \le D < 4$.
In the same way, the four-fermion operators
in the gauged NJL model become relevant/marginal:
\(
  \dim \left( (\bar \psi \psi)^2 \right) = 2(3 - \gamma_m)
  = 2 (2-\sqrt{1-\alpha/\alpha_c}) \le 4
\), while other operators become irrelevant:
\(
  \dim \left( (\bar \psi \psi)^4 \right) = 4(2-\sqrt{1-\alpha/\alpha_c}) >4
\),
\(
  \dim \left( \partial_\mu (\bar\psi\psi) \partial^\mu (\bar\psi\psi) \right)
   = 2 (3 - \sqrt{1-\alpha/\alpha_c}) > 4
\), etc. for $\alpha_c \geq \alpha >0$.\footnote{
These higher dimensional operators can be explicitly shown to be suppressed
in part by the power of the cutoff  in the
low energy physics through renormalization.\cite{kn:MTY92}
}

At this point it should be emphasized that our renormalization breaks down
at the pure NJL limit $\alpha \rightarrow 0$. In that limit our renormalized
effective potential \EQ{eq:(5.19)} has a singularity of $1/\alpha$,\footnote{
The $1/\alpha$ singularity instead of the logarithmic divergence is an
artifact of the inadequate approximation taking only the first
two dominant terms in the solution of SD equation for $\alpha > 0$.
 Taking account of the third dominant
term, we can recover the correct NJL limit\cite{kn:NSY89,kn:BL92}.
}
a signal of the appearance of the logarithmic
divergence for the four-point vertex of the local auxiliary fields. Such
a divergence can only be removed by introduction of eight-fermion operator
(``$\lambda \phi^4$ counter term'') which now becomes a marginal operator;
\(
  \dim \left( (\bar \psi \psi)^4 \right) = 4(2-\sqrt{1-\alpha/\alpha_c})
  \rightarrow 4
\) at $\alpha \rightarrow 0$.
The $1/\alpha$ singularity also appears in the
auxiliary field propagator \EQ{eq:(5.20)}, another signal of the
logarithmic divergence in the induced kinetic term of the auxiliary field.
This again can only be removed by introduction of derivative-type four-fermion
operator (``counter term for kinetic term
$\partial_\mu \phi \partial^\mu \phi$'') which also becomes marginal;
\(
  \dim\left( \partial_\mu (\bar \psi \psi) \partial^\mu (\bar \psi \psi)\right)
  = 2 (3 - \sqrt{1-\alpha/\alpha_c})
  \rightarrow 4
\) at $\alpha \rightarrow 0$.
Without such extra ``higher dimensional'' operators, the pure NJL
model in four dimensions cannot be renormalized even in the nonperturbative
sense of $1/N$ expansion. {\em The presence of gauge interaction turns these
``higher dimensional'' operators into irrelevant ones and hence make the
renormalization possible without such additional ``counter
terms''}.\cite{kn:Yama90-,kn:KHKD90,kn:KSY91}
Similar comments also apply to the \NJLd: The above $1/\alpha$ singularity
corresponds to the $1/(D-4)$ singularity in \EQ{eq:(5.8)} and
\EQ{eq:(5.12)} of \NJLd.

Such a formal resemblance between the renormalization of \NJLd\ and that of the
gauged NJL model may not be a mere accident.
As is well known, the renormalizable \NJLd\ model is equivalent to the
Yukawa model
with the Yukawa coupling lying on the nontrivial IR fixed point in the
$1/N$ leading order.
Actually, the $\beta$ function of dimensionless Yukawa coupling $y$ is
given by
\begin{equation}
  \beta (y) = \left(\frac{D}{2}-2\right)y
               + 4N \dfrac{\Gamma (3-D/2)\left(\Gamma (D/2)\right)^2}
                   {(4\pi )^{D/2} \Gamma (D-1)} y^3,
\label{eq:(8.1)}
\end{equation}
which indeed has a nontrivial infrared (IR) fixed point
\begin{equation}
  y^2
= \dfrac{(4\pi )^{D/2} \Gamma (D-1)}{4N \Gamma (2-D/2)\left(\Gamma
(D/2)\right)^2},
\label{eq:(8.2)}
\end{equation}
corresponding to the \NJLd\ model.
Although two parameters
(Yukawa coupling and scalar boson mass) are needed in order for the
Yukawa model to be renormalizable, the number of parameters is now reduced
by this constraint to that of \NJLd.
This is the essence of the renormalizability of \NJLd. This argument obviously
breaks down at $D=4$ where the nontrivial IR fixed point disappears. This
situation reflects the fact that the NJL model is not renormalizable
in four dimensions as an interacting field theory even in $1/N$
expansion.

However, {\em in the presence of gauge interaction} ($\alpha \ne 0$)
the Yukawa model shows a similar structure to \EQ{eq:(8.1)}
even in four dimensions:
\begin{equation}
  \beta_y (\alpha,y) =  - \dfrac{6C_F \alpha }{4\pi } y
+ \dfrac{N}{8\pi ^2}y^3,
\end{equation}
where $C_F$ is the quadratic Casimir of the fermion representation and
we have ignored the graphs containing scalar particle loop.
Now, the $\beta$ function of gauge coupling may be parameterized as
\begin{equation}
  \beta_{\alpha} (\alpha,y) = - \dfrac{3C_F/\pi }{A} \alpha ^2,
\end{equation}
where $A(=18C_F/(11N-2N_f)$ for $N_f$-flavored $SU(N)$ gauge theory) is the
measure of the running speed of the gauge coupling, i.e., large $A (\gg 1)$
means ``walking'' (slowly running) gauge coupling and $A \rightarrow \infty$
corresponds to the non-running (``standing'') case studied in this paper. For
$A>1$ the above $\beta$ functions lead to the RG invariant IR stable subspace
of the couplings:
\begin{equation}
  y^2 = \frac{12\pi C_F}{N} \frac{A-1}{A} \alpha ,
\label{eq:(8.5)}
\end{equation}
in analogue to \EQ{eq:(8.2)}.
This is another expression of a suggestion\cite{kn:KSY91,kn:Kra92}
that {\em the presence of gauge
interactions with $A >1$ may make the theory renormalizable}.
The explicit renormalization procedure in the present paper is actually
a concrete example of this suggestion in the special case
$A \rightarrow \infty$. Such a possibility in the case with
running/walking gauge coupling $A < \infty$ will
be further studied in the subsequent paper\cite{kn:KSTY92}.

\section*{Acknowledgements}
We would like to thank Tom Appelquist, Bill Bardeen, Howard Georgi,
Peter Hasenfratz, Kiyoshi Higashijima, Bob Holdom, Yoshio Kikukawa
and Volodya Miransky
for interesting
discussions. Two of the authors (K.Y. and K.-I.K.) are supported in part by
the Takeda Science Foundation and the Ishida
Foundation, and by the International Collaboration Program of the Japan
Society for the Promotion of Science. K.-I.K. is also supported in part by
the Grant-in-Aid for Scientific Research from the Ministry of
Education, Science and Culture (\#04740138).
Part of this work was done
during the stay of K.-I.K. and K.Y. at Aspen Institute for Physics in 1992
summer.

\newpage
\appendix
\section{\NJLd\ model}\label{sec-appA}
In this appendix, we derive the effective potential and
the auxiliary field propagators  in \NJLd\ model.

The auxiliary field technique enables us to rewrite the  original
lagrangian \EQ{eq:(5.1)} into an equivalent one:
\begin{equation}
  {\cal L} = \bar\psi i\fsl{\partial} \psi
- \bar\psi (\sigma + \pi i\gamma _5)\psi
          - V_{\rm (cl)}(\sigma ,\pi ),
\label{eq:(A.1)}
\end{equation}
where classical potential of auxiliary fields $\sigma ,\pi $ is given by
\begin{equation}
  V_{\rm (cl)}
= \frac{N}{G} \left[\frac{1}{2}\left(\sigma ^2 + \pi ^2\right)
- m_0 \sigma \right].
\label{eq:(A.2)}
\end{equation}
The effective action is evaluated solely from the fermion integral
within the $1/N$ leading approximation:
\begin{equation}
  \Gamma [\sigma ,\pi ]
  = -i \ln \det \left(i\fsl{\partial}-\sigma -i\gamma _5 \pi \right)
      -\int d^D x V_{\rm (cl)}(\sigma ,\pi ).
\end{equation}
The effective potential $V$ is calculated from the effective action
\begin{equation}
  V(\sigma ,\pi ) \equiv -
  \dfrac{\Gamma [ \sigma ={\rm const}, \pi ={\rm const}]} {\Omega},
\end{equation}
where $\Omega$ is the space-time volume.
Thus, we obtain
\begin{equation}
  V(\sigma ,\pi ) = V_{\rm (cl)}(\sigma ,\pi ) + V_{\rm (qu)}(\sigma ,\pi),
\end{equation}
where
\begin{eqnarray}
  V_{\rm (qu)}(\sigma ,\pi )
    &\equiv & -2N\int ^{\Lambda } \dfrac{d^D p}{(2\pi )^D i}
\ln\left(1+\dfrac{\sigma ^2+\pi ^2}{-p^2}\right)
    \\
    &=&  -\dfrac{2N}{(4\pi )^{D/2} \Gamma (D/2)}
\int _0^{\Lambda ^2} d\Ep (\Ep)^{D/2-1}
          \ln\left(1+\dfrac{\sigma ^2+\pi ^2}{\Ep}\right),
\label{eq:(A.7)}
\end{eqnarray}
with the momentum integral being regularized by the ultraviolet cutoff
$\Lambda$.
Using the following formula
\begin{equation}
  \int_0^\infty dt t^{z-1}
    \left[ \ln\left(1+\frac{1}{t}\right) - \frac{1}{t} \right]
  = -\frac{1}{z} B(z-1,2-z),
\end{equation}
it is rather straightforward to integrate \EQ{eq:(A.7)}:
\begin{eqnarray}
\lefteqn{
   \dfrac{(4\pi)^{D/2} \Gamma(D/2)}{4N} V_{\rm (qu)}(\sigma, \pi )
} \nonumber\\
  &=& - \dfrac{\Lambda ^{D-2}}{g^*}\dfrac{\sigma ^2+\pi ^2}{2}
      - \dfrac{\coeAd}{2-D/2}\dfrac{\left(\sigma ^2+\pi ^2\right)^{D/2}}{D}
      - \sum_{n=0}^{\infty } \dfrac{\Lambda ^D (-1)^n }{(D-4-2n)(n+2)}
        \left(\dfrac{\sigma ^2+\pi ^2}{\Lambda ^2}\right)^{n+2}
  \hspace{-1.5em},
  \hspace{2.5em}
\end{eqnarray}
with  $g^*\equiv D/2-1$ and $\coeAd \equiv B(D/2-1,3-D/2)$.
Finally, we find the effective potential
\begin{eqnarray}
\lefteqn{
 - \dfrac{(4\pi )^{D/2} \Gamma (D/2)}{4N \Lambda ^D} V(\sigma ,\pi )
} \nonumber\\
  &=& \dfrac{1}{g} \dfrac{m_0 \sigma }{\Lambda ^2}
   +\left(\dfrac{1}{g^*}-\dfrac{1}{g}\right)
\dfrac{\sigma ^2+\pi ^2}{2\Lambda ^2}
  - \dfrac{1}{2-\frac{D}{2}} \dfrac{\coeAd}{D}
    \left(\dfrac{\sigma ^2+\pi^2}{\Lambda ^2}\right)^{D/2}
 + {\cal O}\left(\left(\frac{\sigma ^2+\pi ^2}{\Lambda^2}\right)^2\right),
   \hspace{2em} \phantom{a}
\label{eq:(A.10)}
\end{eqnarray}
with the dimensionless four-fermion coupling $g$ being defined by
\begin{equation}
  g \equiv \dfrac{4\Lambda ^{D-2}}{(4\pi )^{D/2} \Gamma (D/2)}G.
\end{equation}
It is easy to see that the last term $\sim {\cal O}(\Lambda ^{D-4})$
in \EQ{eq:(A.10)} becomes negligible for sufficiently large $\Lambda$
and will be neglected
in the following calculation of the propagators of auxiliary fields.
Note that $g^*\rightarrow0$ and $\coeAd\rightarrow\infty$ as $D\rightarrow2$,
so that the divergences in the second and third term cancel each other
to give a well-known logarithmic potential in the Gross-Neveu model
\cite{kn:Coleman}:
\begin{equation}
  -\frac{\pi}{N} V(\sigma,\pi)
   =   \frac{1}{g} m_0 \sigma - \frac{1}{g} \frac{\sigma^2 + \pi^2}{2}
      +\frac{\sigma^2+\pi^2}{2}
       \left(1-\ln\left(\frac{\sigma^2+\pi^2}{\Lambda^2}\right)\right)
      + \Lambda^2 {\cal O}(\left(\frac{\sigma^2+\pi^2}{\Lambda^2}\right)^2).
\end{equation}

Let us now consider solution of the gap equation
$\partial V/\partial \sigma = 0$ for \EQ{eq:(A.10)}:
\begin{equation}
 0 = \frac{1}{g} \dfrac{m_0}{\Lambda}
   + \left( \dfrac{1}{g^*}-\dfrac{1}{g} \right) \frac{\sigma}{\Lambda}
   - \dfrac{\coeAd}{2-\frac{D}{2}} \left(\frac{\sigma}{\Lambda}\right)^{D-1}
   + \cdots.
\label{eq:(B.5)}
\end{equation}
It is convenient to parameterize the solution of the gap equation
$\sigma_{\rm sol}$ as:
\begin{equation}
  \sigma_{\rm sol} = \sigma_{\rm spont} + \sigma_{\rm explicit},
\end{equation}
with $\sigma_{\rm spont}$ being the solution in the chiral limit ($m_0=0$):
\begin{equation}
  \dfrac{\coeAd}{2-D/2} \sigma_{\rm spont}^{D-2}
  = \begin{cases}
       \left( \dfrac{1}{g^*} - \dfrac{1}{g} \right) \Lambda^{D-2} + \cdots
         & for \SxSB\ vacuum  ($g>g^*$) \cr\cr
       \hspace*{1em} \frac{\strut}{\strut} 0
         & for symmetric vacuum
    \end{cases}.
\label{eq:(B.7)}
\end{equation}
Actually it is easy to see from \EQ{eq:(A.10)} that
the symmetric vacuum $\VEV{\sigma }=\VEV{\pi }=0$ becomes unstable
in the strong coupling region $g>g^*$.
In this vacuum the fermion acquires the dynamical mass
\begin{equation}
  \Sigma_{\rm dyn}(-p^2) = \Gamma_S(p,q=0) \sigma _{\rm spont}
 = \sigma_{\rm spont},
\end{equation}
where $\Gamma_S(p,q)$ is the Yukawa-type vertex defined in an analogous way
to \EQ{eq:(2.10)} and calculated as
\begin{equation}
  \Gamma_S(p,q)=1.
\end{equation}

The nonvanishing bare fermion mass ($m_0\ne0$)
causes non-zero $\sigma_{\rm explicit}$:
\begin{equation}
  \sigma _{\rm explicit}
  = \sigma _{\rm sol} - \sigma _{\rm spont}
  =   \left. \dfrac{\partial \sigma _{\rm sol}}{\partial m_0}
\right|_{m_0=0}  m_0  + \cdots .
\end{equation}
We can evaluate $\sigma_{\rm explicit}$ as a Taylor series
from \EQ{eq:(B.5)} in $m_0$:
\begin{equation}
  \sigma_{\rm explicit}
  = \begin{cases}
       \dfrac{1}{D-2} \dfrac{m_0}{\frac{g}{g^*}-1} + \cdots
         & for \SxSB\ vacuum \cr\cr
       \dfrac{m_0}{1-\frac{g}{g^*}} + \cdots
         & for symmetric vacuum
    \end{cases}.
\label{eq:(B.8)}
\end{equation}

Hence the current mass of the fermion may be written as
\begin{equation}
  \Sigma_{\rm explicit}(-p^2) = \Gamma _S(p,q=0) \sigma_{\rm explicit}
                             = \sigma_{\rm explicit}.
\label{eq:(A.17)}
\end{equation}
Thus, the current mass $\Sigma_{\rm explicit}$ has
the same (constant) high energy behavior as that of the dynamical mass
in the \NJLd\ model.

The auxiliary field propagators $D_{\sigma \sigma }^{-1}$
and $D_{\pi \pi }^{-1}$
are given by the second derivative of the effective action:
\begin{subequations}
\begin{eqnarray}
  iD_{\sigma \sigma }^{-1}(-q^2)
    &=& \FT \left. \dfrac{\delta ^2\Gamma [\sigma ,\pi ]}
{\delta \sigma (x)\delta \sigma (0)}
            \right|_{\sigma =\sigma _{\rm sol}, \pi =0},
    \\
  iD_{\pi \pi }^{-1}(-q^2)
    &=& \FT \left. \dfrac{\delta ^2\Gamma [\sigma ,\pi ]}
{\delta \pi (x)\delta \pi (0)}
            \right|_{\sigma =\sigma _{\rm sol},\pi =0}.
\end{eqnarray}
\end{subequations}
They are evaluated as
\begin{subequations}
\begin{eqnarray}
  iD_{\sigma \sigma }^{-1}(-q^2)
    &=& -\int \dfrac{d^D k}{(2\pi )^D i}
         \tr\left[i\dfrac{i}{\fsl{k}-\sigma _{\rm sol}}
              i\dfrac{i}{\fsl{k}-\fsl{q}-\sigma _{\rm sol}}\right]
        -\left. \dfrac{\partial^2 V_{\rm (cl)}(\sigma ,\pi )}
{\partial \sigma ^2}
         \right|_{\sigma =\sigma _{\rm sol}, \pi =0},
    \\
  iD_{\pi \pi }^{-1}(-q^2)
    &=& -\int \dfrac{d^D k}{(2\pi )^D i}
         \tr\left[\gamma _5 \dfrac{i}{\fsl{k}-\sigma _{\rm sol}}
    \gamma _5 \dfrac{i}{\fsl{k}-\fsl{q}-\sigma _{\rm sol}}\right]
        -\left. \dfrac{\partial^2 V_{\rm (cl)}(\sigma ,\pi )}
{\partial \pi ^2}
         \right|_{\sigma =\sigma _{\rm sol}, \pi =0}.
\end{eqnarray}
\end{subequations}

The subtraction at zero momentum makes these expressions
finite in the $\Lambda \rightarrow \infty $ limit:\footnote{
  Thus, we do not need to worry about keeping
  the regularization chiral invariant
  for the auxiliary field propagator with non-vanishing
  momentum $q^2\not=0$.
}
\begin{subequations} \label{eq:(A.21)}
\begin{eqnarray}
\lefteqn{
  iD_{\sigma \sigma }^{-1}(-q^2)-iD_{\sigma \sigma }^{-1}(0)
} \nonumber\\
  &=&
  8N \dfrac{\Gamma (2-D/2)}{(4\pi )^{D/2}}
     \left\{\left(\dfrac{q^2}{4}-\sigma _{\rm sol}^2\right)
       \int _0^1 dx\left[\sigma _{\rm sol}^2-x(1-x)q^2\right]^{D/2-2}
+\sigma _{\rm sol}^{D-2}
      \right\}
\hspace{2em}
\label{eq:(A.21a)}
  \\
\lefteqn{
  iD_{\pi \pi }^{-1}(-q^2)-iD_{\pi \pi }^{-1}(0)
} \nonumber\\
  &=&
  8N \dfrac{\Gamma (2-D/2)}{(4\pi )^{D/2}}
     \left(\dfrac{q^2}{4}\right)
       \int _0^1 dx\left[\sigma _{\rm sol}^2-x(1-x)q^2\right]^{D/2-2}.
\label{eq:(A.21b)}
\end{eqnarray}
\end{subequations}
The second derivative of the effective potential gives the auxiliary
field propagator at zero momentum:
\begin{subequations} \label{eq:(A.22)}
\begin{eqnarray}
  iD_{\sigma \sigma }^{-1}(-q^2=0)
  &=& - \left. \dfrac{\partial^2}{\partial\sigma ^2} V(\sigma ,\pi )
      \right|_{\sigma =\sigma _{\rm sol},\pi =0},
\\
  iD_{\pi \pi }^{-1}(-q^2=0)
  &=& - \left. \dfrac{\partial^2}{\partial\pi ^2} V(\sigma ,\pi )
      \right|_{\sigma =\sigma _{\rm sol},\pi =0}.
\end{eqnarray}
\end{subequations}
For the symmetric vacuum $\sigma _{\rm sol}=0$ we find
\begin{equation}
 iD_{\sigma \sigma }^{-1} (-q^2=0) = iD_{\pi \pi }^{-1} (-q^2=0) =
   -\dfrac{4N}{(4\pi )^{D/2}} \dfrac{\Lambda ^{D-2}}{\Gamma (D/2)}
   \left(\dfrac{1}{g}-\dfrac{1}{g^*}\right).
\label{eq:(A.23)}
\end{equation}

Combining \EQ{eq:(A.21)} with \EQ{eq:(A.23)}, we find
\begin{subequations} \label{eq:(A.24)}
\begin{eqnarray}
  -\dfrac{(4\pi )^{D/2}}{4N} \Gamma (D/2) iD_{\sigma \sigma }^{-1}
(-q^2)
  &=& \left(\dfrac{1}{g}-\dfrac{1}{g^*}\right)\Lambda ^{D-2}
+ \dfrac{\coeCd}{2-D/2}(-q^2)^{D/2-1},
  \\
  -\dfrac{(4\pi )^{D/2}}{4N} \Gamma (D/2) iD_{\pi \pi }^{-1}(-q^2)
  &=& \left(\dfrac{1}{g}-\dfrac{1}{g^*}\right)\Lambda ^{D-2}
+ \dfrac{\coeCd}{2-D/2}(-q^2)^{D/2-1},
\end{eqnarray}
\end{subequations}
with $\coeCd$ defined by
\begin{equation}
  \coeCd\equiv \dfrac{B(3-D/2,D/2-1)}{\Gamma (D-1)}.
\end{equation}
For the S$\chi$SB  vacuum $\sigma _{\rm sol}\not=0$ in the
strong coupling region $g>g^*$,  the auxiliary field propagators at
zero-momentum are evaluated as
\begin{equation}
  iD_{\sigma \sigma }^{-1}(-q^2=0)
  = -\dfrac{8N}{(4\pi )^{D/2}}\Gamma (2-D/2)\sigma _{\rm sol}^{D-2},
  \qquad
  iD_{\pi \pi }^{-1}(-q^2=0) = 0,
\end{equation}
which lead to
\begin{subequations}
\begin{eqnarray}
  -\dfrac{(4\pi )^{D/2}}{4N}\Gamma (D/2)iD_{\sigma \sigma }^{-1}(-q^2)
  &=& 2 \left(\sigma _{\rm sol}^2-\dfrac{q^2}{4}\right)
        \int _0^1 dx\left[\sigma _{\rm sol}^2-x(1-x)q^2\right]^{D/2-2},
        \hspace{2em}
  \\
  -\dfrac{(4\pi )^{D/2}}{4N}\Gamma (D/2)iD_{\pi \pi }^{-1}(-q^2)
  &=& 2 \left(-\dfrac{q^2}{4}\right)
        \int _0^1 dx\left[\sigma _{\rm sol}^2-x(1-x)q^2\right]^{D/2-2}.
\end{eqnarray}
\end{subequations}
Thus, the auxiliary field $\sigma$ acquires a pole at
$q^2=4\sigma_{\rm sol}^2=4m^2$, while $\pi$ becomes massless
NG boson.  Note that $m_\sigma=2m$ is independent of $D$.

\section{OPE in \NJLd\ model} \label{sec-appO}
Let us consider the time ordered bilocal operator
$T\left[\psi(x) \bar \psi(0) \right]$.
The OPE for it reads
\begin{equation}
  -i \FT \TOP{\psi(x)\bar\psi(0)}
      =   c_{\unit}(p;g_R, m_R, \mu ) \unit
        + c_{\bar\psi\psi}(p;g_R, m_R, \mu)
         \left[ (\bar\psi\psi)_R +\gamma_5(\bar\psi\gamma_5\psi)_R \right]
      + \cdots,
\label{eq:(B.1)}
\end{equation}
It is useful to expand $c_{\unit}$ by $m_R$:
\begin{equation}
  c_{\unit}(p; g_R, m_R, \mu )
  = \frac{\fsl{p}}{p^2} + m_R c_{\unit}'(p; g_R ; \mu ) + \cdots.
\end{equation}
Corresponding to \EQ{eq:(6.6)}, we have
\begin{equation}
  \Sigma_{\rm explicit}(-p^2)
  = p^2 m_R(\mu) c_{\unit}'(p; g_R ; \mu ) + \cdots.
\label{eq:(B.**)}
\end{equation}
 From \EQ{eq:(A.17)} and \EQ{eq:(B.8)} we obtain
\begin{equation}
  \Sigma_{\rm explicit}(-p^2)
  = \begin{cases}
        \dfrac{1}{D-2} \dfrac{Z_m m_R}{\frac{g}{g^*}-1}
     + \cdots
        & for \SxSB\ vacuum \cr\cr
     \dfrac{Z_m m_R}{1-\frac{g}{g^*}} + \cdots
        & for symmetric vacuum
    \end{cases},
\label{eq:(B.9)}
\end{equation}
where $Z_m \equiv m_0/m_R$ is given by \EQ{eq:(5.7)}
\begin{equation}
  Z_m \equiv \dfrac{m_0}{m_R}
      = \dfrac{g}{g_R} \left( \dfrac{\mu}{\Lambda} \right)^{D-2}.
\label{eq:(B.10)}
\end{equation}

\EQ{eq:(B.10)} and \EQ{eq:(5.6)} lead to
\begin{equation}
  \dfrac{Z_m}{1-\frac{g}{g^*}} = \dfrac{1}{1-\frac{g_R}{g_R^*}}.
\label{eq:(B.***)}
\end{equation}
Comparing \EQ{eq:(B.**)} with \EQ{eq:(B.9)} and \EQ{eq:(B.***)}, we obtain
\footnote{This agrees with Ref.\cite{kn:KY90} except for the factor
$1/(D-2)$ in the \SxSB\ vacuum.}
\begin{equation}
  c_\unit'(p; g_R, \mu)
    = \begin{cases}
        \dfrac{1}{p^2} \dfrac{1}{D-2} \dfrac{1}{\frac{g_R}{g_R^*}-1}
        & for \SxSB\ vacuum \cr\cr
        \dfrac{1}{p^2} \dfrac{1}{1-\frac{g_R}{g_R^*}}
        & for symmetric vacuum
      \end{cases}.
\end{equation}
Note that the Wilson coefficient $c'_\unit$ in the \SxSB\ vacuum
remains finite in two dimensions, while $c'_\unit$ calculated in the
symmetric vacuum vanishes in $D\rightarrow2$.
This is consistent with the one phase structure in two dimensions.

Let us next consider the coefficient function of
$\bar\psi\psi$, $c_{\bar\psi\psi}$.
This can be calculated from the fermion four-point function
by taking $x\rightarrow 0$ limit:
\begin{equation}
  -i\VEV{ T[\psi(x) \bar\psi(0) \psi(y) \bar\psi(z)]}_{\rm connected}
   = c_{\bar \psi \psi }(x)
     \VEV{T (\bar \psi \psi )_R(0) \psi (y) \bar \psi (z)}_{\rm connected}
    + \cdots.
\label{eq:(B.12)}
\end{equation}
The Wilson coefficients of the unit operator do not appear
in the RHS of \EQ{eq:(B.12)}, since they only contribute
to disconnected diagrams.

In the following calculation $x,y,z$ are Fourier transformed to
$p,q,k$, respectively.
The calculation at $q=k$ is sufficient to determine $c_{\bar\psi\psi}$.
The LHS of \EQ{eq:(B.12)} is evaluated:
\begin{equation}
  S(p) i\Gamma_S(-p^2) S(p) \dfrac{1}{- V''(\sigma_{\rm sol})} i \Gamma_S(-q^2)
 = \frac{1}{p^2} \dfrac{1}{- V''(\sigma_{\rm sol})} \Gamma_S(-q^2) + \cdots,
\label{eq:(B.13)}
\end{equation}
where the legs for $q,k$ are amputated and
the scalar vertex is defined by $\Gamma_S \equiv 1$.

The RHS of \EQ{eq:(B.12)} can be evaluated by Fig.\ref{fig:OPE}
which reads
\begin{eqnarray}
\lefteqn{
 Z_m c_{\bar\psi\psi}(p; g_R, m_R, \mu) \left[
    1 -i V_{\rm (qu)}''(\sigma_{\rm sol}) D_{\sigma\sigma}(0)
  \right] \Gamma_S(-q^2)
} \nonumber \\
 & &
 = -c_{\bar\psi\psi}(p; g_R, m_R, \mu)
  V''_{\rm (cl)}(\sigma_{\rm sol}) \dfrac{Z_m}{-V''(\sigma_{\rm sol})}
  \Gamma_S(-q^2),
\label{eq:(B.14)}
\end{eqnarray}
where in the last line we have used \EQ{eq:(A.22)}.

By comparing \EQ{eq:(B.13)} and \EQ{eq:(B.14)}, we obtain the
$c_{\bar \psi \psi }$ \cite{kn:KY90}:
\begin{eqnarray}
  c_{\bar\psi \psi }(p; g_R, m_R=0, \mu )
    &=& -\dfrac{1}{p^2}
        \frac{Z_m^{-1}}{V''_{\rm (cl)}(\sigma_{\rm sol} )}
    \nonumber\\
    &=& - \dfrac{1}{p^2} \dfrac{(4\pi )^{D/2} \Gamma (D/2) }{4N}
          \dfrac{g_R}{\mu ^{D-2}}
\label{eq:(B.15)}
    \\
    &=& - \dfrac{1}{p^2} \dfrac{G_R}{N},
    \nonumber
\end{eqnarray}
where $G_R$ is defined by
$G_R\equiv (4\pi )^{D/2}\Gamma (D/2) g_R \mu ^{2-D}/4$
and we have used
\[
  \dfrac{1}{V_{\rm (cl)}''} = \frac{G}{N}
  = \dfrac{(4\pi)^{D/2} \Gamma(D/2)}{4N} \dfrac{g}{\Lambda^{D-2}}.
\]

\section{Renormalization Group Functions from the Solution of Ladder SD Gap
Equation}\label{sec-appB}
Let us consider the RG  of {\em bare} parameters
which was first discussed by Miransky \cite{kn:Mira85} in the analysis of the
ladder SD equation of dynamical mass of fermion in the quenched
QED\@.
In this argument the bare parameters of the theory are required to
depend on the cutoff $\Lambda $ so as to fix the fermion dynamical mass
$M_d$,
with such a flow of bare parameters being identified as the
RG evolution.
\par
Solving the ladder SD gap equation of the gauged NJL model,
Kondo, Mino and Yamawaki \cite{kn:KMY89}
and independently Appelquist, Soldate, Takeuchi and Wijewardhana
\cite{kn:ASTW88}
obtained the critical line
\begin{equation}
 g = g^{*} \equiv {1 \over 4}(1+\omega)^2,
\ \omega = \sqrt{1-\alpha/\alpha_c},
\end{equation}
and the scaling relation near the critical line:
\begin{equation}
  \dfrac{M_d}{\Lambda } \sim
  \begin{cases}
     \left(-\dfrac{C_1}{D_1} \dfrac{\frac{1}{g^*}-\frac{1}{g}}
                                {\frac{1}{\tilde g^*}-\frac{1}{g}}
                                \right)^{1/2\omega }
   & ($0<\alpha<\alpha_c$)
   \\
     \exp \left[1+\delta_0 - \dfrac{8}{\frac{1}{g^*}-\frac{1}{g}} \right]
   & ($\alpha=\alpha_c$)
   \\
      \exp \left[ \delta
       - \dfrac{ n\pi+\tan^{-1}\omega' +\tan^{-1}({\omega'/2 \over
                                      g-(1-\omega'^2)/4})}{\omega'}
     \right]
   & ($\alpha>\alpha_c$)
  \end{cases},
\label{eq:(C.2)}
\end{equation}
with $n=1$ being the ground state solution,
where $\tilde g^*$ is defined by $\tilde g^* \equiv (1-\omega )^2/4$.

The usual SD equation \cite{kn:BLL86} is obtained from the stationary
condition of $V[\Sigma,\Sigma_5]$ \EQ{eq:(2.24)}:
\begin{eqnarray}
0 &=& \dfrac{\delta}{\delta\Sigma(\Ep)}V[\Sigma,\Sigma_5],
\label{eq:(C.3)}
  \\
0 &=& \dfrac{\delta}{\delta\Sigma_5(\Ep)}V[\Sigma,\Sigma_5],
\label{eq:(C.4)}
\end{eqnarray}
which read
\begin{subequations} \label{eq:(C.5&6)}
\begin{eqnarray}
 \Sigma(\Ep) &=& m_0
    +{g \over \Lambda^2}
 \int  _0^{\Lambda^2} d\Ep \dfrac{\Ep
\Sigma(\Ep)}{\Ep+\Sigma^2+\Sigma_5^2}
+\int  _0^{\Lambda^2}d\Ek\dfrac{\Ek \Sigma(\Ek)}{\Ek+\Sigma^2+\Sigma_5^2}
K(\Ep,\Ek),
\label{eq:(C.5)}
  \\
 \Sigma_5(\Ep) &=&
  {g \over \Lambda^2}\int  _0^{\Lambda^2} d\Ep \dfrac{\Ep
\Sigma_5(\Ep)}{\Ep+\Sigma^2+\Sigma_5^2}
+\int  _0^{\Lambda^2}d\Ek \dfrac{\Ek \Sigma_5(\Ek)}{\Ek+\Sigma^2+\Sigma_5^2}
K(\Ep,\Ek).
\label{eq:(C.6)}
\end{eqnarray}
\end{subequations}
Since $\Sigma_5$ can be rotated away by the chiral symmetry,
it is sufficient to consider the SD equation
\begin{eqnarray}
 \Sigma(\Ep) &=& m_0
    +{g \over \Lambda^2}
 \int  _0^{\Lambda^2} d\Ep \dfrac{\Ep
\Sigma(\Ep)}{\Ep+\Sigma^2}
+\int  _0^{\Lambda^2}d\Ek\dfrac{\Ek \Sigma(\Ek)}{\Ek+\Sigma^2}
K(\Ep,\Ek),
\label{eq:(C.7)}
\end{eqnarray}
which is equivalent to the differential equation
\begin{equation}
  \left[
     \Ep \left(\dfrac{d}{d\Ep}\right)^2
   + 2\dfrac{d}{d\Ep}
    +\frac{3C_F}{4\pi} \dfrac{\alpha}{\Ep+\Sigma^2(\Ep)}
  \right]\Sigma(\Ep) = 0,
\label{eq:(C.8)}
\end{equation}
with IR BC:
\begin{equation}
  \lim_{\Ep \rightarrow 0} p_{\scriptscriptstyle E}^4
  \dfrac{d}{d\Ep} \Sigma(\Ep) = 0,
\label{eq:(C.9)}
\end{equation}
and UV BC:
\begin{equation}
  \left.\left[1+\left(1+{g \over 3C_F\alpha/4\pi}\right)
  \Ep \dfrac{d}{d\Ep} \right]\Sigma(\Ep)\right|_{\Ep=\Lambda^2}
= m_0.
\label{eq:(C.10)}
\end{equation}
Irrespective of presence or absence of the explicit fermion mass $m_0$, the
solution of this differential equation exhibits the same asymptotic behavior in
the
high energy region:
\begin{equation}
  \Sigma(\Ep) = M \left[
       c_1  \left(\dfrac{\Ep}{M^2}\right)^{-(1-\omega )/2}
      +d_1  \left(\dfrac{\Ep}{M^2}\right)^{-(1+\omega )/2}
      + {\cal O} (\left(\dfrac{\Ep}{M^2}\right)^{-3(1-\omega)/2-1})
      \right],
\label{eq:(C.11)}
\end{equation}
where this time $M$ is not a field but a constant having
the dimension of mass and gives the mass scale of the solution.
For details on the behavior of this solution, see section~2.
\par
First of all, we calculate the mass renormalization constant $Z_m$
{\it \'{a} la} Miransky \cite{kn:Mira85}.
 From the UV BC, the equation of state is obtained:
\begin{equation}
 {m_0 \over \Lambda} = - \left({M \over \Lambda} \right)^2
\left[ {g-g^* \over g^*} C_1
  \left({M \over \Lambda} \right)^{-\omega}
+ {g-\tilde g^* \over \tilde g^*} D_1
  \left({M \over \Lambda} \right)^{\omega} \right].
\end{equation}
Then we obtain
\begin{equation}
 {\partial m_0 \over \partial M} = - \left({M \over \Lambda} \right)
\left[ (2-\omega){g-g^* \over g^*} C_1
  \left({M \over \Lambda} \right)^{-\omega}
+ (2+\omega){g-\tilde g^* \over \tilde g^*} D_1
  \left({M \over \Lambda} \right)^{\omega} \right].
\end{equation}
In the chiral limit $m_0=0$, the dynamical fermion mass $M_d$ obeys
the scaling law \EQ{eq:(C.2)}
\begin{equation}
 {M_d \over \Lambda}
= \left[- {\tilde g^* C_1 \over g^* D_1}
{g-g^* \over g-\tilde g^*} \right]^{{1 \over 2\omega}},
\ (0<\alpha<\alpha_c).
\end{equation}
Then the renormalization constant in the region $0<\alpha<\alpha_c$ is obtained
as
\begin{equation}
 Z_m \equiv  {\partial m_0 \over \partial M}\Big|_{M=M_d}
= 2 {M \over \Lambda}
\sqrt{ - {\omega^2 C_1 D_1 \over \tilde g^* g^*} (g-\tilde g^*)(g-g^*)},
\end{equation}
Thus we obtain the mass renormalization constant:
\begin{equation}
 Z_m  =
    \begin{cases}
       2 \tA \dfrac{\sqrt{(g-\tilde g^*)(g-g^*)}}{\sqrt{1-\omega^2}}
       \dfrac{M_d}{\Lambda}
     & ($0<\alpha<\alpha_c$)
     \\
       2 \tA_0 (g-{1 \over 4}) \dfrac{M_d}{\Lambda}
     & ($\alpha=\alpha_c$)
     \\
       2 \tA'\dfrac{\sqrt{(g-\bar g^c)(g-g^c)}}{\sqrt{1+\omega'^2}}
       \dfrac{M_d}{\Lambda}
     &($\alpha>\alpha_c$)
\end{cases},
\label{eq:(C.16)}
\end{equation}
where $g^c \equiv (1+i\omega')^2/4$ and $\bar g^c \equiv (1-i\omega')^2/4$.

Since the bare chiral condensation $\VEV{\bar \psi \psi}$ is given by
\begin{eqnarray}
  \VEV{\bar \psi \psi}
  &\equiv& - \frac{N}{4\pi^2} \int_0^{\Lambda^2} d\Ep
                              \dfrac{\Ep \Sigma(\Ep)}{\Ep+\Sigma^2(\Ep)}
  \nonumber\\
  &=& \begin{cases}
         -\dfrac{N \tA}{4\pi^2\sqrt{1-\omega^2}}
          \dfrac{\Lambda M_d^2}{ \left[(g-\tilde g^*)(g-g^*) \right]^{1/2}}
      & ($0<\alpha<\alpha_c$)
      \\
         - \dfrac{N\tA_0}{4\pi^2} \dfrac{\Lambda M_d^2}{g-{1 \over 4}}
      & ($\alpha=\alpha_c$)
      \\
        -\dfrac{N \tA'}{4\pi^2\sqrt{1+\omega'^2}}
          \dfrac{\Lambda M_d^2}{\left[(g-\bar g^c)(g-g^c) \right]^{1/2}}
     & ($\alpha>\alpha_c$)
      \end{cases},
\label{eq:(C.17)}
\end{eqnarray}
the renormalized condensate $\VEV{(\bar \psi \psi)_R}$
near the critical line is calculated as
\begin{equation}
 \VEV{(\bar \psi \psi)_R} = Z_m \VEV{\bar \psi \psi} =
    \begin{cases}
       - \dfrac{N\tA^2}{2\pi^2(1-\omega^2)} M_d^3
     & ($0<\alpha<\alpha_c$)
     \\
       - \dfrac{N\tA_0^2}{2\pi^2} M_d^3
     & ($\alpha=\alpha_c$)
     \\
       - \dfrac{N\tA'^2}{2\pi^2(1-\omega^2)} M_d^3
     &($\alpha>\alpha_c$)
     \end{cases}.
\label{eq:(C.18)}
\end{equation}

 This shows that $\VEV{(\bar\psi \psi)_R}$ is $g$-independent and the
RG flow can be identified with the fixed-$\alpha$ line
(upward direction)\cite{kn:KMY89,kn:NSY89,kn:KHKD90}.

Once the RG flow is so identified,
the $\beta$ function of bare four-fermion coupling was explicitly calculated
from \EQ{eq:(C.2)} in the gauged NJL model \cite{kn:BLL89,kn:IMO89}:
\begin{equation}
  \beta _g(g,\alpha) \equiv \Lambda {\partial g \over \partial \Lambda}
\Big|_{\alpha,M_d}
= -2 (g-\tilde g^*)(g-g^*), \ (g > g^{*}).
\label{eq:(C.19)}
\end{equation}
The anomalous dimension is obtained \cite{kn:MY89}:
\begin{equation}
  \gamma _m(g,\alpha)
\equiv - \Lambda {\partial \ln Z_m \over \partial \Lambda} \Big|_{\alpha,M_d}
 \equiv - \Lambda {\partial \ln m_0 \over \partial \Lambda} \Big|_{\alpha,M_d}
=  1+\omega, \ (g = g^{*}).
\label{eq:(C.20)}
\end{equation}
Now we calculate the anomalous dimension above the critical line $g>g^*$ from
the
solution of the SD equation. The scaling law leads to
\begin{equation}
   g-\tilde g^* = {1 \over 1-F} (g^*-\tilde g^*), \quad
   g-g^*        = {F \over 1-F} (g^*-\tilde g^*),
\end{equation}
with $F$ being defined by
\[
  F \equiv -{g^* \over \tilde g^*} {D_1 \over C_1}
          \left({M_d \over \Lambda}\right)^{2\omega}.
\]
Then we get
\begin{equation}
{\partial \ln (g-\tilde g^*) \over \partial \ln \Lambda}
= -{\Lambda \over M_d}{\partial \ln (1-F) \over \partial ({\Lambda \over M_d})}
= -2\omega {F \over 1-F}
= -2\omega {g-g^* \over g^*-\tilde g^*} = - 2(g-g^*).
\end{equation}
Similarly, we get
\begin{equation}
{\partial \ln (g-g^*) \over \partial \ln \Lambda}
= {\Lambda \over M_d}{\partial [\ln F - \ln (1-F)] \over \partial
({\Lambda \over M_d})} = -2\omega-2\omega {F \over 1-F}
= -2\omega - 2(g-g^*).
\end{equation}
Accordingly, we obtain the anomalous dimension:
\begin{eqnarray}
 \gamma_m
  &\equiv& - {\partial \ln Z_m \over \partial \ln \Lambda}
  = 1 -{1 \over 2}
       {\partial \over \partial \ln \Lambda}[\ln (g-\tilde g^*)+\ln (g-g^*)]
  \nonumber\\
 &=& 1+\omega+2(g-g^*)=2g+\dfrac{\alpha }{2\alpha _c}.
\label{eq:(C.24)}
\end{eqnarray}
This is shown to be valid also in the region
$\alpha>\alpha_c$ and coincides with the anomalous dimension numerically
obtained in the gauged NJL-model with running gauge coupling \cite{kn:KSY91}.

\section{Renormalization Group of Bare Parameters}\label{sec-appC}
In the previous appendix, we have defined and calculated the RG
function of bare parameters. It is evident, however, that
such a definition of RG function is applicable only
in the $S\chi SB$ vacuum of strong
coupling phase $g>g^{*}$ where fermion acquires non-vanishing
dynamical   mass.

Thus, we define here the RG of the bare parameters
by using the effective potential expressed in terms of mass scale
of the fermion $M$.
This enables us to obtain the RG functions also in
the symmetric phase.

We first consider the case of $m_0=0$.
The effect of explicit chiral symmetry breaking term will be discussed
later.
In section~2, we have obtained the effective potential:
\begin{eqnarray}
  - \frac{8\pi^2}{N} \dfrac{V(M)}{\Lambda ^4}
  &=& \left(\dfrac{1}{g^*}-\dfrac{1}{g}\right)C_1\left[
        C_1\left(\dfrac{M}{\Lambda }\right)^{4-2\omega }
+\dfrac{2+\omega }{2} D_1\left(\dfrac{M}{\Lambda }\right)^4
       \right]
  \nonumber\\
  & & +\left(\dfrac{1}{\tilde g^*}-\dfrac{1}{g}\right)D_1\left[
        D_1\left(\dfrac{M}{\Lambda }\right)^{4+2\omega }
+\dfrac{2+\omega }{2} C_1\left(\dfrac{M}{\Lambda }\right)^4
       \right].
\label{eq:(D.1)}
\end{eqnarray}
Actually, the stationary condition of \EQ{eq:(D.1)}
$\partial V/\partial M=0$
has a nontrivial solution $M=M_d \not=0$ for strong coupling region
$g>g^*$ and leads to the correct scaling relation:
\begin{equation}
  \left(\dfrac{M_d}{\Lambda }\right)^{2\omega }
      = -\dfrac{C_1}{D_1} \dfrac{\frac{1}{g^*}-\frac{1}{g}}
                                {\frac{1}{\tilde g^*}-\frac{1}{g}}.
\label{eq:(D.2)}
\end{equation}

Let us define the RG flow of the bare four-fermion coupling $g$
so as to make \EQ{eq:(D.1)} independent of $\Lambda $:
\footnote{It is evident, however, that the effective potential
\EQ{eq:(D.1)} cannot be made $\Lambda$-independent for all the
region of $M$. Thus, we need to specify the value of $M$ to define the
RG flow of bare parameters.
The plausible choice of $M$ is the value of the stationary condition.
In this sense,
the RG flow of the bare parameters depends on the
choice of vacuum.}
\begin{eqnarray}
  0 = -\dfrac{8\pi^2}{N\Lambda^3} \dfrac{d V(M)}{d\Lambda }
  &=& 2\omega \left\{\left(\dfrac{1}{g^*}-\dfrac{1}{g}\right)C_1^2
\left(\dfrac{M}{\Lambda }\right)^{4-2\omega }
       -\left(\dfrac{1}{\tilde g^*}-\dfrac{1}{g}\right)D_1^2
\left(\dfrac{M}{\Lambda }\right)^{4+2\omega }
          \right\}
  \nonumber\\
  & & + \dfrac{\beta _g}{g^2} \left(\dfrac{M}{\Lambda }\right)^4
         \left[C_1 \left(\dfrac{M}{\Lambda }\right)^{-\omega }
+ D_1 \left(\dfrac{M}{\Lambda }\right)^{\omega }\right]^2,
\end{eqnarray}
where $\beta_g$ is defined by
$
 \beta_g(g,\alpha) =  \Lambda {\partial g \over \partial \Lambda}.
$
Namely, the $\beta$ function reads
\begin{equation}
  \beta _g(g,\alpha )
  = -2\omega g^2
       \dfrac{
         \left(\dfrac{1}{g^*}-\dfrac{1}{g}\right)
         C_1^2 \left(\dfrac{M}{\Lambda }\right)^{-2\omega }
        -\left(\dfrac{1}{\tilde g^*}-\dfrac{1}{g}\right)
         D_1^2 \left(\dfrac{M}{\Lambda }\right)^{2\omega }
       }{\left[
           C_1 \left(\dfrac{M}{\Lambda }\right)^{-\omega }
         + D_1 \left(\dfrac{M}{\Lambda }\right)^{\omega }
         \right]^2}.
\label{eq:(D.4)}
\end{equation}

We first consider the symmetric vacuum $M=0$.
In this case, \EQ{eq:(D.4)} becomes
\begin{equation}
  \beta _g(g,\alpha ) = 2\omega g\left(1-\dfrac{g}{g^*}\right),
\label{eq:(D.5)}
\end{equation}
which corresponds to the {\em phase-independent}
RG evolution of the {\em renormalized} four-fermion coupling
 discussed in section~5.

Another choice of $M$  is  the solution of the gap equation
\EQ{eq:(D.2)} in the strong
coupling region $g>g^*$, which gives
\begin{equation}
  \beta _g(g,\alpha ) = -2(g-\tilde g^*)(g-g^*).
\label{eq:(D.6)}
\end{equation}
\EQ{eq:(D.6)} agrees with the original definition
of the RG flow of the bare parameters
\EQ{eq:(C.19)}.

These two results \EQ{eq:(D.5)} and \EQ{eq:(D.6)} give
the same result for the region $g\simeq g^*$ where $M$ is sufficiently
smaller than cutoff $\Lambda $.
On the other hand, the deviation becomes significant
when the four-fermion coupling $g$ is far beyond the critical line $g^*$
and $M/\Lambda $ in \EQ{eq:(D.2)} is not negligible.

For the anomalous dimension $\gamma _m$, we need to evaluate the
explicit chiral symmetry breaking term
proportional to the bare mass of fermion $m_0$ in the effective
potential: \begin{equation}
  V_{\rm explicit}(M) \equiv - {N \Lambda^2 \over 4\pi^2} {m_0 \sigma \over g}.
\end{equation}
The result is:
\begin{equation}
  - \frac{8\pi^2}{N} \dfrac{V_{\rm explicit}(M)}{\Lambda ^4} =
    \dfrac{2m_0}{\Lambda g}
    \left[C_1\left(\dfrac{M}{\Lambda }\right)^{2-\omega }
+D_1\left(\dfrac{M}{\Lambda }\right)^{2+\omega }\right].
\end{equation}
The anomalous dimension
$\gamma _m m_0 = -\Lambda \partial m_0 / \partial \Lambda $
can be obtained in the same way as the $\beta$ function;
${\partial V_{\rm explicit} \over \partial \Lambda}=0$.
We find
\begin{equation}
  \gamma _m(g,\alpha )=
    1 + \omega \dfrac{
             \left(2\frac{g}{g^*}-1\right)C_1^2
\left(\dfrac{M}{\Lambda}\right)^{-2\omega }
            -\left(2\frac{g}{\tilde g^*}-1\right)D_1^2
\left(\dfrac{M}{\Lambda }\right)^{2\omega }
           }{\left[C_1\left(\dfrac{M}{\Lambda }\right)^{-\omega }
+D_1\left(\dfrac{M}{\Lambda }\right)^{\omega }\right]^2}.
\label{eq:(D.9)}
\end{equation}
Again, the definition of the anomalous dimension $\gamma _m$ depends
on the choice of $M$.
The symmetric vacuum $M=0$ gives the anomalous dimension
\begin{equation}
  \gamma _m(g,\alpha )= 1-\omega + 2\omega \dfrac{g}{g^*},
\label{eq:(D.10)}
\end{equation}
which takes the same form as \EQ{eq:(5.22)}, the anomalous
dimension for the renormalized coupling in the symmetric renormalization.
It should be  noted that
the anomalous dimension $\gamma _m$ is continuous across
the critical line $g=g^*$.
On the other hand, using the value of $M$ in S$\chi$SB
 vacuum \EQ{eq:(D.2)}, we find
\begin{equation}
  \gamma _m(g,\alpha ) = 2g + \dfrac{\alpha }{2\alpha _c}.
\label{eq:(D.11)}
\end{equation}
\EQ{eq:(D.11)} for $g>g^{*}$ can also be derived by use of the SD gap
equation, as shown in  Appendix~\ref{sec-appB}.
\EQ{eq:(D.10)} and \EQ{eq:(D.11)}
give the same anomalous dimension
\EQ{eq:(C.20)} at $g=g^{*}$.

\newpage
\begin{singlespace}

\end{singlespace}
\newpage
\section*{Figure Captions}
\begin{enumerate}
\item   Critical line in $(\alpha,g)$ plane.
        It separates the spontaneously broken phase (\SxSB) and
        the unbroken phase ($\chi$Sym) of the chiral symmetry.
\label{fig:phase}
\item
        The lowest order diagram in the two-particle irreducible part
        $\kappa^{\rm 2PI}[S]$ of the CJT potential of
        the gauged NJL model written in terms of
        auxiliary field Eq.(2.2).
        The wavy line and the solid line with shaded blob represent the
        bare gauge boson propagator $D_{\mu\nu}$ and the full fermion
        propagator $S$, respectively.
\label{fig:CJT}
\item   The lowest order diagram in the two-particle irreducible part of
        the CJT potential of the gauged NJL model (without auxiliary fields),
        Eq.(2.1).
\label{fig:CJT2}
\item   Momentum assignment of Yukawa type vertex $\Gamma_S(p,q)$.
        Dashed line represents auxiliary field propagator.
\label{fig:yukawa}
\item   Amputated multi-point Green function at zero momentum of $\sigma$.
        External fermion lines are amputated.
        The shaded blob stands for the induced $\sigma$ vertex
        ($V^{(n)}(\sigma)$).
\label{fig:amputated}
\item   Auxiliary field propagator.
        The solid line with shaded blob represents the full fermion propagator.
\label{fig:auxprop}
\item   The second derivative of the auxiliary field propagator.
        A slash represents once derivative with respect to $q_\mu$.
\label{fig:auxprop2}
\item   Self-consistent equation for the Yukawa-type vertex $\Gamma_S(p,q)$.
\label{fig:yukawaSD}
\item   Another form for the second derivative of the auxiliary field
propagator.
\label{fig:auxprop3}
\item   Four-point fermion Green function:
        (a) $\sigma$-exchange diagram,
        (b) pure ladder diagram.
\label{fig:fourfunc}
\item   $(\bar\psi\psi)$ inserted Green function.
\label{fig:OPE}
\item   RG functions in the symmetric and the $\tM$-dependent renormalizations:
        (a) $\beta$ function,
        (b) anomalous dimension.
        Solid line is for the symmetric renormalization ($\tM=0$), while
        dotted line is for $\tM=M_d$.
\label{fig:rgefunc}
\end{enumerate}
\end{document}